\documentclass[aps,prl,a4paper,10pt,twocolumn,showpacs,longbibliography,floatfix,superscriptaddress,amsmath,amsfonts,amssymb,preprintnumbers,nofootinbib]{revtex4-2}
\usepackage{graphicx}
\usepackage{calc}
\usepackage{array}
\usepackage{amsmath, amssymb}
\usepackage{natbib}
\usepackage{hyperref}
\usepackage{overpic}
\usepackage{comment}
\usepackage[dvipsnames]{xcolor}

\DeclareMathOperator{\dd}{\mathrm{d}\!}

\DeclareMathOperator{\ii}{\mathrm{i}}

\newcommand{\ket}[1]{\left| #1 \right>}
\newcommand{\bra}[1]{\left< #1 \right|}

\begin{document}

\title{Quantum size effects on Andreev transport in Nb/Au/Nb Josephson junctions: \\
A combined ab-initio and experimental study}

\author{Hiroki Yamazaki}
\affiliation{RIKEN, Wako, Saitama 351-0198, Japan}

\author{G\'abor Csire}
\email{csire.gab@gmail.com}
\affiliation{Materials Center Leoben Forschung GmbH, Roseggerstraße 12, 8700 Leoben, Austria}
\affiliation{Wigner Research Centre for Physics, Institute for Solid State Physics and Optics, H-1525 Budapest, Hungary}
\affiliation{H.H. Wills Physics Laboratory, University of Bristol, Tyndall Avenue, BS8-1TL, United Kingdom}

\author{N\'ora Kucska}
\affiliation{Wigner Research Centre for Physics, Institute for Solid State Physics and Optics, H-1525 Budapest, Hungary}

\author{Nic Shannon}
\affiliation{H.H. Wills Physics Laboratory, University of Bristol, Tyndall Avenue, BS8-1TL, United Kingdom}
\affiliation{Theory of Quantum Matter Unit, Okinawa Institute of Science and Technology Graduate University,
Onna-son, Okinawa 904-0495, Japan}

\author{Bal\'azs Gy\H{o}rffy}
\affiliation{H.H. Wills Physics Laboratory, University of Bristol, Tyndall Avenue, BS8-1TL, United Kingdom}

\author{Hidenori Takagi}
\affiliation{Department of Physics, The University of Tokyo, Bunkyo, Tokyo 113-8656, Japan}
\affiliation{Max-Planck-Institute for solid state research, Heisenbergstrasse 1, 70569 Stuttgart,
Germany}

\author{Bal\'azs \'Ujfalussy}
\affiliation{Wigner Research Centre for Physics, Institute for Solid State Physics and Optics, H-1525 Budapest, Hungary}

\date{\today}
\begin{abstract}
We have measured the critical current density, superconducting coherence length, and 
superconducting transition temperature of single-domain, epitaxially-grown Nb(110)/Au(111)/Nb(110) trilayers, 
all of which show a non-monotonic dependence on the thickness of the Au layer.
These results are compared with the predictions of a relativistic, ab-initio theory, which incorporates 
superconducting correlations.
We find good agreement with experiment, coming from a rich interplay between superconducting proximity-- 
and quantum size effects, mediated by Andreev bound states.  
These results suggest that quantum size effects could provide a systematic method of controlling 
the transport properties of superconducting multilayers.
\end{abstract}

\maketitle

The characteristic feature of superconductors (SCs) is the macroscopic
occupation of bound pairs of electrons, known as Cooper
pairs, in the same quantum-coherent ground state
described by a complex superconducting order parameter (SOP).
In homogeneous $s$-wave SCs this is accompanied by a fully gapped quasiparticle spectrum.
However, if the SOP varies in space, lower energy, subgap excitations may be formed.
These, so called Andreev Bound States (ABS), are the consequence of
Andreev scattering in which an incoming electron-like state can convert
into an outgoing hole-like one\cite{Andreev1964}.
The classical example of such systems is the superconductor - normal metal - superconductor
(SNS) junctions which is characterized by a non-zero phase difference between the SOP belonging to the two superconductors.
It was predicted that the ABS depends strongly on the phase difference between the SCs \cite{Furusaki1991, Sauls2018}
and it is responsible for the flow of the Josephson supercurrent\cite{Josephson1962}.
Josephson junctions are among the most active areas of research 
in superconducting spintronics\cite{Buzdin2005, Linder2015, Eschrig2015, amundsen2022colloquium, Nadeem2023, Mercaldo2023}.
They are mostly driven by the rapid progress in growth and characterization techniques\cite{Guo2004, Bao2005, Yamazaki2006, Yamazaki2010}.
However, little attention is paid to the actual effects of the realistic band structure,
which is likely to be very significant for short junctions where 
there is a quantum confinement of electrons in the growth direction.
If the characteristic size of the metal is of the order of the bulk Fermi wavelength,
the in-gap band structure (and consequently the Fermi surface as well) splits into a series of subbands, known as quantum-well states (QWSs), leading to multiband superconductivity.

In this paper we show that, in epitaxial Nb(110)/Au(111)/Nb(110) trilayers, the critical current density, the superconducting coherence length and the superconducting transition temperature are all oscillating as a function of the Au-layer thickness.
Although
0-$\pi$ phase transitions (meaning that the phase of the SOP is inverted between two superconducting leads) are usually connected to a similar oscillatory behavior,
we show that these oscillations are intrinsic to the proximitized QWS in the Au(111) layer.
We develop a detailed microscopic theory based on the Kohn-Sham-Dirac-Bogoliubov-de~Gennes (KS-DBdG) equations, which allow us to quantitatively predict the superconducting properties including the quasiparticle spectrum and the Josephson supercurrent by involving
the details of the electronic structure together with spin-orbit coupling (SOC).
This theory reveals that, although the Au layer is disorder-free in the ab-initio description, 
the layer-dependent electronic structure causes extra scattering - similarly to impurities - and transforms the Josephson supercurrent from the ballistic to the tunneling limit.
Moreover, it predicts the spontaneous appearance of a non-uniform phase, 
which can be associated with crossing QWS near the Fermi level, and yet again is the consequence of multiband behavior.

The starting point of our study 
is a series of experiments performed on Nb(110)/Au(111)/Nb(110) trilayer samples with various Au thicknesses ($t_{\textrm{Au}}$)
prepared on single crystals of Al$_2$O$_3 (1 1 \overline{2} 0)$ using molecular-beam-epitaxy.
One atomic gold monolayer (1~ML) corresponds to $t_{\textrm{Au}} =0.2355$~nm.
Epitaxial layer-by-layer growth occurred without structural change for  $0 < t_{\textrm{Au}} \leq 2.10$~nm.
The details of the sample preparation and structural characterization can be found in the Supplemental Material~\cite{SupMat}.
For the superconducting transition temperature,
$T_c^\textrm{resistive}$ shall be defined as the temperature at 50\% of the residual resistivity,
and $T_c^\textrm{magnetic}$ is as the onset point of the diamagnetic transition.
Typical temperature dependencies of normalized resistivity and of normalized magnetic susceptibility can be found in the Supplemental Material~\cite{SupMat}.

\begin{figure}[tb]
    \centering
    \includegraphics[width=0.75\linewidth]{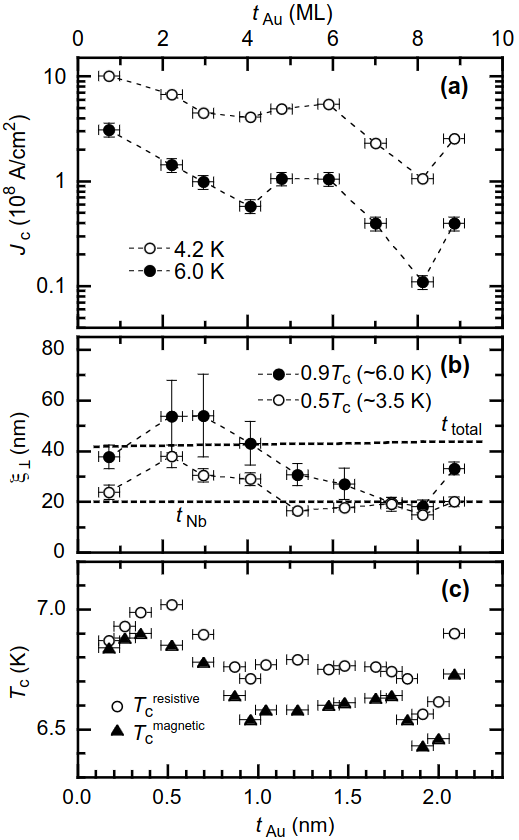}
    \caption{
    Superconducting properties of epitaxially-grown Nb/Au/Nb trilayers as a function of 
    Au-layer thickness, $t_{\sf Au}$.
    (a) Critical current density $J_c$ in zero field.  %at $0~\text{Oe}$.
    (b) Perpendicular coherence length $\xi_\perp$. 
    (c) Transition temperature $T_c$.  
    A single atomic monolayer (ML) of Au corresponds to $\Delta t_{\textrm{Au}} = 0.2355~\text{nm}$. 
    The thickness of the Nb layers, $t_\textrm{Nb} = 20.0~\text{nm}$, and the thickness 
    of the entire sample, $t_\textrm{total} = 20.0 + t_{\textrm{Au}} + 20.0 + 1.74~\text{nm}$, 
    are indicated with dashed lines in (b).
    Vertical error bars on $\xi_\perp$ are reflect uncertainty in measurements of 
    $H_{c2}^\perp$ and $H_{c2}^\parallel$.
    }
    \label{fig:experiment}
\end{figure}

The average electron mean free paths \mbox{$l_{\textrm{Au}} = 140 \pm 40$}~nm and 
\mbox{$l_{\textrm{Nb}} = 3 \pm 1$}~nm were calculated from the residual resistivity as a function of $t_{Au}$
by performing a theoretical fit using a parallel register model. 
We also estimated the superconducting coherence lengths $\xi^{\parallel}$ and $\xi^{\perp}$
from the measurements of the upper critical fields $H_{c2}^\perp$  and $H_{c2}^\parallel$ 
using the expressions:
\mbox{$\xi^{\parallel} = \sqrt{\Phi_0 / (2 \pi H_{c2}^\perp)}$} and
\mbox{$\xi^{\perp} = \Phi_0 / (2 \pi \xi^{\parallel} H_{c2}^\parallel)$}
where $\Phi_0$ is the flux quantum and the directions, $\parallel$ and $\perp$, 
are referred to with respect to the surface. 
The superconducting coherence length of the Nb layer extrapolated to zero Kelvin was found to be 
$\xi^{\parallel}_\textrm{Nb}(T=0~\textrm{K}) \approx \xi^{\perp}_\textrm{Nb}(T=0~K) \approx 20$~nm.

This result highlights a key fact about the Nb/Au/Nb trilayers: 
while the geometry of the sample is layered and highly anisotropic, its superconducting properties 
are governed by Nb layers of thickness greater than or equal to the correlation length, 
and are therefore fully three--dimensional.
As theoretical calculations (described below) confirm, the Au layer is fully proximitized, 
and therefore also three--dimensional in its superconductivity.
Consequently, the magnetic properties of the trilayer closely resemble those
of an isotropic, bulk superconductor.
Indeed, the detailed theoretical calculations presented in the Supplemental Material\cite{SupMat} (Sec.~V) indicate that the anisotropy in the critical current density is exceedingly small and effectively negligible for practical considerations.

%%%%%%%%%%%%%%%%%%%%%%%%%%%%

Taken together, these results imply that the charge transport through the Au spacer layer is in the clean limit \mbox{($l_{\textrm{Au}} \gg t_{\textrm{Au}}$)} and the superconductivity
of the Nb layers is in the dirty limit \mbox{($\xi_{\textrm{Nb}}(T=0~\textrm{K}) > l_{\textrm{Nb}}$)}. 
We also point out that the sample total
thickness, \mbox{$t_{\textrm{total}} =20.0+t_{\textrm{Au}} +20.0+1.74$~nm} (excluding the substrate) 
shown in Fig.~\ref{fig:experiment}(b), is comparable to the field-penetration depth 
of \mbox{$\lambda(T=0~K) \approx 39$}~nm in bulk Nb.
The superconducting critical current density $J _c$ in zero magnetic field was estimated 
from the $M$-$H$ hysteresis curve based on the Bean model~\cite{SupMat}.
Since superconductivity of the trilayer is three--dimensional, and in the dirty limit $\xi > l_{Nb}$, 
$J_c$ can be taken to be isotropic, and the simplest version of the Bean model applies, leading 
to the results shown in Fig.~\ref{fig:experiment}(a).

For samples with $t_{\textrm{Au}} < 0.5$~nm ($\approx$2~ML) disorder was 
more pronounced, however in case of samples with $t_{\textrm{Au}} \gtrapprox$~2~ML
over a range of temperatures $\sim$3.5~K$<T<T_c$, the
temperature dependence of $J_c$ is well-described by the expression $J_c(T)=J_c(0) (1-T/T_c)^\alpha$, with
$\alpha=1.4 \pm 0.1$. This is in good agreement with the exponent $\alpha=1.39$ obtained for critical current
at 0~Oe in an 120~nm thick Nb film\cite{Pan1998}. Ginzburg-Landau theory predicts $\alpha=3/2$ near $T_c$~\cite{Rusanov2004}. 
The values of \mbox{$J_c=(0.1-10)\times 10^8\ \textrm{A/cm}^2$} seen in Fig.~\ref{fig:experiment}(a) 
are extremely high, relative to the SC/FM/SC (FM: ferromagnet) systems 
studied so far\cite{Buzdin2005}.

The proximity effect in these Nb(110)/Au(111)/Nb(110) trilayers also displays a number
of features. As shown in Fig.~\ref{fig:experiment}, neither $J_c$, nor $\xi^\perp$ or $T_c$ are smooth, monotonic functions of the
thickness $t_{\textrm{Au}}$. In particular, there are sharp features at $t_{\textrm{Au}}=0.94$~nm ($\approx$4 ML) and 1.88~nm
($\approx$8 ML). 
We wish to emphasize that, at these thicknesses, there are no changes in crystal structure and
the observed modulation in $J_c(t_{\textrm{Au}})$ 
and $T_c(t_{\textrm{Au}})$ is therefore intrinsic to Nb(110)/Au(111)/Nb(110) trilayers. Such systematic, non-monotonic
variations in $J_c$ and $T_c$ as functions of the nonmagnetic-layer thickness have not been reported before,
and we infer that a series of epitaxially grown multilayer systems of equivalently high quality are
necessary for it to be observed.

\begin{figure*}[htb]
    \begin{overpic}[width=1.\linewidth]{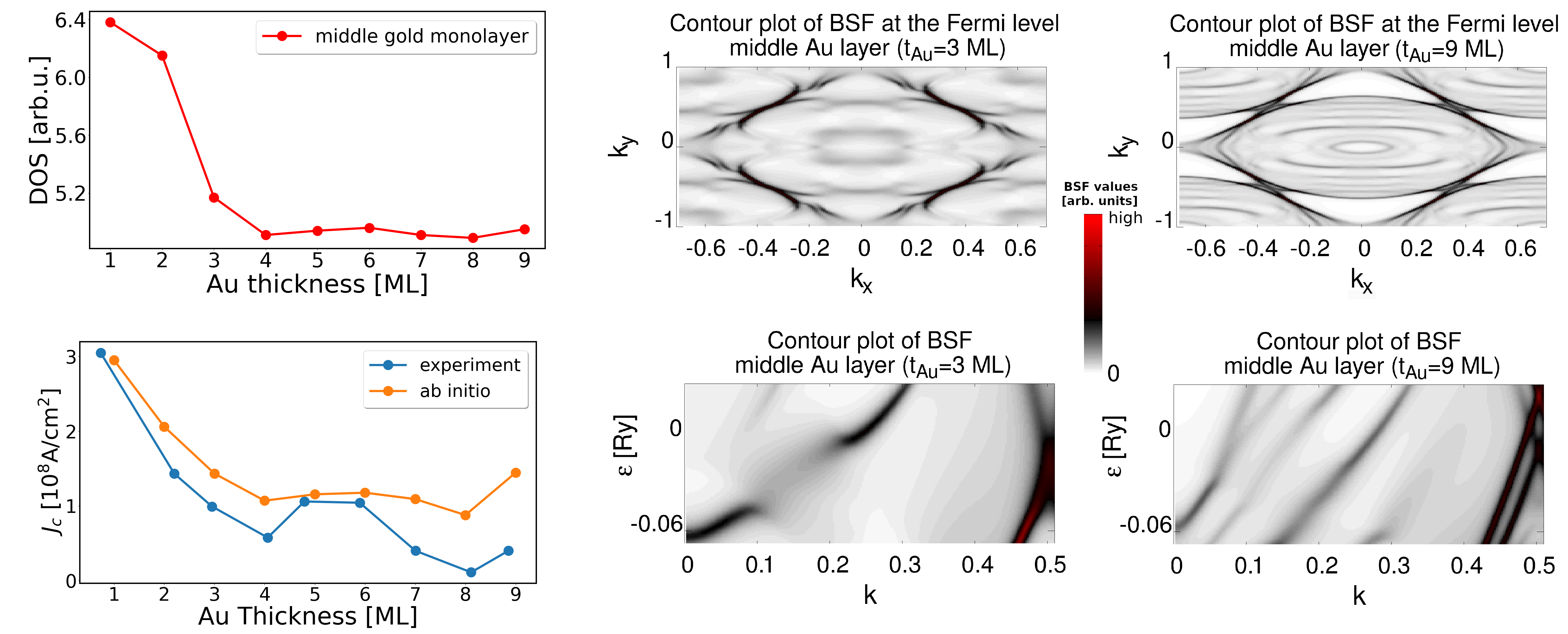}
        \put (0,39) {\normalsize \textbf{(a)}}
        \put (38,39) {\normalsize \textbf{(b)}}
        \put (71,39) {\normalsize \textbf{(c)}}
        \put (0,18) {\normalsize \textbf{(d)}}
        \put (38,18) {\normalsize \textbf{(e)}}
        \put (71,18) {\normalsize \textbf{(f)}}
    \end{overpic}
    \caption{\textbf{The relationship of the normal state electronic structure to the critical current}
    \textbf{a)} The middle Au monolayer's Density of States (DOS) at the Fermi level as a function of the Au's thickness.
    \textbf{b-c)} Fermi surface of the middle Au monolayer for 3 ML and 9 ML Au thickness, respectively. Details of the layer-dependent electronic structure, which are fully incorporated in the theory, are provided in the Supplemental Material\cite{SupMat}.
    \textbf{d)} Critical current as a function of the Au's thickness at $T=6$~K compared to experimental results.
    \textbf{e-f)} Contour plot of Bloch Spectral Function (BSF) along the high-symmetry line $k_x=\sqrt{2} k_y$ ($k$ is measured in $\pi/a$ units)
                    for 3 ML and 9 ML Au thickness, respectively.}
    \label{fig:bandstruct}
\end{figure*}

\begin{figure*}[htb]
    \begin{overpic}[width=0.95\linewidth]{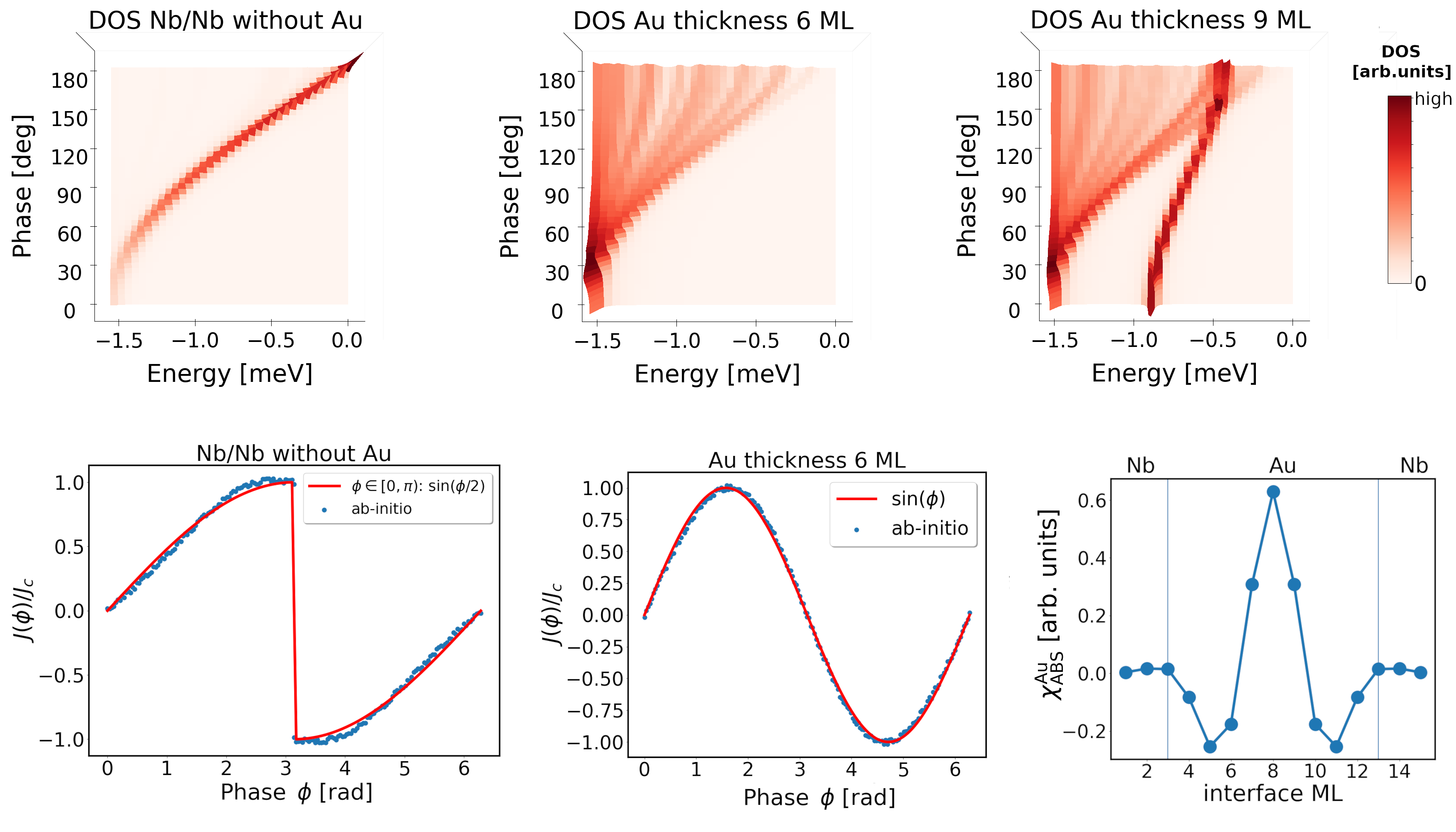}
        \put (1,55) {\normalsize \textbf{(a)}}
        \put (35,55) {\normalsize \textbf{(b)}}
        \put (66,55) {\normalsize \textbf{(c)}}
        \put (0,25) {\normalsize \textbf{(d)}}
        \put (36,25) {\normalsize \textbf{(e)}}
        \put (70,25) {\normalsize \textbf{(f)}}
    \end{overpic}
    \caption{\textbf{Josephson tunneling in Nb/Au/Nb junctions.}
    \textbf{a-c)} Quasiparticle Density of States as a function of the phase difference $\phi$ between the two Nb lead
                  for different Au thicknesses: 0~ML, 6~ML, 9~ML, respectively.
    \textbf{d-e)} Normalized Josephson current at $T=0$~K as a function of the phase difference $\phi$ for 
                  for 0~ML, 6~ML Au thickness and corresponding analytical fit for the ballistic and tunneling limit, respectively.
    \textbf{f)} The layer dependence of the anomalous density of states (energy resolved singlet pairing amplitude) for the Andreev bound state ($\phi=0$)
                shown in plot \textbf{c)}.
      }
    \label{fig:josephson}
\end{figure*}

On the one hand, theoretically, transport phenomena are usually studied in terms of a quasiclassical approach which
typically involves either Eilenberger equations\cite{Eilenberger1968}
for ballistic systems or Usadel equations\cite{Usadel1970} to describe the limit of strong disorder.
On the other hand, the quasi-classical picture is valid only if the Fermi wavelength is much smaller than any other length scale of the problem.  
Because certain information has been integrated over in the equations, they represent a less sophisticated theory compared to the BdG equations.
Moreover, it has recently been shown that the first-principles treatment of the electronic structure is necessary even for understanding the
supercurrent decay in magnetic Josephson junctions~\cite{Ness2022}.
Therefore, to properly account for the realistic electronic structure including the multiorbital nature of the problem and simultaneously account for spin-orbit coupling on the same footing,
we solve the following fully relativistic KS-DBdG Hamiltonian written in Rydberg units
\begin{equation}
H_{\text{DBdG}}=
 \begin{pmatrix}
   H_D & \Delta_{\text{eff}} \\
   \Delta_{\text{eff}}^\dagger & -H_D^*
 \end{pmatrix},
\end{equation}
where $ H_D(\vec r)=c\vec{\boldsymbol{\alpha}} \vec{p} + \left( \boldsymbol{\beta}-\mathbb{I}_4 \right) c^2/2+ \left( V_\text{eff}(\vec r)-E_F \right) \mathbb{I}_4 + \vec{\boldsymbol{\Sigma}}\vec{B}_\text{eff}(\vec r)$, with
$\vec{\boldsymbol \alpha} = \boldsymbol \sigma_x \otimes \vec{\boldsymbol \sigma}$, 
$\boldsymbol \beta = \boldsymbol \sigma_z \otimes \mathbb I_2$, 
$\vec{\boldsymbol \Sigma} = \mathbb I_2 \otimes \vec{\boldsymbol \sigma}$, 
$\vec{\boldsymbol \sigma} $ denotes the Pauli-matrices, 
and $\mathbb I_n $ being the identity matrix of order $n$. $V_{\text{eff}}(\vec r)$ and $\vec{B}_\text{eff}(\vec r)$  are the effective potential and the exchange field, respectively. $\Delta_\text{eff}(\vec r)$ is the effective $4 \times 4$ pairing potential matrix due to the four component Dirac spinors. 
The KS-DBdG equations should be solved self-consistently, which, for layered systems can be achieved most efficiently by the Screened Korringa-Kohn-Rostoker Green's function (SKKR-GF) method\cite{Csire2015, Csire2018}
by assuming that the superconducting host has isotropic $s$-wave spin-singlet pairing.
The description of the formalism and computational details can be found in the Supplemental Material\cite{SupMat}.

In order to execute the above program, first we carried out normal state band structure calculations using the layered SKKR-GF method for various thicknesses of the Au layer (0-12~ML)
between two semi-infinite Nb. As shown in Fig.~\ref{fig:bandstruct} the electronic structure contains several interesting and very relevant features.
Based on the DOS shown in Fig.~\ref{fig:bandstruct}(a) we argue that the electronic structure changes the most up to 4~ML Au thickness due to charge transfer at the interface.
Figure~\ref{fig:bandstruct}(b-c, e-f) show the Fermi surfaces and band structure of the middle Au monolayer for 3 and 9 ML Au thicknesses along a high symmetry line in the Brillouin zone (BZ). The figure nicely demonstrates that QWSs are formed with increasing thickness and the appearance of the Au's relative band gap in Fig.~\ref{fig:bandstruct}(c).
These QWSs are standing electron waves quantized in the direction perpendicular to the
surface, while parallel to the surface the bands typically show a free electron-like dispersion.
When a new subband related to this QWS behavior intersects the Fermi level,
the electronic properties can change quite dramatically.
We mention that the oscillatory behavior of the superconducting properties due to QWS has been reported for
superconducting lead thin films\cite{Guo2004} (including the critical temperature\cite{Shanenko2007} and the critical field\cite{Bao2005, Wjcik2014}). 
Another important observation is the appearance of a Dirac-like state in the middle of the Au layer at the edge of the BZ which has an intersection point
for $t_{\textrm{Au}}\geq$9~ML on the Fermi surface. 
In principle, the electronic structure suggests that interband scattering mechanisms can be at play, 
including specular interband Andreev reflection\cite{Beenaker2006} and
interband pairing leading to Fulde-Ferrel-Larkin-Ovchinnikov (FFLO)\cite{Fulde_1964, Larkin1964} type non-uniform phase\cite{Wojcik2019}.

One can also observe that some states in the Au layer are smeared out: this happens in the $k$-space
regions where the DOS is high in the Nb, since the appropriate electrons can scatter more easily onto the other side
of the interface. Another interesting problem is the effect of SOC.
One may expect a Rashba-type SOC and corresponding splitting of the states due to
the interfaces of Nb/Au/Nb\cite{Yakovkin2018, Yakovkin2020}. However, no SOC-induced band splitting can be observed in the band structure around the Fermi level. This would clearly not be the case for a pure surface, where the symmetry breaking would be more apparent. 
We argue that the existence of a surface/interface itself in an inversion symmetric junction is not a sufficient condition to lift Kramers’ degeneracy. In fact, free-standing Au(111) slabs have been reported to not show band splitting in the repeated-slab model, since there is no asymmetry
between the two surfaces\cite{Yakovkin2018, Yakovkin2020}.

Finally, by solving the KS-DBdG equations we obtained the quasiparticle spectrum for various Au thicknesses (0-12~ML) which in turn allows the quantitative prediction of current-phase relationship (CPR) corresponding to the Josephson supercurrent based on the formula \cite{Beenakker1992} written in Ry units:
\begin{equation}
    J(\phi) = - 2\sqrt{2} \sum_{n > 0} \int_{\textrm{BZ}} \!\!\! 
                     \dd  \mathbf{k_\parallel} \frac{\partial \varepsilon_n(\mathbf{k_\parallel, \phi}) }{ \partial \phi} 
                     \tanh \frac{\varepsilon_n( \mathbf{k_\parallel, \phi}) }{ 2k_B T}.
\end{equation}
The critical current is associated with the maximum of the Josephson supercurrent that can exist through the junction. 
In a real experimental situation, one passes a current across the Josephson junction which reacts by changing the phase difference between the superconducting leads.
The values obtained for the critical current $J_c$ are shown in Fig.~\ref{fig:bandstruct}(d).
One can see that there is a good qualitative agreement between the results obtained by the first-principles-based calculations
and those by experimental study. In what follows, we further analyze the features of the superconducting state to explain the obtained behavior.

In general, CPR can be expanded in Fourier harmonics as $J(\phi) = \sum_n J_n \sin (n\phi)$.
Although in many cases, the CPR can be reasonably described by the first harmonics, the sinusoidal shape is not always preserved.
At a microscopic level, the supercurrent through a short junction is determined by bound states forming in the confinement due to Andreev reflection.
In the short junction limit assuming scattering events in the conduction channel with a
transmission $t$, the Andreev levels are given by\cite{Beenakker1991} $\varepsilon_n(\phi) = \pm \Delta \sqrt{1-t \sin^2(\phi/2)}$
which yields the well-known sinusoidal shape\cite{Sauls2018}.
In fact, the sinusoidal shape was obtained in the tunneling limit (meaning the transmission $t << 1$) first by Ambegaokar\cite{Ambegaokar1963}:
$J(\phi)=J_c \sin (\phi)$ with $J_c=\pi \Delta/eR_N$ where $R_N$ is the normal state resistivity.
However, in the ballistic limit ($t \rightarrow 1$) at 0~K, the Josephson current\cite{Kulik1978} is
$J(\phi)=2J_c \sin (\phi/2) \mathrm{sign}(\sin \phi)$ which has twice as large amplitude compared to the tunneling limit, and
it shows an anomalous discontinuity at $\phi=\pi$.

The quasiparticle DOS and the supercurrents as a function of $\phi$ are shown in Fig.~\ref{fig:josephson}(a-e).
In Fig.~\ref{fig:josephson}(a) we show a hypothetical scenario where two semi-infinite Nb with a phase difference $\phi$
are touched together. The quasiparticle spectrum shows a well-defined $\phi$-dependent ABS 
recovering the $t \rightarrow 1$ result: $\varepsilon_n(\phi) = \pm \Delta \cos(\phi/2)$ and
yields the supercurrent (Fig.~\ref{fig:josephson}(d)) showing discontinuity at $\phi=\pi$ at 0~K which corresponds to the analytical result of Ambegaokar\cite{Ambegaokar1963}.
In Fig.~\ref{fig:josephson}(b) the quasiparticle DOS is shown for an Nb/Au/Nb junction with $t_{\textrm{Au}}=6$~ML.
Striking contrast can be observed when it is compared to the case where only two semi-infinite Nb were joined together.
Several ABS can be observed changing with $\phi$ and in fact they form an almost continuous spectrum within the superconducting gap region.
Although the Au layers can be considered in the clean limit (absence of impurities),
the different band structures of Nb and Au can cause a similar effect as the impurities
with varying strength of transmission $t$ for different wavenumbers $\mathbf{k}_\parallel$.
We stress that this picture is also supported by the findings in Ref.~\onlinecite{Ness2022} where it was
shown that the electronic structure has similar effects on the FFLO state as expected for impurities.
Corresponding to this picture (tunneling limit), in Fig.~\ref{fig:josephson}(e) the supercurrent ($t_{\textrm{Au}}=6$~ML) shows the usual sinusoidal shape.
Therefore, we argue that the anomalous changes in the superconducting properties around $t_{\textrm{Au}}=4$~ML in Fig.~\ref{fig:experiment}
are related to the rapid change of the electronic structure from the ballistic to the tunneling limit
(see also the Supplemental Material\cite{SupMat}) which causes a drop in the critical current, superconducting transition temperature, and coherence length.
For $t_{\textrm{Au}} \leq 2$~ML interface disorder occurred in the region of $\sim$2 ML thickness between the lower-Nb and the Au layer, therefore,
the $T_c$ increases slightly up to  $\approx$2 ML as the disorder continuously eliminated with growing thickness.
However, as we have seen in Fig.~\ref{fig:bandstruct} the DOS at the Fermi level significantly decreases up to $\sim$4 ML
(corresponding to the change from the ballistic to the tunneling limit) which lowers the critical temperature since $T_c \sim \exp{-1/\Lambda D(E_F)}$ where $\Lambda$ is the electron-phonon interaction and 
$D(E_F)$ is the DOS at the Fermi level. Similarly, the changes in the measured perpendicular coherence length also fit this theory.
In the ballistic limit (up to 4~ML) one experiences larger coherence length compared to the tunneling limit
(for $t_{\textrm{Au}} \leq 2$~ML interface disorder tends to drive the system into the tunneling limit).

Interestingly, in Fig.~\ref{fig:experiment} the experimental $J_c$ shows an anomalous behavior at $t_{\textrm{Au}}=8-9$~ML as well.
This feature can be explained based on the quasiparticle DOS for $t_{\textrm{Au}}=9$~ML 
in Fig.~\ref{fig:josephson}(c), where an additional feature can be observed: even at $\phi=0$ a well-defined striking ABS is present which also changes in energy with the
variation of $\phi$ giving a kick to the critical current (shown in Fig.~\ref{fig:bandstruct}(d)). This new ABS is present only for $t_{\textrm{Au}} \geq 9$~ML in the calculations.
We argue that this ABS is related to the QWS obtained at the edge of the BZ (see Fig.~\ref{fig:bandstruct}(f)),
where several QWSs get close to each other and crossings are observed in the vicinity of the Fermi level.
To better understand the nature of this state the normalized energy-resolved singlet order parameter (details in the Supplemental Material\cite{SupMat})
is shown Fig.~\ref{fig:josephson}(f): one can clearly observe the non-uniform structure
(change in sign within the interface region) with the maximum value obtained in the middle of the Au layer.
This is a clear signature of the spontaneous appearance of Cooper pairs with nonzero center-of-mass
momentum in the Au layer which was also suggested in Ref.~\onlinecite{Wojcik2019}.
The interesting feature of this state is that it is formed without any magnetic field in the system.
Therefore, the discontinuous behavior obtained around $t_{\textrm{Au}}=8-9$~ML is related to the appearance of a co-existing
non-uniform superconducting phase in the Au layer causing larger critical current and transition temperature.

In summary we have studied the effect of the electronic structure on the proximity effect in Nb(110)/Au(111)/Nb(110) trilayers
as a function of the Au layer thickness. We have demonstrated
similar behavior of the superconducting properties between the experimental results and the ab-initio based BdG theory
which emphasizes the crucial role of the electronic structure in understanding the physics of superconducting heterostructures.

\emph{Acknowledgement} $-$
The authors are pleased to acknowledge helpful conversations with James Annett, J\'ozsef Cserti, Robert Joynt, Mario Cuoco and Ilya Vekhter.
N.S. acknowledges support through the Theory of Quantum Matter Unit, OIST, and EPSRC Grant No. EP/C539974/1.
G.Cs. acknowledges support from Austrian Science Fond (FWF) project No. 33491-N “ReCALL”.
K.N. and B.U. acknowledge financial support by the National Research, Development, and Innovation Office (NRDI Office) of Hungary under Project Nos. K131938 and K142652.

\bibliography{josephson}
%\bibliography{main}

\clearpage
\appendix
\section*{Supplementary Material}
\setcounter{section}{0} % optional, to reset numbering

% Optional: reset figure/table counters
\renewcommand{\theequation}{S\arabic{equation}}
\setcounter{equation}{0}
\renewcommand{\thefigure}{S\arabic{figure}}
\setcounter{figure}{0}
\renewcommand{\thetable}{S\arabic{table}}
\setcounter{table}{0}

\section{Sample preparation and analysis}

\begin{figure}[htb]
    \centering
    \includegraphics[width=0.375\linewidth]{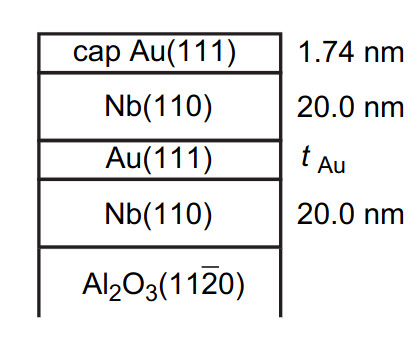}
    \caption{Schematic diagram of a vertical section of the sample and layer thicknesses.}
    \label{fig:schematic}
\end{figure}

Nb(110)/Au(111)/Nb(110) trilayers were prepared on single crystals of Al$_2$O$_3 (1 1 \overline{2} 0)$ using molecular-beam-epitaxy (MBE). 
A schematic diagram of the sample structure is shown in Fig.~\ref{fig:schematic} Details of preparation,
structural characterization and measurements are essentially the same as those for
Nb(110)/Au(111)/Fe(110)\cite{Yamazaki2006} and Nb(110)/Au(111)/Co(0001)\cite{Yamazaki2010}.
First, a Nb layer of 20.0~nm($=t_{\textrm{Nb}}$) was grown on the substrate at $T_s =400^\circ$C ($T_s$: the substrate temperature). The
condition $T_s \leq 450^\circ$C keeps the oxygen of the Al$_2$O$_3$ substrate from diffusing into the Nb layer\cite{Srgers1992}.
Subsequently, a Au layer with a thickness of $t_{\textrm{Au}}=$0.17-4.00~nm and a Nb layer with a
thickness of $t_{\textrm{Nb}}$ were deposited at $T_s=200^\circ$C. The trilayer was finally capped with a Au layer
of 1.74~nm at $T_s=100^\circ$C in order to avoid oxidization. One atomic monolayer (1~ML) of
Au(111) corresponds to $t_{\textrm{Au}}=$0.2355~nm.

\begin{figure}[htb]
    \centering
    \includegraphics[width=0.75\linewidth]{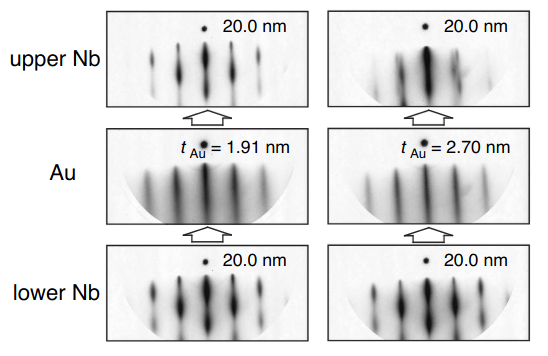}
    \caption{Reversal images of RHEED patterns obtained in the growth process of the
Nb/Au[$t_{\textrm{Au}}$]/Nb trilayer for $t_{\textrm{Au}}$=1.91~nm (left column; typical of the samples for
$0.40\leq t_{\textrm{Au}} \leq 2.10$~nm) and $t_{\textrm{Au}}$=2.70~nm (right column; typical of the samples for $t_{\textrm{Au}} \geq$2.25~nm).
The direction of the incident electron beam is parallel to $<1\overline{1}0>$ of the Nb(110) layer.}
    \label{fig:rheed}
\end{figure}

\begin{figure}[htb]
    \centering
    \includegraphics[width=0.75\linewidth]{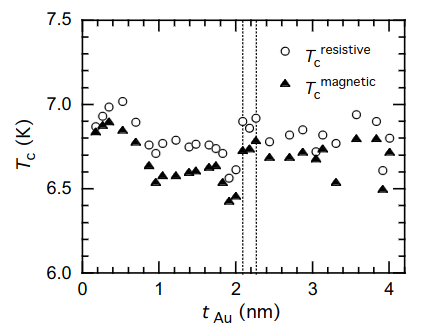}
    \caption{Superconducting transition temperature $T_c$: $T_c^\textrm{resistive}$ and $T_c^\textrm{magnetic}$ as a function of $t_{\textrm{Au}}$,
where $T_c^\textrm{resistive}$ was defined as the temperature at 50\% of the residual resistivity and $T_c^\textrm{magnetic}$
was as the onset point of the diamagnetic transition. The error of $T_c$ is within each symbol.
The broken lines are drawn at 2.10 and 2.25~nm, indicating a region of structural transition
for the upper Nb layer.}
    \label{fig:tc}
\end{figure}

\begin{figure}[htb]
    \centering
    \includegraphics[width=0.75\linewidth]{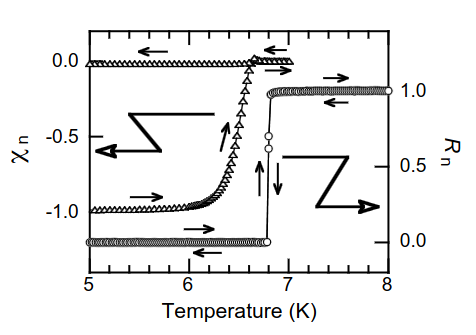}
    \caption{Typical temperature dependences of normalized magnetic susceptibility $\chi_n (H=0.1$~Oe, H $\perp$ surface)
    and of normalized resistivity $R_n (H=0$~Oe) for the $t_{\textrm{Au}}=$1.48 nm sample. A
    transition width of 0.04~K (10-90\% criterion) is seen for $R_n$. Magnetic susceptibility
    measurements were performed after zero-field cooling, using a superconducting quantum
    interference device magnetometer (Quantum Design MPMSXL). Resistivity measurements
    were carried out in a standard four-terminal configuration using low-frequency ac technique.}
    \label{fig:susceptibility}
\end{figure}

During the sample growth, reflection high-energy electron diffraction (RHEED) patterns
were measured for the surface of each layer. The patterns in the left and right columns of
Fig.~\ref{fig:rheed} are typical of the samples for $0.40\leq t_{\textrm{Au}} \leq 2.10$~nm and $t_{\textrm{Au}} \geq 2.25$~nm, respectively.
The Au-layer surface presented fine streak pattern for $t_{\textrm{Au}} \geq 0.40$~nm, indicating that epitaxial
layer-by-layer growth occurs from the early stages of the Au-layer growth. For the upper Nb
layer, the RHEED pattern changed from that of single domain ($0\leq t_{\textrm{Au}} \leq 2.10$~nm) to that of
twinned domain ($t_{\textrm{Au}} \geq 2.25$~nm) structure with a narrow interval (2.10-2.25~nm) of transition.
This structural change is due to the lattice mismatch between Au and Nb, and the structure of
the upper Nb layer depends entirely upon $t_{\textrm{Au}}$. The twinning is accompanied by a $\pm 120^\circ$
rotation of the in-plane $<001>$ axis around the $<110>$ axis that is perpendicular to the
surface. Off-axial X-ray diffraction measurements support this result. For the samples of
$t_{\textrm{Au}}>2.10$~nm, therefore, it is due to the presence of the twinned domains (i.e., a form of
structural disorder) in the upper Nb layer that we cannot see a systematic change of
the superconducting transition temperature $T_c$
($T_c^\textrm{resistive}$ and $T_c^\textrm{magnetic}$) as a function of $t_{\textrm{Au}}$ (see Fig.~\ref{fig:tc}). Here $T_c^\textrm{resistive}$ is defined as the
temperature at 50\% of the residual resistivity, and $T_c^\textrm{magnetic}$ is as the onset point of the
diamagnetic transition. Typical temperature dependences of normalized resistivity $R_n (H=0$~Oe)
and of normalized magnetic susceptibility $\chi_n(H=0.1~\textrm{Oe}$, H $\perp$ surface) are shown in Fig.~\ref{fig:susceptibility}.
In the article we will focus on the single-domain samples ($0 < t_{\textrm{Au}} \leq 2.10$~nm) whose
superconducting properties are uniquely determined by $t_{\textrm{Au}}$.

\begin{figure}[htb]
    \centering
    \includegraphics[width=0.75\linewidth]{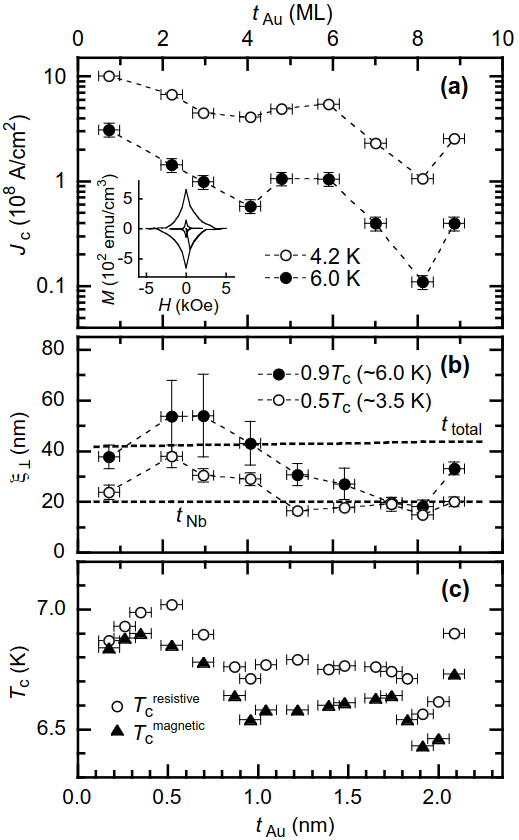}
    \caption{Superconducting properties as functions of $t_{\textrm{Au}}$ for the Nb/Au[$t_{\textrm{Au}}$]/Nb trilayers: (a)
critical current density $J_c$ at 0~Oe , (b) perpendicular coherence length $\xi^\perp$ (the vertical error
bars are due to the uncertainty in $H_{c2}^\perp$ and $H_{c2}^\parallel$), and (c) transition temperature $T_c$. 
One atomic monolayer (ML) corresponds to $t_{\textrm{Au}}$=0.2355~nm. In (b), the horizontal tNb
line ($\xi^\perp$=20.0~nm) corresponds to the thickness of a single Nb layer; the inclined $t_\textrm{total}$ line
shows a thickness of the whole sample (excluding the substrate): $t_\textrm{total}=20.0+t_{\textrm{Au}}+20.0+1.74$~nm.
Inset in (a): Typical $M-H$ curves ($H \parallel$ sample plane) measured at 4.2~K (outer) and 6~K (inner) for the
$t_{\textrm{Au}}=1.13$~nm ($\sim$5 ML) sample. Diamagnetic contribution from the substrate has been subtracted.}
    \label{fig:experiment}
\end{figure}

The superconducting critical current density $J_c$ in zero magnetic field was estimated from the width 
($\Delta M$) 
of the M-H hysteresis curve for $H$ parallel to the sample plane by using the 
result $\Delta M = J_c t /20$ derived for a Bean model of a long slab of thickness $t$\cite{Gyorgy1989}.
Here, $J_c$ is measured in A/cm$^2$, $t_{\textrm{total}}$ is taken for $t$ [cm], and $\Delta M$ [gauss] is
measured at 0~Oe.
Our results for $J_c$ are shown in Fig.~\ref{fig:experiment}(a), together with a typical $M-H$
curve (inset).
% in a sample with $t_{\textrm{Au}}=1.13$~nm ($\sim$5 ML). 
The vertical error bars on $J_c$ reflect the
fact that the magnetization decays exponentially to a saturation value in a time of order $10^2$
sec when the system is held at 0~Oe, with a $\sim$1\% ($\sim$10\%) reduction at 4.2~K (6.0~K).
The values of $J_c=(0.1-10)\times 10^8 \textrm{A/cm}^2$ seen in Fig.~\ref{fig:experiment}(a) are extremely high.
In fact, these $J_c$ values are close to the depairing
current density $J_{\textrm{dp}}$ observed in Nb thin films using transport measurements\cite{Rusanov2004}. As
theoretical studies have shown that, in parallel fields, the $J_c$ in films with thickness $\leq \lambda$ can
be comparable to $J_{\textrm{dp}}$\cite{Carneiro1998, Stejic1994, Mawatari1994}.
Especially in our samples, for $|H| \rightarrow 0$ ($H^\parallel$ film plane), vortices
are trapped in and along the Au layer between the Nb layers, because the Au layer is placed
at the middle of the trilayer and behaves itself as pinning centers.

\section{Methodological and computational details}

\subsection{Self-consistent Kohn-Sham-Dirac-Bogoliubov-de Gennes equations}

The microscopic theory of inhomogeneous superconductors is based on the Bogoliubov-de~Gennes (BdG) equations. The relativistic generalization -- called Dirac--Bogoliubov--de~Gennes (DBdG) equations -- was combined with density functional theory in Ref.~\onlinecite{Csire2018}
based on the idea of superconducting density functional theory~\cite{OGK1988, Suvasini}.
The relativistic order parameter -- with assuming a contact potential for the interaction -- is given by
\begin{equation}
 \chi(\vec{r})=\left< {\Psi}^T(\vec{r}) \boldsymbol{\boldsymbol{\eta}} {\Psi}(\vec{r})\right>,
\end{equation}
with the time-reversal matrix
\begin{equation}
 \boldsymbol{{\eta}}=
 \begin{pmatrix}
   0 & 1 & 0 & 0 \\
   -1 & 0 & 0 & 0 \\
   0 & 0 & 0 & 1\\
   0 & 0 & -1 & 0
 \end{pmatrix},
\end{equation}
and ${\Psi}(\vec{r})$ represents the four-component Dirac spinor field operator. The proper relativistic generalization leads to the following KSDBdG Hamiltonian written in Rydberg units ($\hbar=1$, $m=1/2$, $e^2=2$)
\begin{equation}
H_{\text{DBdG}}=
 \begin{pmatrix}
   H_D & D_{\text{eff}}(\vec{r}) \boldsymbol{\boldsymbol{\eta}} \\
   D_{\text{eff}}^\ast(\vec{r}) \boldsymbol{\boldsymbol{\eta}}^T & -H_D^\ast
 \end{pmatrix} \, ,
\end{equation}
where
\begin{align}
\nonumber
 H_D =& c \vec{\boldsymbol{\alpha}}\vec{p} + \left( \boldsymbol{\beta}-\mathbb{I}_4 \right) c^2/2+ \left( V_{\text{eff}}(\vec r)-E_F \right) \mathbb{I}_4 
 \\ &+ \boldsymbol{\beta}  \vec{\boldsymbol{\Sigma}}\vec{B}_{\text{eff}}(\vec r) + e \vec{\boldsymbol{\alpha}}\vec{A}_{\text{eff}}(\vec r), \,
\end{align}
\begin{equation}
 \vec{\boldsymbol \alpha} =
 \begin{pmatrix}
  0 & \vec{\boldsymbol \sigma} \\
  \vec{\boldsymbol \sigma} & 0
  \end{pmatrix} \, , \qquad
 \boldsymbol \beta=
 \begin{pmatrix}
  \mathbb I_2 & 0 \\
  0 & -\mathbb I_2
  \end{pmatrix} \, , \qquad
 \vec{\boldsymbol \Sigma} =
 \begin{pmatrix}
   \vec{\boldsymbol \sigma} & 0 \\
    0&  \vec{\boldsymbol \sigma}
  \end{pmatrix} \, ,
\end{equation}
and $\vec{\boldsymbol \sigma} $ denotes the vector of the Pauli-matrices. By adopting the simple semi-phenomenological parametrization of the exchange-correlation functional described in Refs.~\onlinecite{Csire2015}, the effective electrostatic potential, exchange field and pairing potential can be written as
\begin{subequations}
\begin{eqnarray}
  V_{\text{eff}}(\vec r)   &=&  V_{\text{ext}}(\vec r) +
  \int \frac{\rho(\vec r')}{|\vec r - \vec r'|}  \dd^{~\!3}\!r' +
  \frac{\delta E^0_{xc}[\rho,\vec m]}{\delta \rho(\vec r)},\\
  \vec{B}_{\text{eff}}(\vec r)&=& \vec{B}_{\text{ext}}(\vec r) +
                              \frac{\delta E^0_{xc}[\rho,\vec m]}{\delta \vec m(\vec r)},\\
  D_{\text{eff}}(\vec r) &=& \Lambda \, \chi(\vec r),
\end{eqnarray}
\end{subequations}
where $\Lambda$ is the strength of the interaction responsible for superconductivity (which can be treated as an adjustable semi-phenomenological parameter), $V_{\text{ext}}(\vec r)$ is the external potential (e.g. the Coulomb attraction from the protons), $\vec{B}_{\text{ext}}(\vec r)$ is the external field, $\rho (\vec r)$ is the charge density, $\vec m (\vec r)$ is the magnetization density, $E^0_{xc}[\rho,\vec m]$ is the usual (local spin density approximation) exchange-correlation energy for normal electrons. The equations describe the relativistic generalization of BCS theory for inhomogeneous superconductors taking into account their realistic band structure involving magnetism and spin-orbit effects.

\subsection{KKR-GF method}

We use a Green's function based approach exploiting the real-space representation of the resolvent of the KSDBdG Hamiltonian, 
\begin{equation}
 \mathcal{G}(z)=\left( z \mathbb{I}
 -H_{\text{DBdG}}\right)^{-1}.
 \label{eq:res}
\end{equation}
%Although the Green's function can be built up from Kohn-Sham orbitals obtained from linear band structure methods, we shall follow a distinct route here. 
The Multiple Scattering Theory (MST), i.e. the Korringa-Kohn-Rostoker (KKR) method gives direct access to the Green's function avoiding the step of calculating Kohn-Sham orbitals. In Ref.~\cite{Csire2018} this technique was generalised for the solution of the KSDBdG equation with layered structure.
In MST, %the potential is treated in the so called muffin-tin approximation, i.e. 
the potential is written as a sum of single-domain potentials centered around each lattice site, $n$, namely $V_{\text{eff}}(\vec r)= \sum_n V_n(r)$, $\vec B_{\text{eff}}(\vec r)= \sum_n \vec B_{n}(r)$, $D_{\text{eff}}(\vec r)= \sum_n D_{n}(r)$. The potentials treated within the atomic sphere approximation (ASA) are zero if $r=|\vec r_n|\geq S_n$, where $S_n$ is the radius of the Wigner-Seitz (WS) sphere that has the same volume as the atomic cell $n$.

In the relativistic case, we search the solutions of the KSDBdG equations in the following form
\begin{equation}
 \Psi(z, \vec r)= \sum_{Q}
 \begin{pmatrix}
  g^e_{Q}(z,r) \chi_{Q} (\hat r) \\
  \ii f^e_{Q}(z,r)\chi_{\overline Q} (\hat r) \\
  g^h_{Q}(z,r) \chi^*_{Q} (\hat r) \\
  - \ii f^h_{Q}(z,r)\chi^*_{\overline Q} (\hat r)
 \end{pmatrix},
\end{equation}
where $Q=(\kappa,\mu)$ and $\overline{Q}=(-\kappa,\mu)$ are the composite indices for the spin-orbit ($\kappa$) and magnetic ($\mu$) quantum numbers; $g^{e(h)}_{Q}(z,r)$ and $f^{e(h)}_{Q}(z,r)$ are the large and small components of the electron (hole) part of the solution, respectively. The spin-angular function is an eigenfunction of the spin-orbit operator $ K = \boldsymbol \sigma L +\mathbb I$,
\begin{equation}
 K \ket{\kappa \mu} = -\kappa \ket{\kappa \mu}.
\end{equation}
For later purposes the following notations are also introduced: $\overline{l}=l-S_k$, and $S_k=\kappa/|\kappa|$ the sign of $\kappa$.
One should mention that in the general case where the $D_{\text{eff}}(\vec r)$ is complex, it is also necessary to 
to distinguish the left hand-side and right hand-side solutions where the left-hand side solutions are defined as
\begin{equation}
  \bra{\Psi^\times(z)} \left( z-H_{DBdG}(z) \right) =0.
\end{equation}

The magnetic field can be rotated to a local frame such that it points along the $\hat z$ direction. 
In the normal state it has two advantages: the least amount of coupling occurs between the states with different $\kappa,\mu$ quantum numbers, and there is no need to distinguish the left-hand-side and right-hand-side solutions of the radial KSDBdG equations.
The details can be found in Ref.~\onlinecite{Csire2018}.

The pairing potential matrix for $s$-wave singlet superconductivity is defined as
\begin{equation}
 \Delta_{Q Q'}(r)=(-1)^{\mu'-\frac{1}{2}} ~S_{\kappa'} ~\delta_{\kappa \kappa'} \delta_{\mu~\! -\mu'} D(r),
\end{equation}
which shows that the pairing interaction couples electrons with $\kappa,\mu$ quantum numbers to holes with $\kappa,-\mu$ quantum numbers. This is a direct consequence of our initial assumption that the pairing acts between Kramers pairs, namely between electrons and their time-reversed pairs (holes). The KSDBdG equations can be solved in a local frame with a predictor-corrector algorithm on logarithmic scale similarly, as it was done for the radial scalar relativistic BdG equations in Ref.~\onlinecite{Csire2015}.

The two most important quantities leading to the Green's function are the single-site $t$-matrix and the structure constants. Physically, the $t$-matrix describes scattering on the single-site potential involving relativistic Andreev-scattering, while the structure constants contain all information about the crystal structure. This is one of the most amazing feature of the multiple scattering theory allowing the separation of scattering events and structural information. We shall denote the irregular and regular solutions of the KSDBdG equations inside the WS spheres by $\mathbf{J}_Q(z,\vec{r})$ and $\mathbf{Z}_Q(z,\vec{r})$ (matrices in electron-hole indices), respectively. The scattering solutions obey the matching conditions described in Refs.~\onlinecite{Csire2015, Csire2018}. 
The following supermatrix formalism can be introduced for the scattering matrices, the matrices of the structure constant and the scattering path operator:
\begin{eqnarray}
 \mathbf{t}(z) &=& \{t^{n,ab}_{QQ'}(z) \delta_{nm} \},
 \label{eq:sh_t}\\
 \mathbf{G}_0(z) &=& \{G^{nm,ab}_{0,QQ'}(z) (1-\delta_{nm}) \delta_{ab} \},
 \label{eq:sh_G_0}\\
 \boldsymbol{\tau}(z) &=& \{\tau^{nm,ab}_{QQ'}(z) \},
 \label{eq:sh_tau}
\end{eqnarray}
where $\boldsymbol{\tau}(z)$ can be determined from the single-site t-matrix and the real space structure constant matrix 
\begin{equation}
 \boldsymbol{\tau}(z)=
 \left(
 \mathbf{t}(z)^{-1} -\mathbf{G}_0(z)
 \right)^{-1} \, .
  \label{eq:sh_tau_t}
\end{equation}

As in the normal state formalism, based on the expansions of the free-particle Green's function, %in the non-relativistic case and in the relativistic case, 
the derivation of the one-particle Green's function  is straightforward and leads to the following formula,
\begin{equation}
\begin{split}
 \mathbf{G}(z ,\vec r,\vec {r'}) &=
   \sum_{QQ'} \mathbf{Z}_Q(z ,\vec r) \boldsymbol{\tau}_{QQ'}(z )
   \mathbf{Z}_{Q'}(z ,\vec{r'})^\times
 \\ &- \theta(r'- r) \sum_Q \mathbf{Z}_Q(z ,\vec r) \mathbf{J}_Q(z ,\vec{r'})^\times
 \\ &- \theta(r - r') \sum_Q \mathbf{J}_Q(z ,\vec r) \mathbf{Z}_Q(z ,\vec{r'})^\times,
\end{split}
  \label{eq:gf_formula}
\end{equation}
where $\theta(x)$ is the step function and
$\mathbf{Z}_Q(z ,\vec{r'})^\times, 
\mathbf{J}_Q(z ,\vec{r'})^\times$
stand for the left-hand side solutions of the KSDBdG equations (see Ref.~\onlinecite{Csire2018}).
%Note that $\mathbf{Z}_Q(z ,\vec r)$ and $\mathbf{J}_Q(z ,\vec r)$ are also matrices in electron-hole space according to Eqs.~(\ref{eq:sh_tau}) and \ref{eq:GF_calcrel}.

The formulas given above can be applied to surfaces and interfaces quite straightforwardly following the idea of the so-called Screened KKR (SKKR) formalism described in Refs.~\onlinecite{Szunyogh, Zeller}. In this formalism, it is made use of the 2D periodicity of the layers by introducing 2D lattice Fourier transformed version for the scattering path operator
\begin{equation}
 \boldsymbol{\tau}(z,\vec k_{||})=
 \left(
 \mathbf{t}(z)^{-1} -\mathbf{G}_0(z,\vec k_{||})
 \right)^{-1}.
  \label{eq:sh_tau_t2}
\end{equation}
To perform the inverse of the KKR matrix, a special reference system is used to obtain structure constants that are localized in real space. In the supermatrix formalism we used above, the screening transformation, described in detail in~\onlinecite{Zeller}, can be written in a way that is formally the same as it was presented in Sec.~III of Ref.~\onlinecite{Szunyogh}. Thus, the formalism can be derived for layered systems with two-dimensional periodicity and applied as the SKKR method prescribes.

\subsection{Computational procedure and details}

Following Ref.~\onlinecite{OGK1988} the KSDBdG equations shall be solved self-consistently by assuming that the superconducting host has isotropic $s$-wave spin-singlet pairing as described by BCS theory\cite{Suvasini}.
Such a solution results in self-consistent charge and magnetization density and effective pairing potential.
Furthermore, in this work, we imposed various phases of the pairing potential on the right Nb layer of the Nb/Au/Nb heterostructure.
It is important to note that semi-infinite geometry is used for the Nb both on the right and left side, while in the interface region we use always 3 Nb monolayers both on the right and left side to model the change in the electrostatic potential and there are various numbers of Au monolayers between the two Nb leads.

First, a series of normal state, conventional DFT calculations are performed within the same Green's function formalism, which is often referred to as Multiple Scattering Theory (MST). We begin with a bulk calculation to obtain self-consistent potentials and the Fermi energy for the bulk host (Nb). Then, a normal state interface calculation is done to obtain potentials for the monolayers in the interface.
The partial waves within MST are treated with an angular momentum cutoff of $\ell_\mathrm{max}=2$. In the self-consistent normal state calculations, we used a Brillouin zone (BZ) integration with 253 $\mathbf{k}$ points in the irreducible wedge of the BZ and a semicircular energy contour on the upper complex plane with 16 points for energy integration.
To examine the band structure we plotted the Bloch Spectral Function (BSF) for monolayer $I$
\begin{equation}
    \mathrm{BSF}_I(\varepsilon, \mathbf{k}) = -\frac{1}{\pi} \Im \mathrm{Tr_Q} G^{II} (\varepsilon, \mathbf{k}).
\end{equation}

Once the normal state calculations are ready, and all the self-consistent potentials are obtained, the same steps are repeated in the superconducting state. In these steps, we obtain the pairing potential, the anomalous charge $\chi$ and the pairing interaction strength $\Lambda$ in the bulk. Then, we assume the same interaction strength $\Lambda$ on the Nb, and zero pairing interaction strength on the Au layer. The singlet pairing amplitude could be derived from the Green function obtained in the $\ket{\kappa \mu}$ basis as described in Ref.~\onlinecite{Csire2018}.
Here we mention that, in Fig.~3(f) of the main text, the real energy and layer resolved singlet pairing amplitude
\begin{widetext}
 \begin{equation}
    \Re \chi_I(\varepsilon) =
      - \frac{1}{\pi} \int \dd \vec{r} \int_{\mathrm{BZ}} \dd^{~ \! 2} \! k_{||} 
       ~\Im~\! \mathrm{Tr}_Q~\! \boldsymbol \eta G^{eh,II}(\varepsilon + \ii 0 , \vec r, \vec{k}_{||}),
  \end{equation}
\end{widetext}
has been used (in Fig.~3(f) the values were taken at the energy of the additional Andreev bound state normalized with the value of Nb's coherence peak and we plotted the layer dependence to show the non-uniform structure).
However, to explore the effect of Josephson tunneling, we also impose various phase differences between the superconducting Nb on the different sides.
The BZ integration for the host Green's function was performed by using an increasing number of $\mathbf{k}$ points with respect to the normal state, including 200$\times$200 points in the irreducible wedge of the 2D BZ. A sufficient energy resolution of the LDOS in the superconducting gap is acquired by considering 101 energy points between $\pm 1.95$~meV with an imaginary part of 13.6~$\mu$eV related to the smearing of the resulting LDOS. 
To calculate the superconducting critical current at finite temperatures (Eq.~(2) in the main text)
the quasi-particle spectrum has to be obtained with the correct superconducting gap.
For the sake of computational simplicity the temperature dependence of the superconducting gap is approximated by the well-known BCS formula
$\Delta(T) = \Delta(T=0\textrm{K}) \sqrt{1-T/T_c}$.

\section{Normal state electronic structure}

\begin{figure*}[htb]
    \centering
    \includegraphics[width=0.375\linewidth]{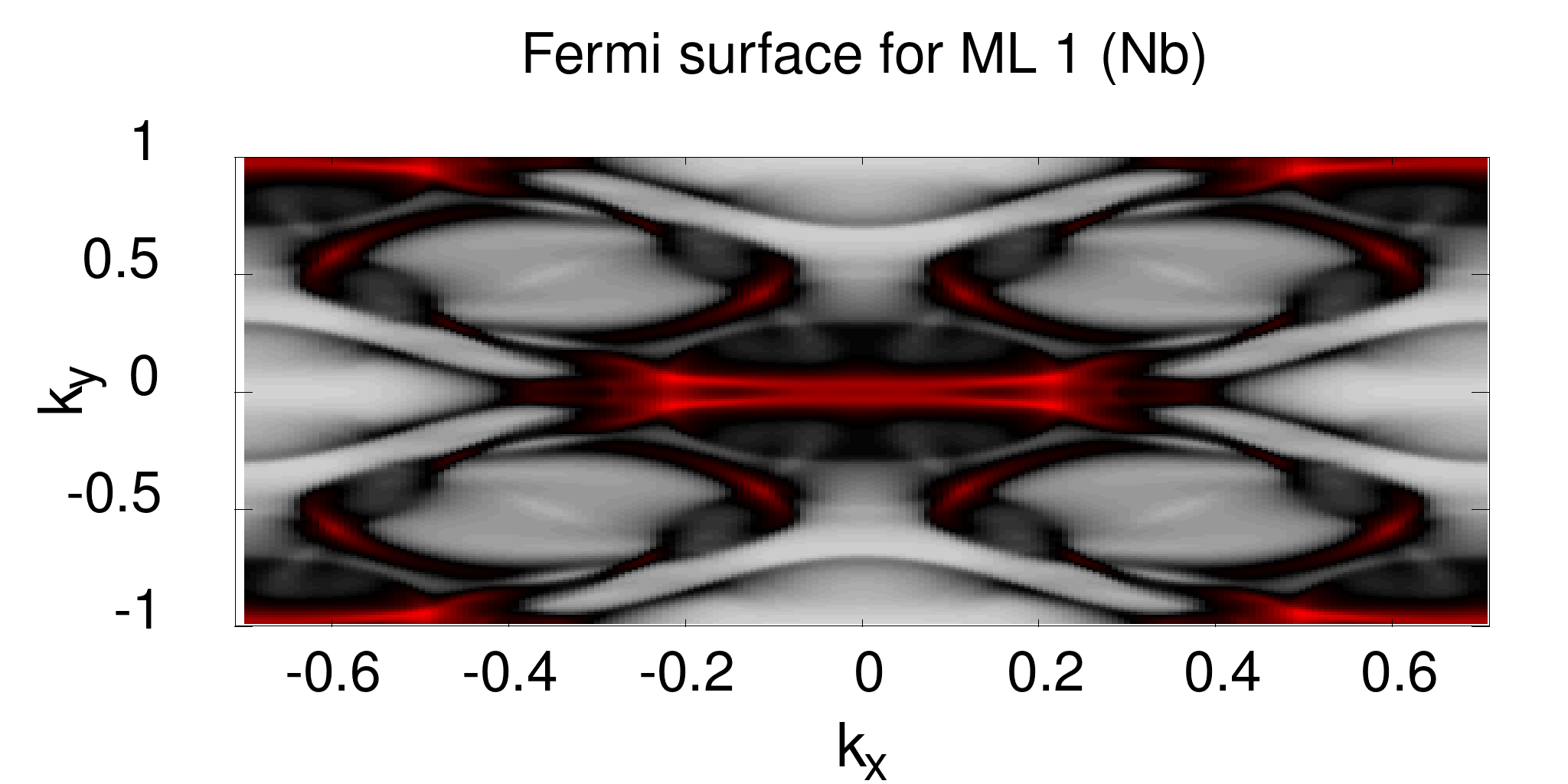}
    \includegraphics[width=0.375\linewidth]{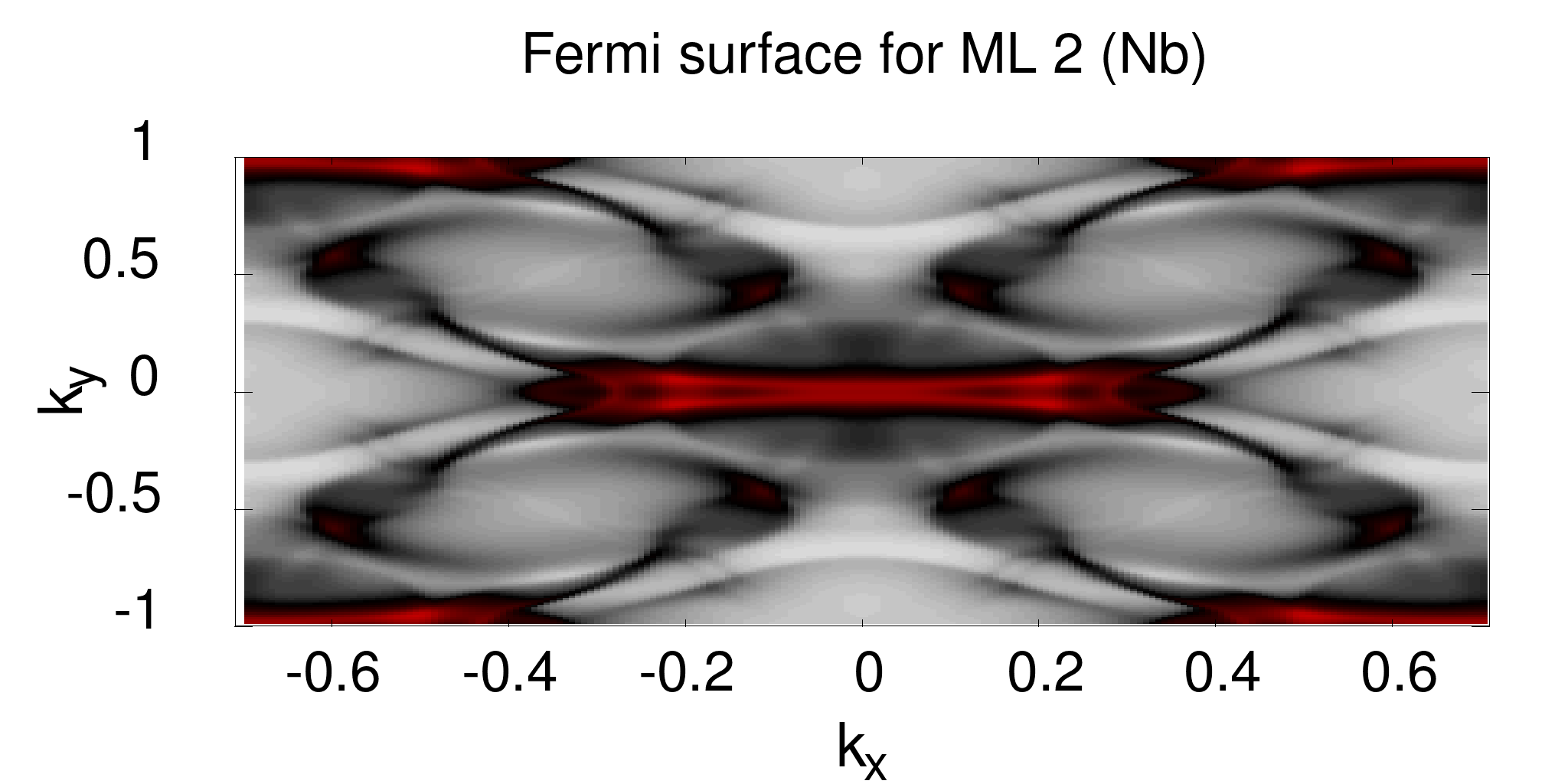}\\
    \includegraphics[width=0.375\linewidth]{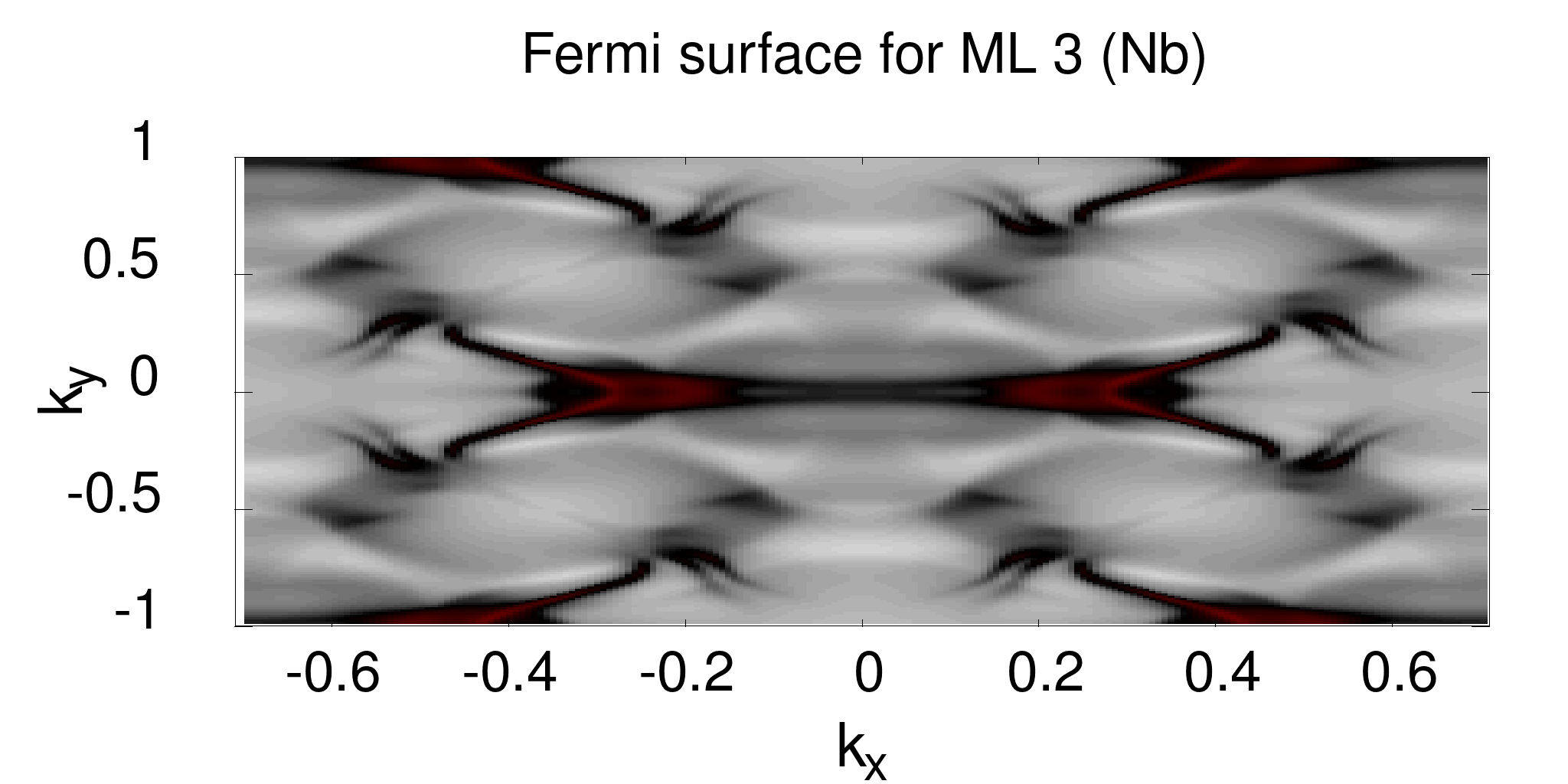}
    \includegraphics[width=0.375\linewidth]{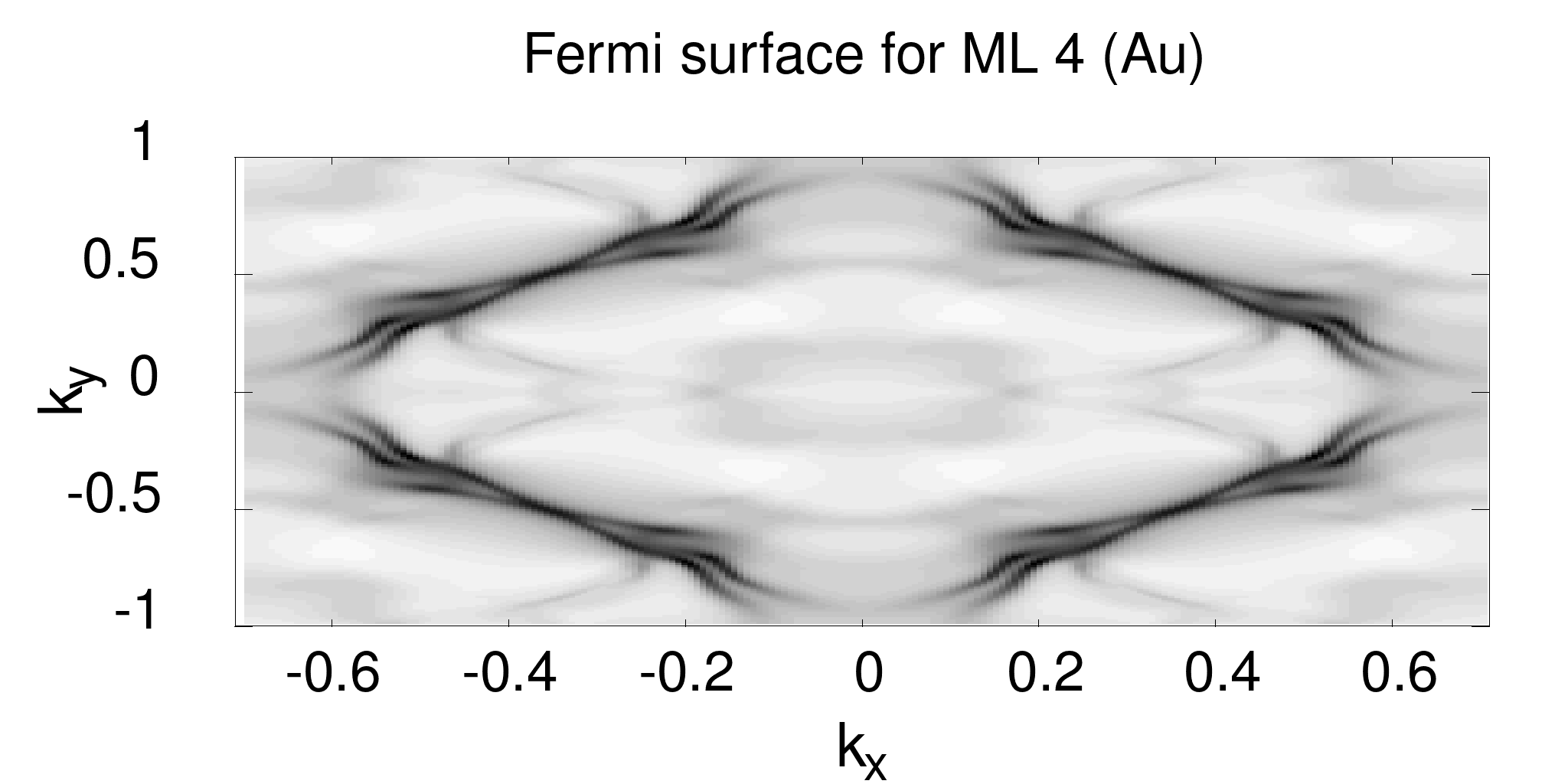}\\
    \includegraphics[width=0.375\linewidth]{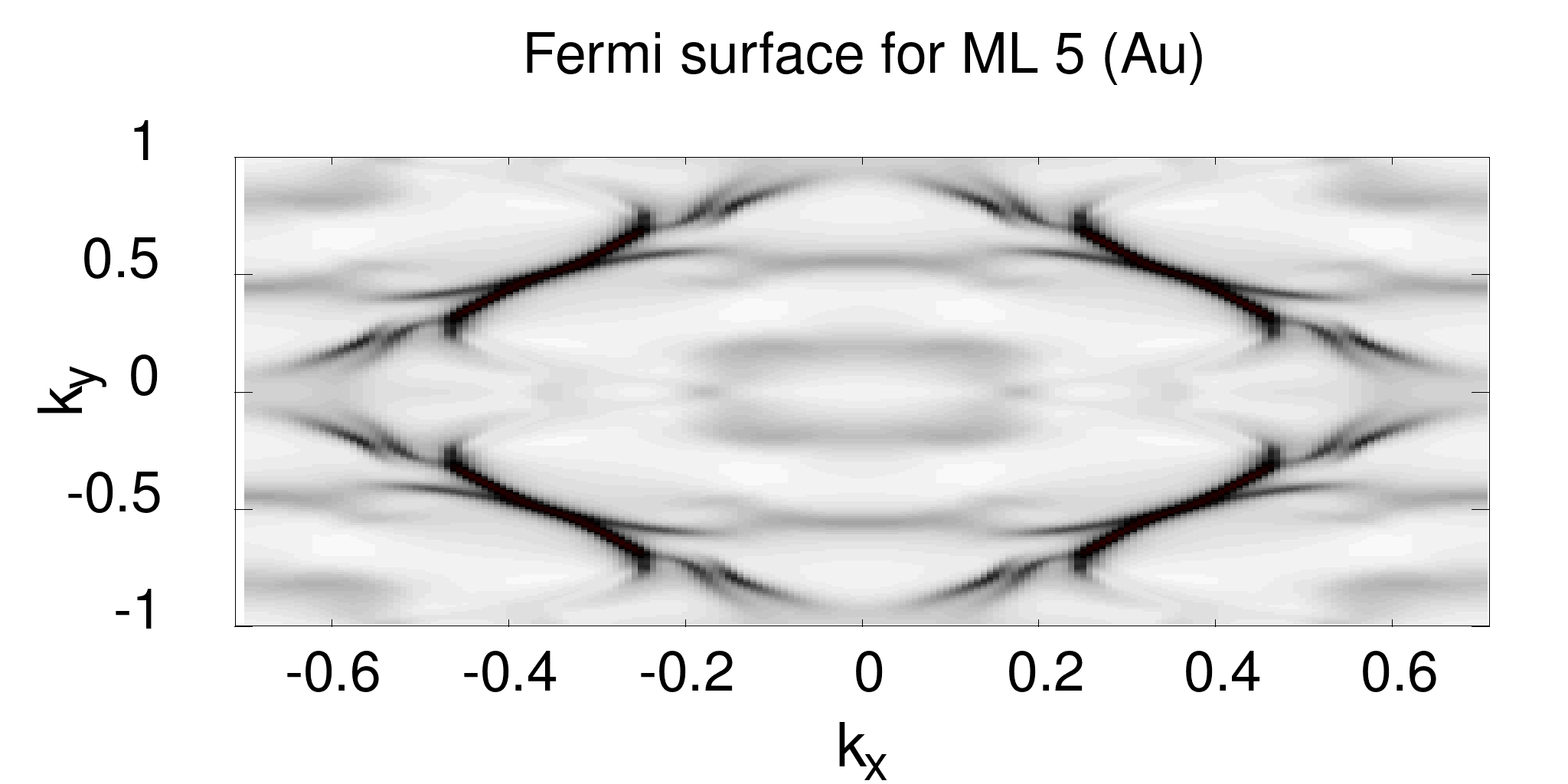}
    \caption{Layer resolved Fermi surface for gold thickness 3~ML}
    \label{fig:fs_l3}
\end{figure*}

\begin{figure*}[htb]
    \centering
    \includegraphics[width=0.375\linewidth]{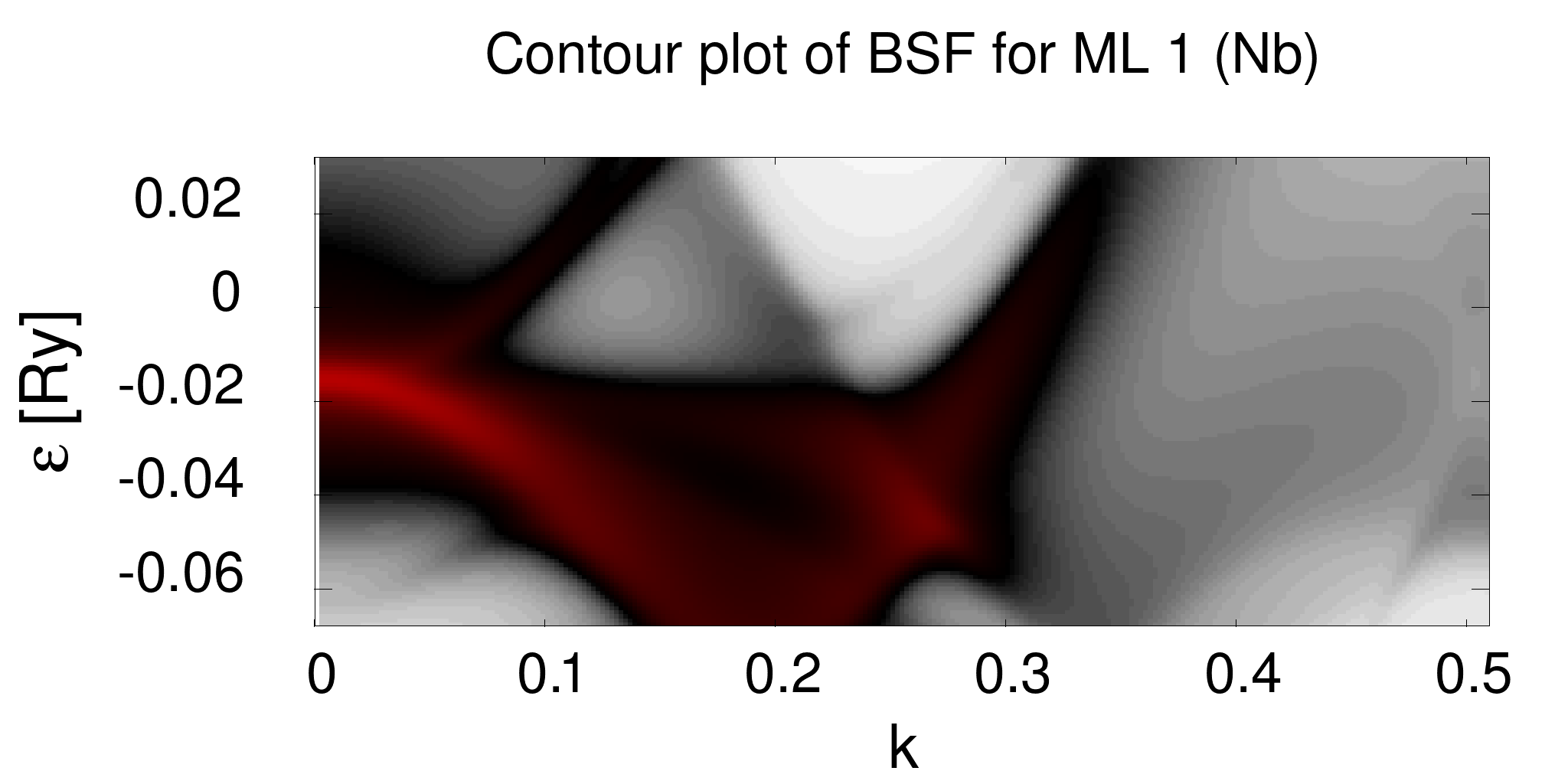}
    \includegraphics[width=0.375\linewidth]{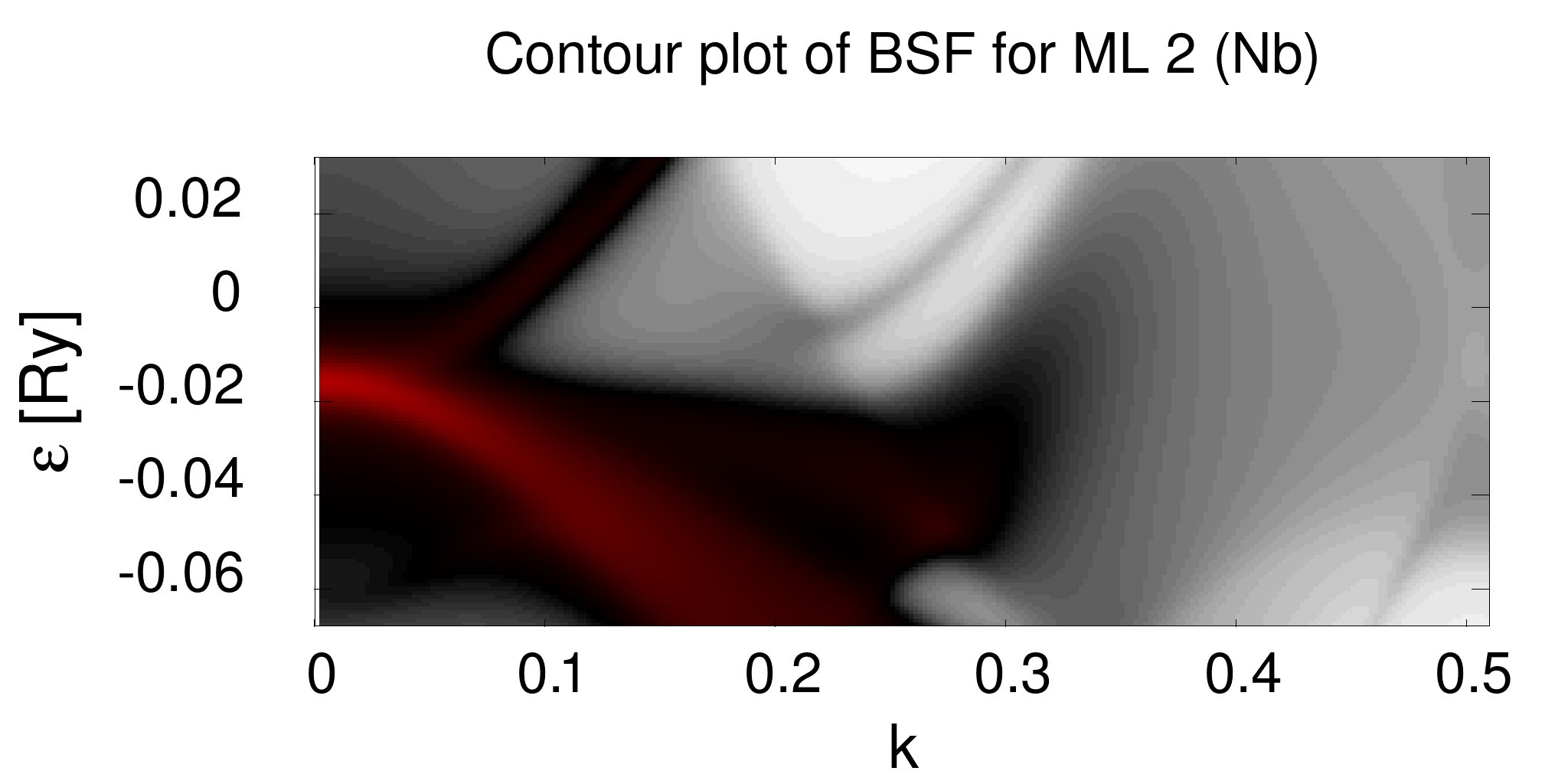}\\
    \includegraphics[width=0.375\linewidth]{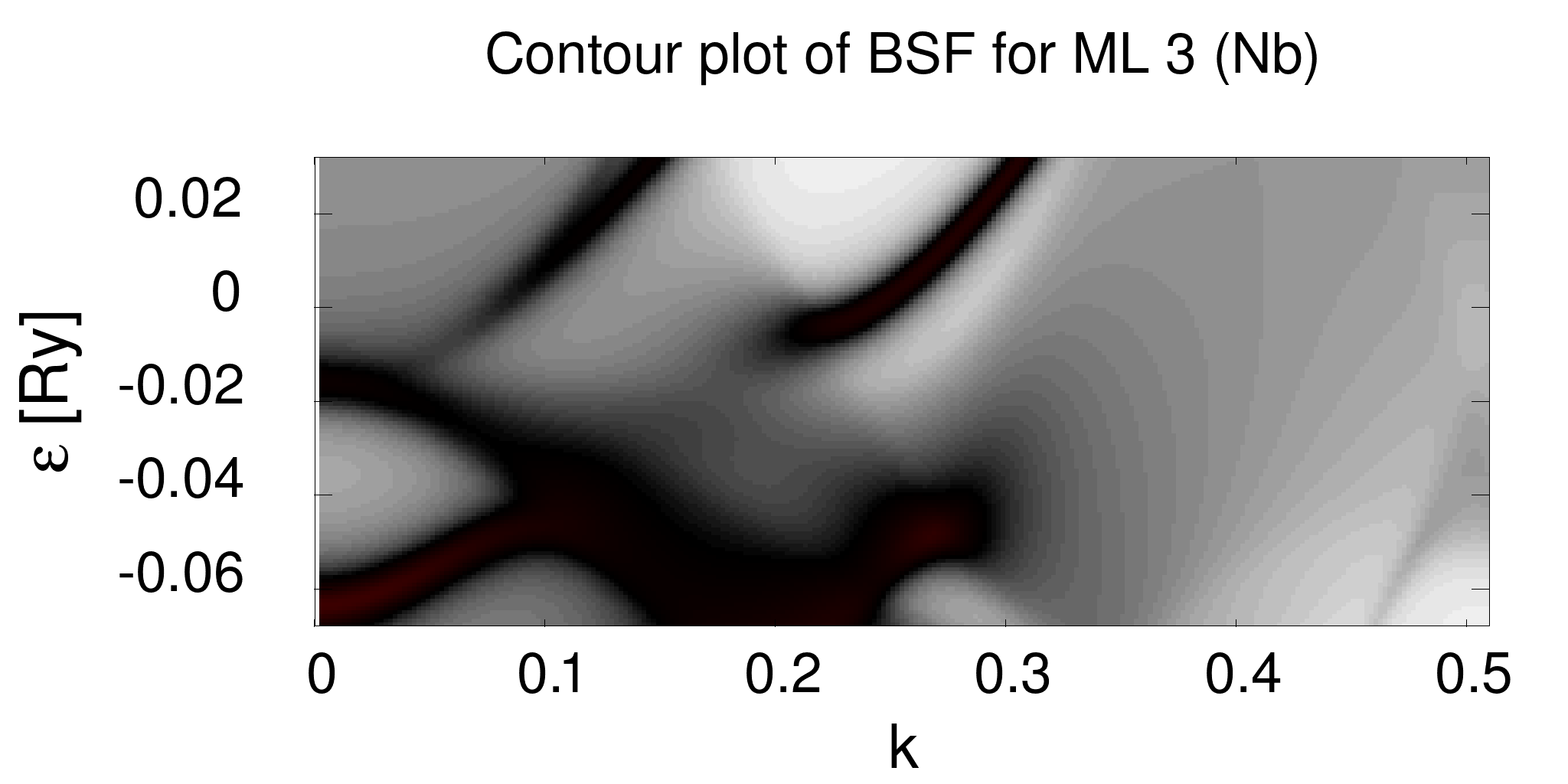}
    \includegraphics[width=0.375\linewidth]{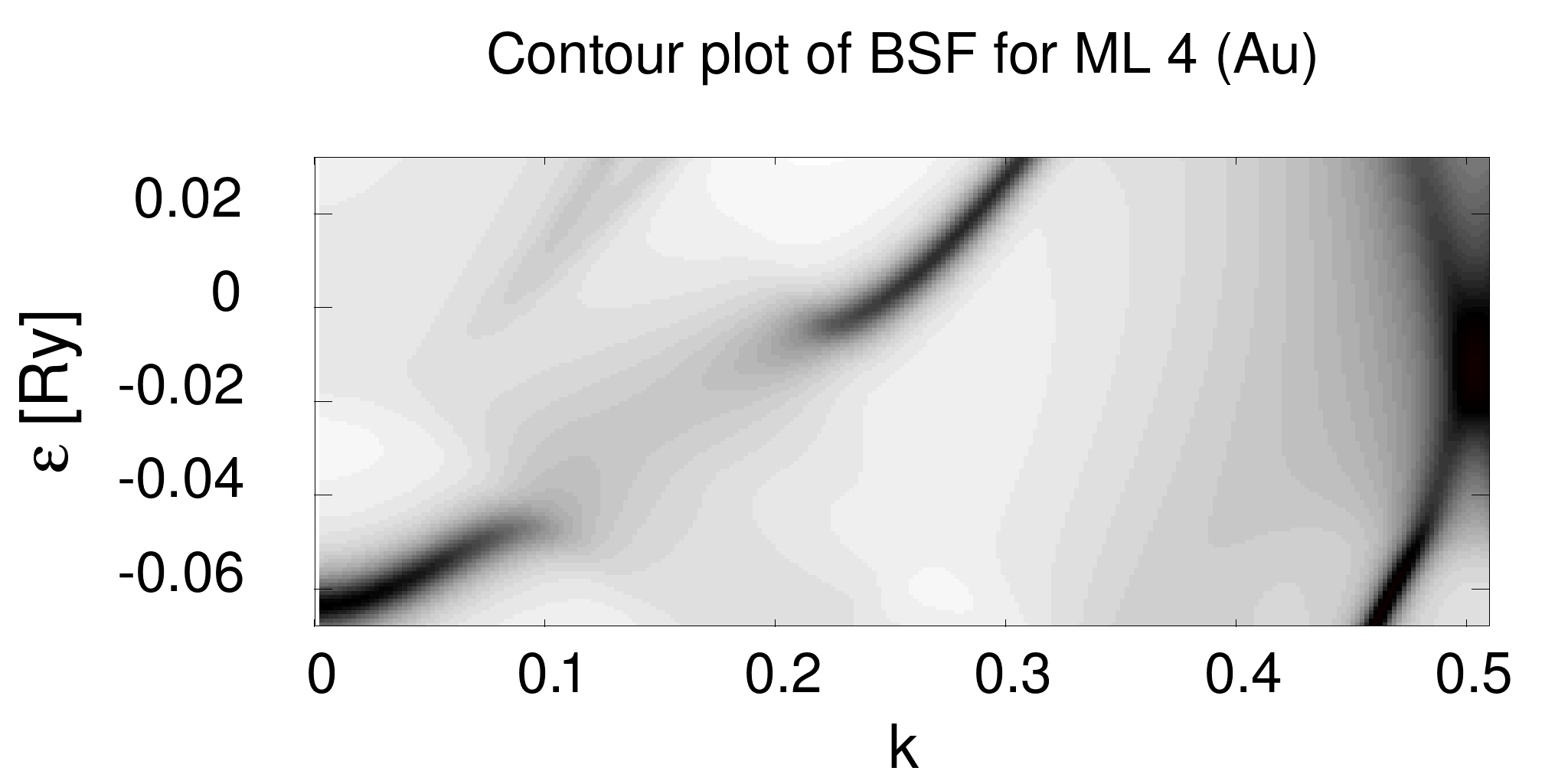}\\
    \includegraphics[width=0.375\linewidth]{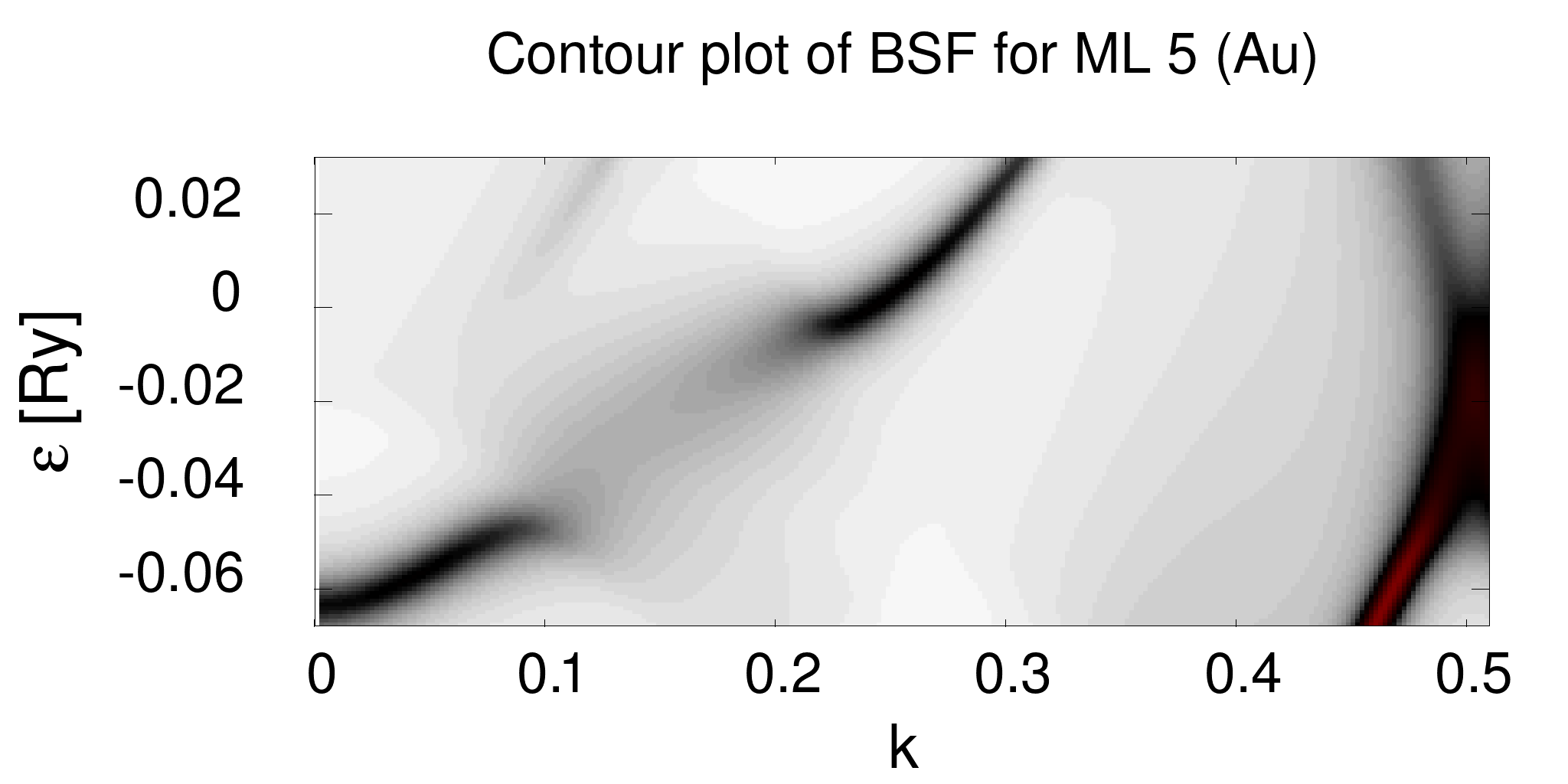}
    \caption{Layer resolved contour plot of Bloch Spectral Function~(BSF) along $k_x=\sqrt{2} k_y$ for gold thickness 3~ML}
    \label{fig:bsf_l3}
\end{figure*}

\begin{figure*}[htb]
    \centering
    \includegraphics[width=0.375\linewidth]{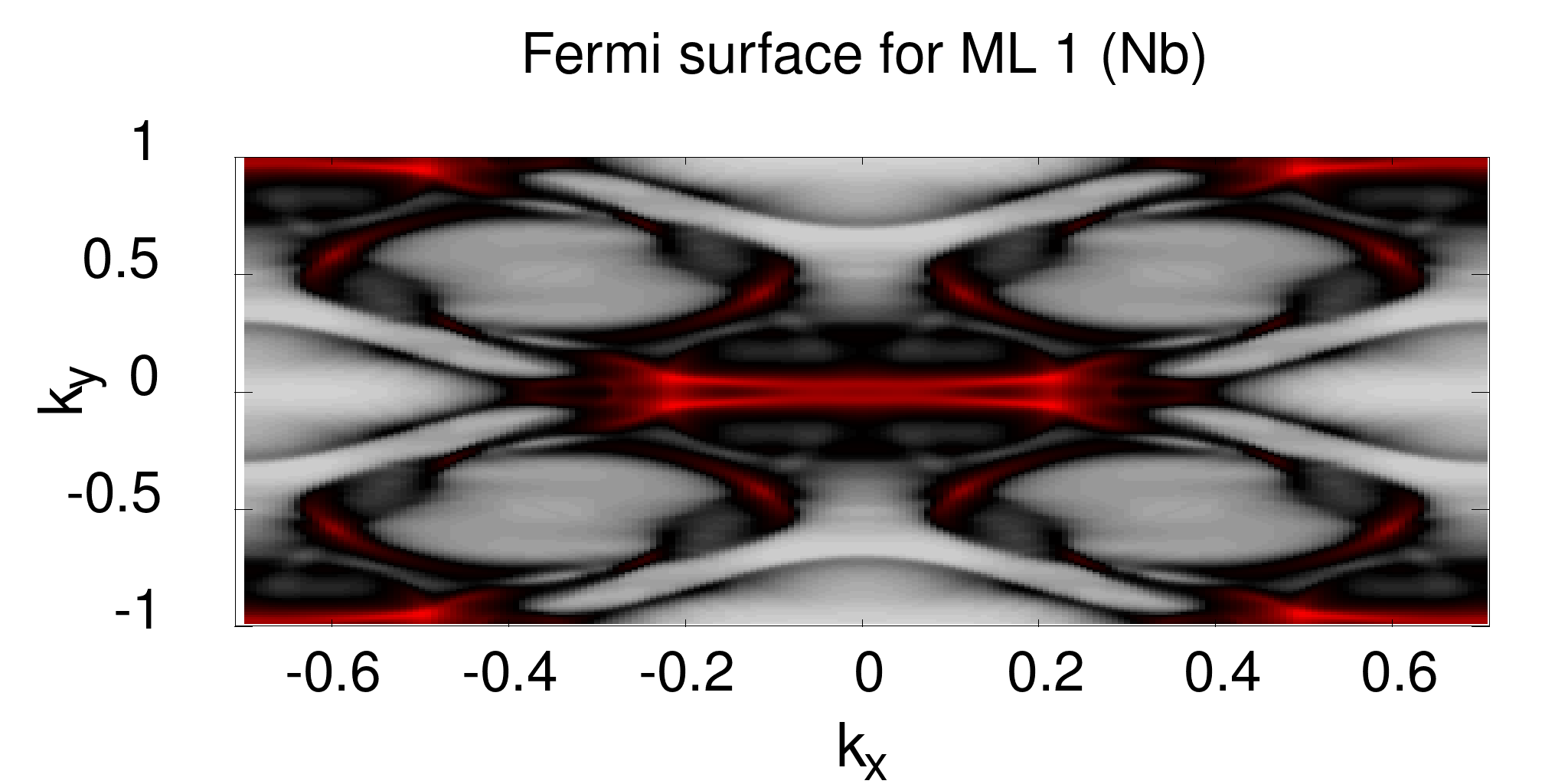}
    \includegraphics[width=0.375\linewidth]{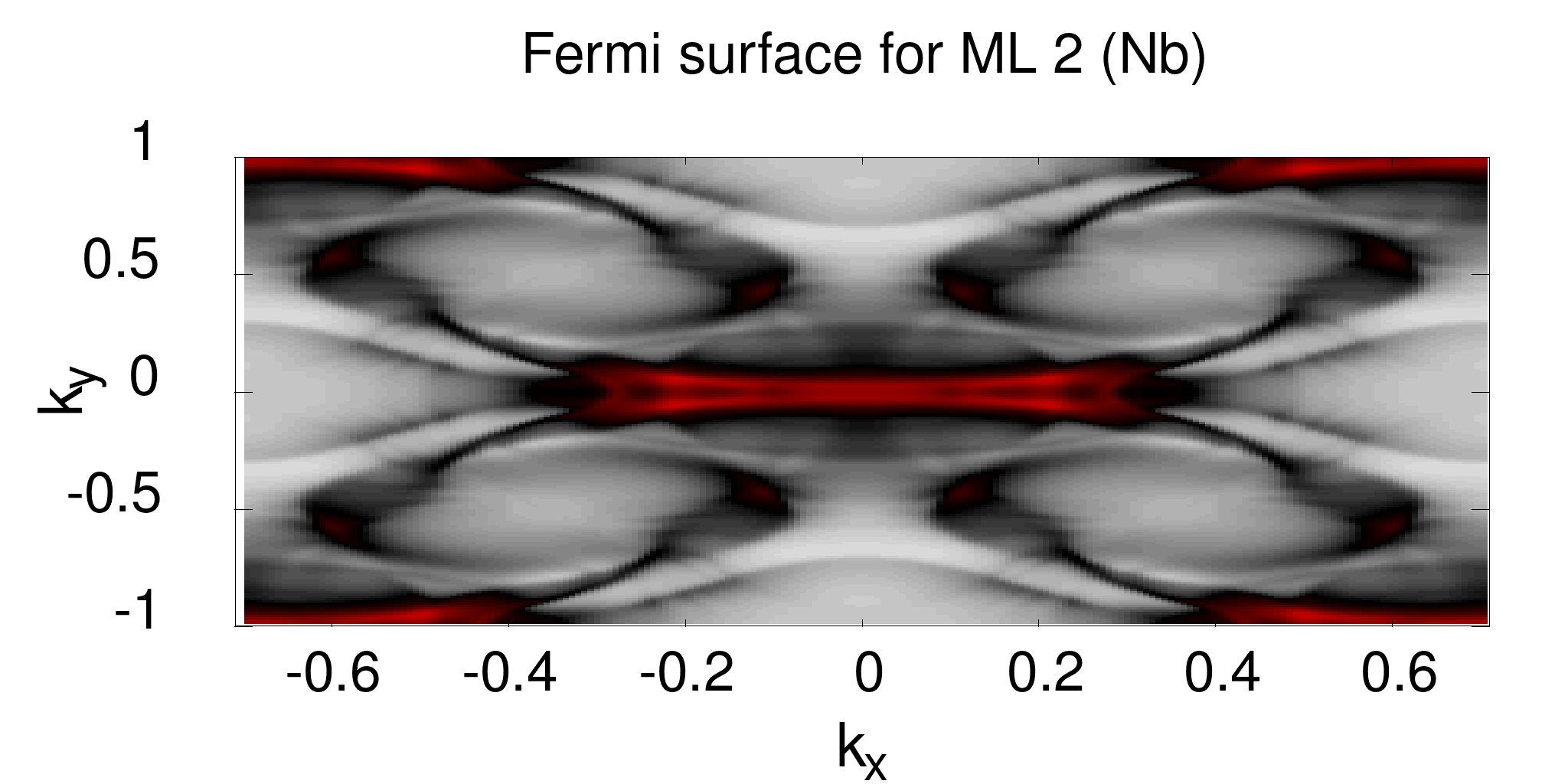}\\
    \includegraphics[width=0.375\linewidth]{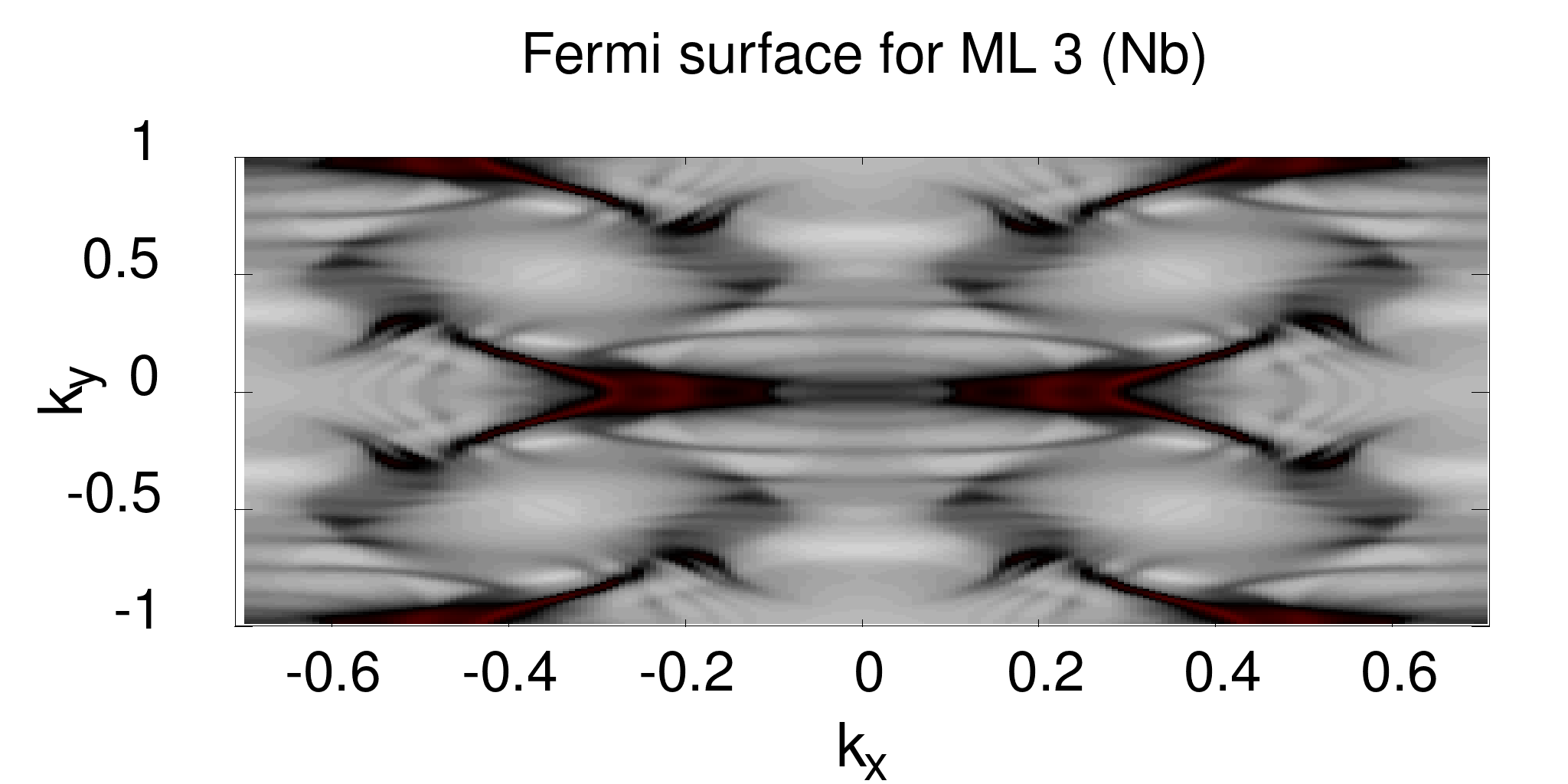}
    \includegraphics[width=0.375\linewidth]{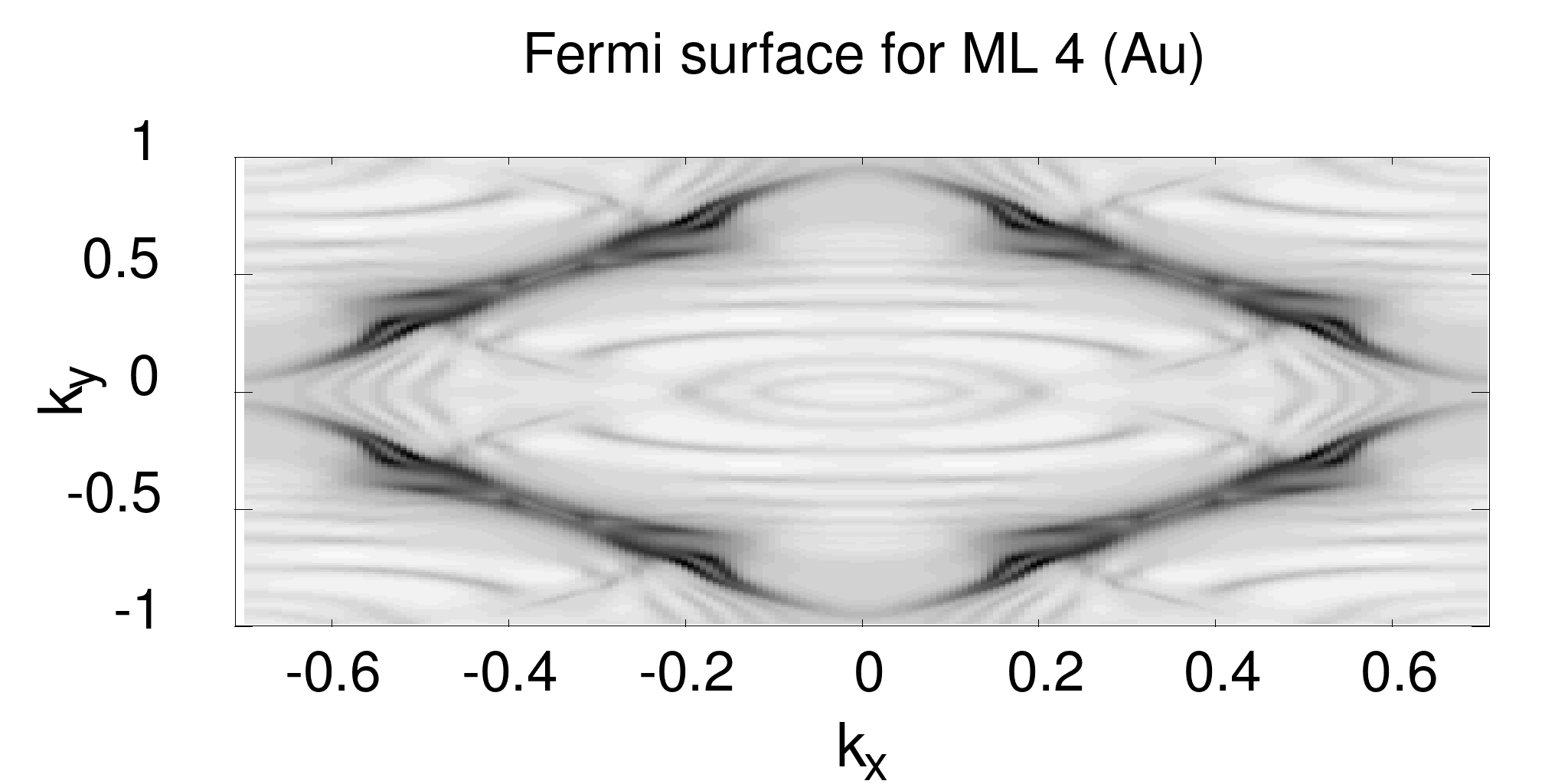}\\
    \includegraphics[width=0.375\linewidth]{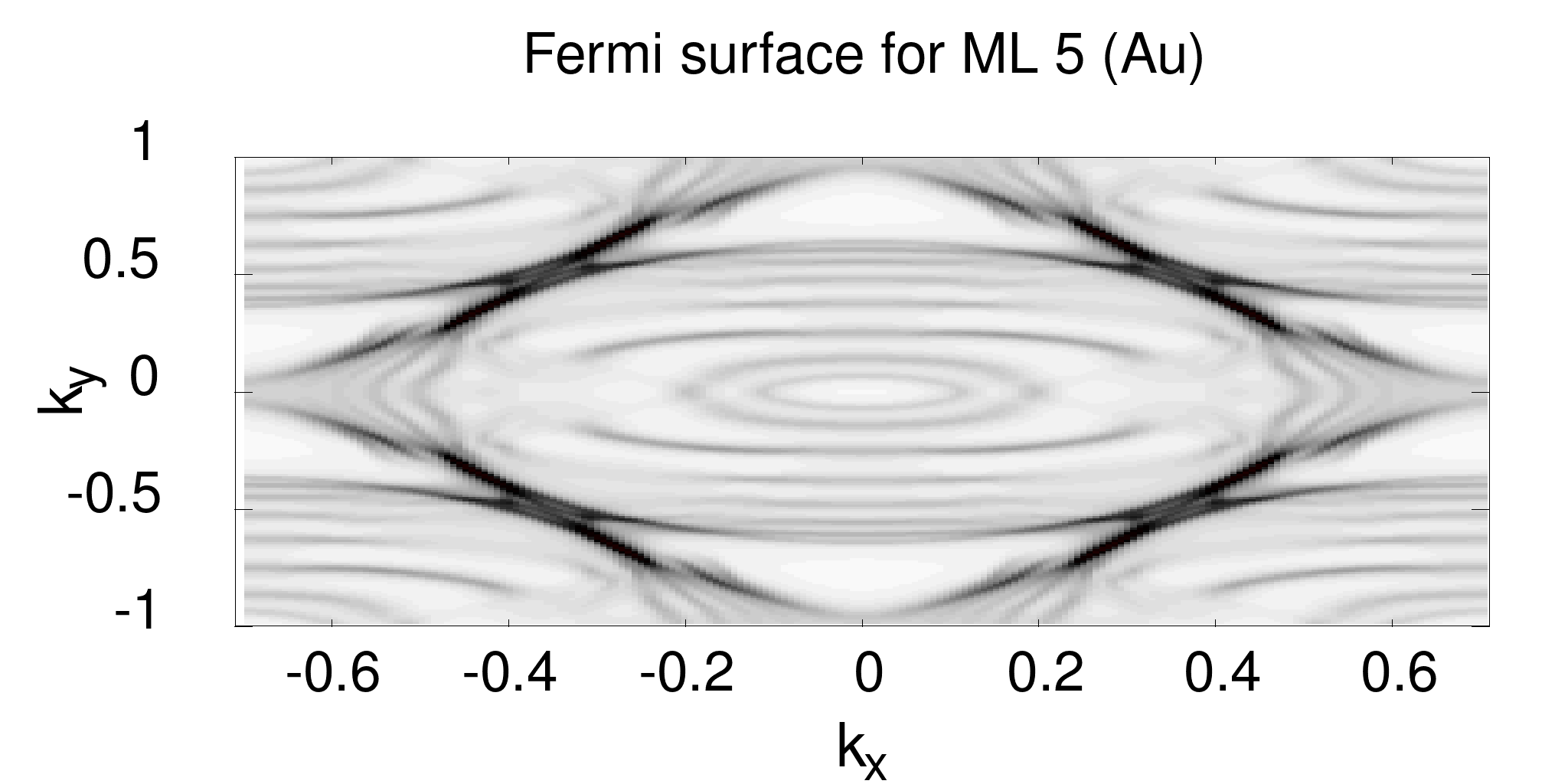}
    \includegraphics[width=0.375\linewidth]{./SupMat_Figures/fs_l9-5.png}\\
    \includegraphics[width=0.375\linewidth]{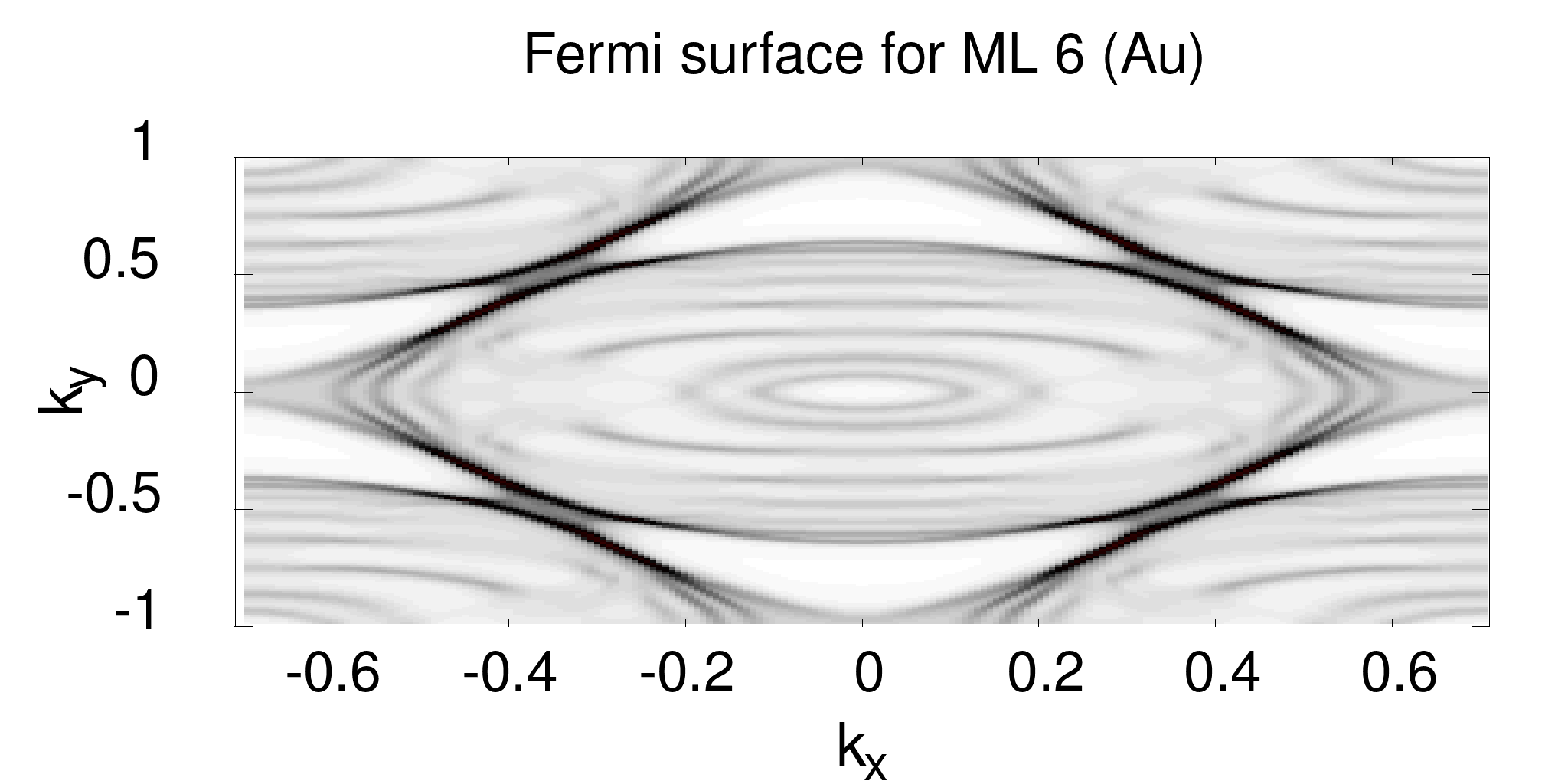}
    \includegraphics[width=0.375\linewidth]{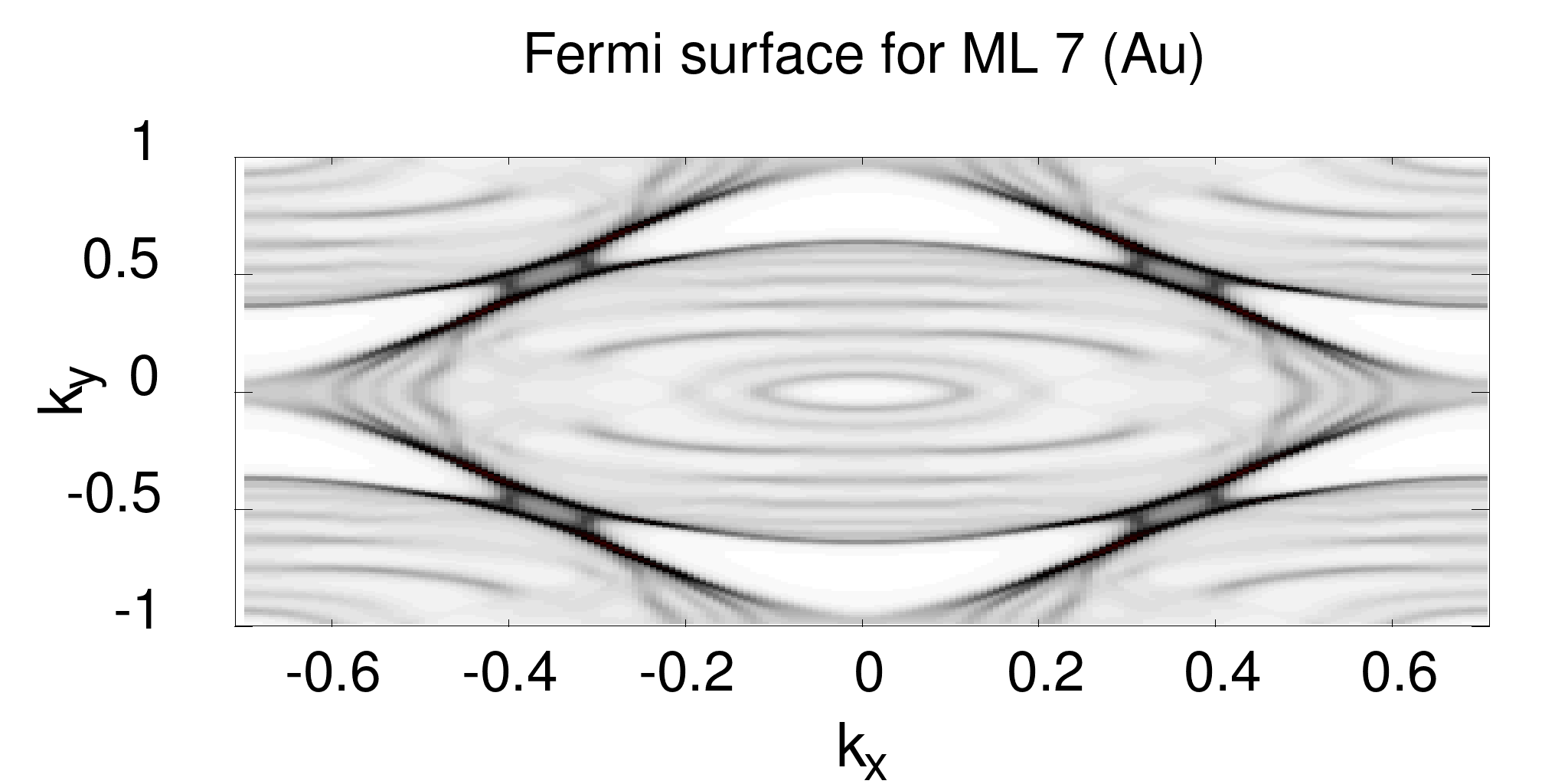}\\
    \includegraphics[width=0.375\linewidth]{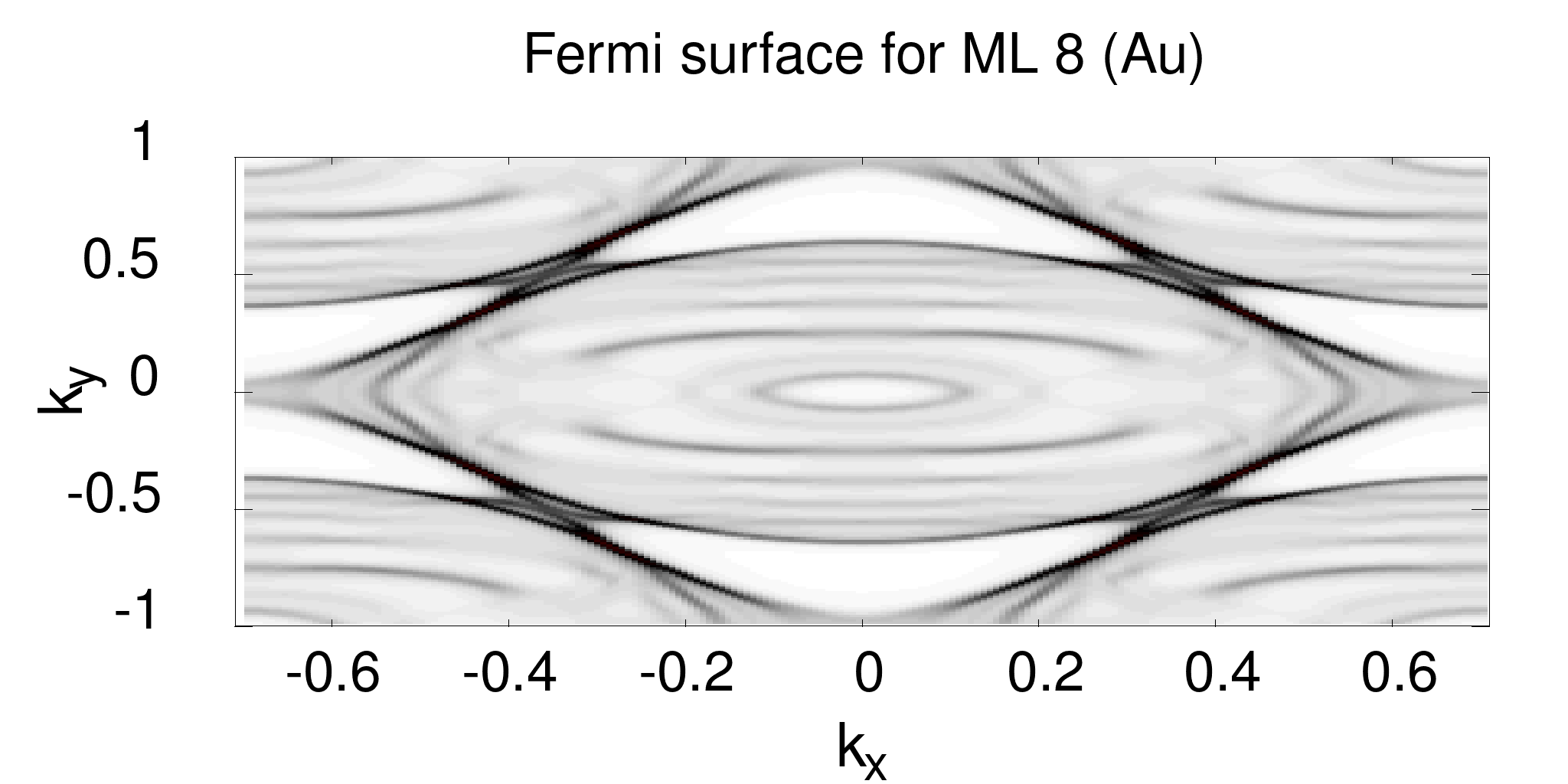}
    \caption{Layer resolved Fermi surface for gold thickness 9~ML}
    \label{fig:fs_l9}
\end{figure*}

\begin{figure*}[htb]
    \centering
    \includegraphics[width=0.375\linewidth]{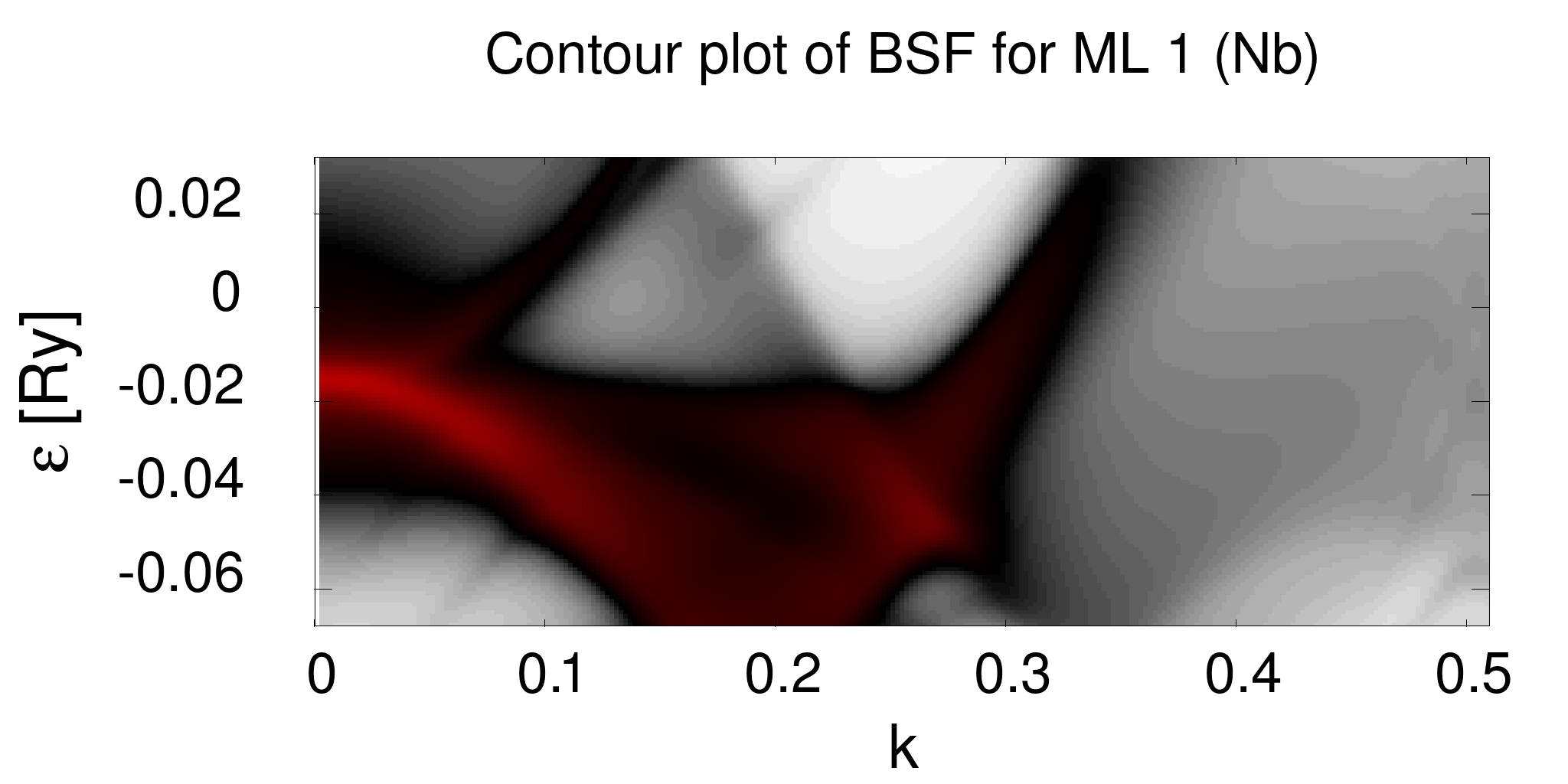}
    \includegraphics[width=0.375\linewidth]{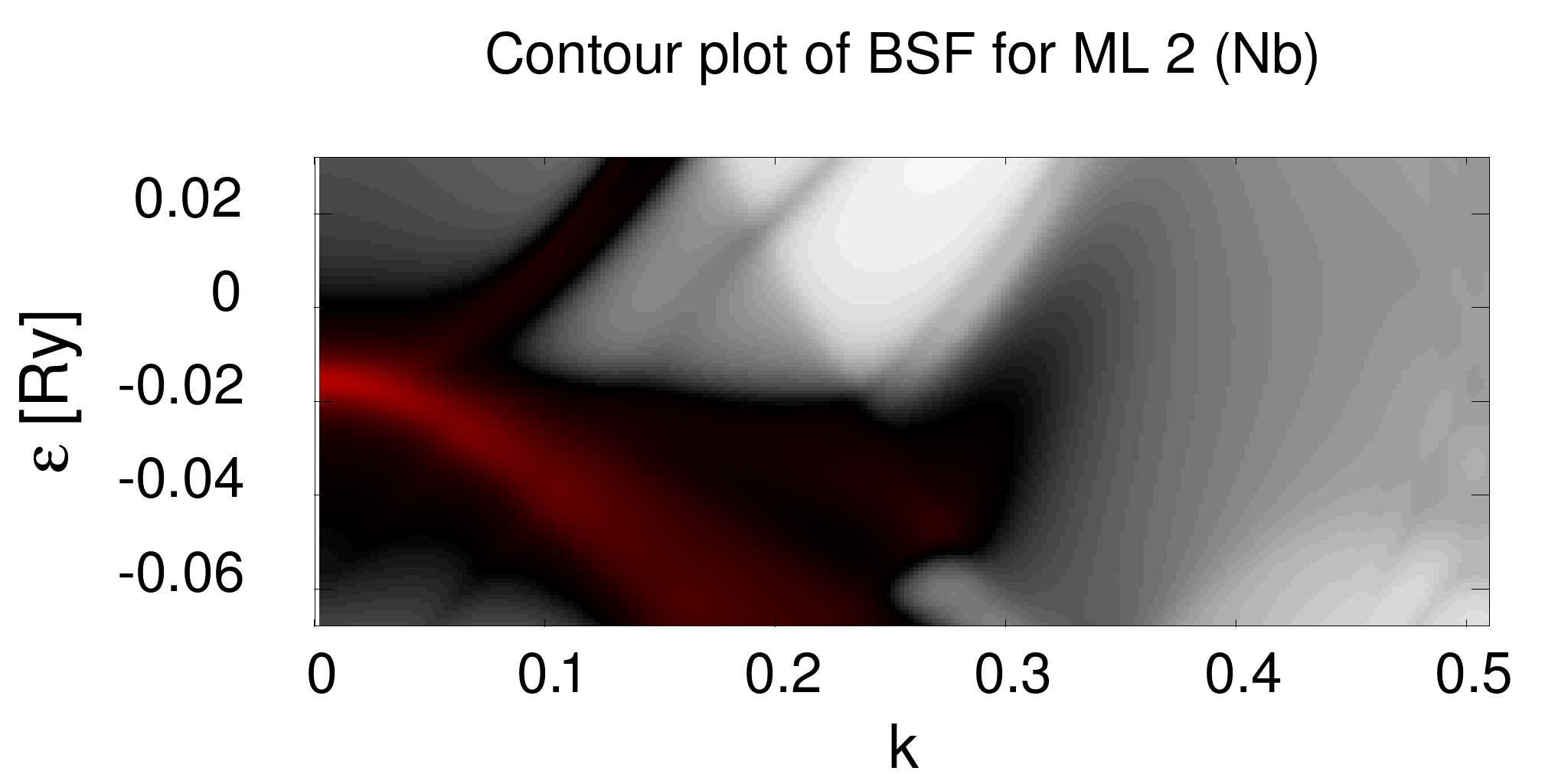}\\
    \includegraphics[width=0.375\linewidth]{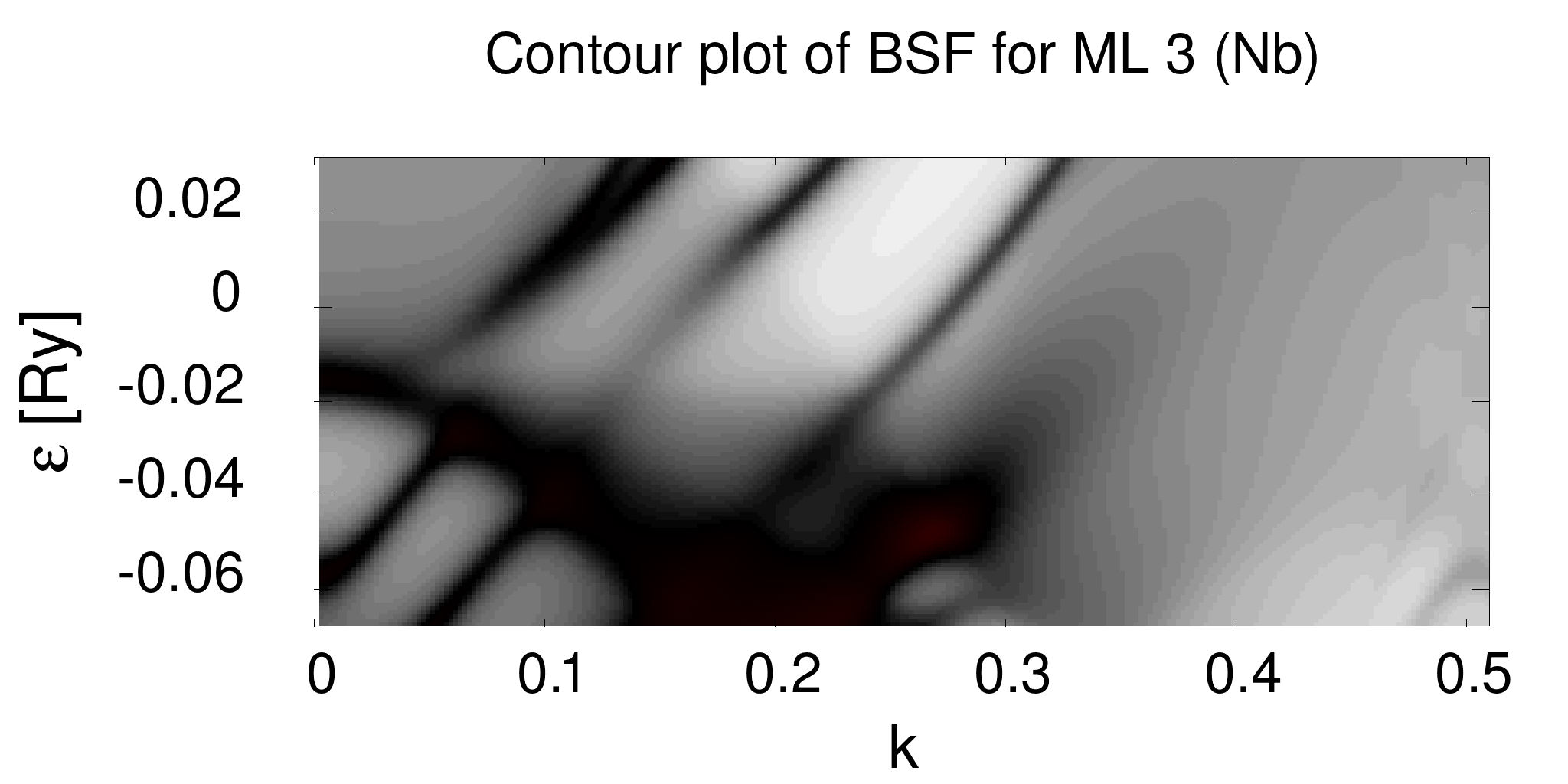}
    \includegraphics[width=0.375\linewidth]{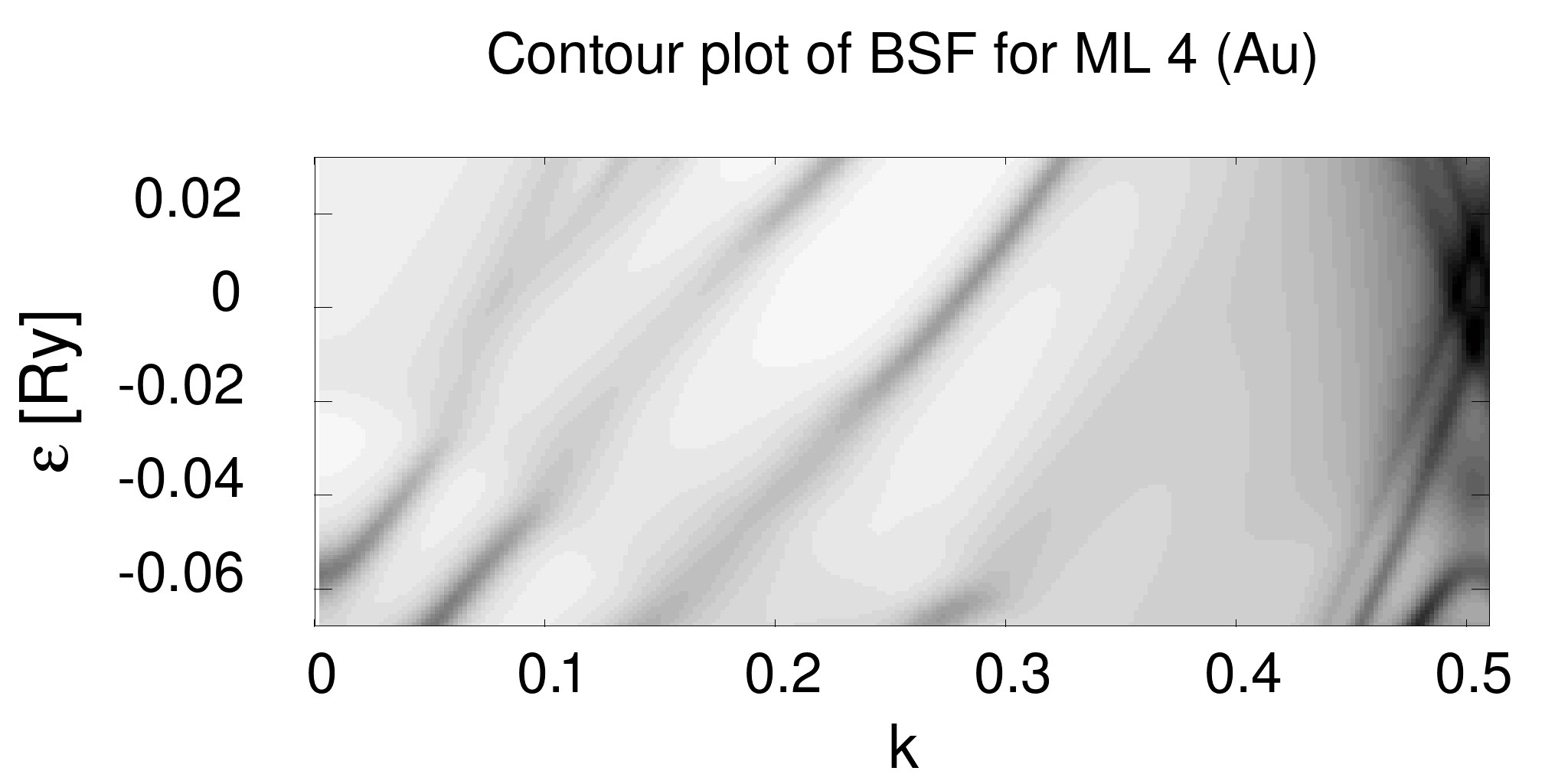}\\
    \includegraphics[width=0.375\linewidth]{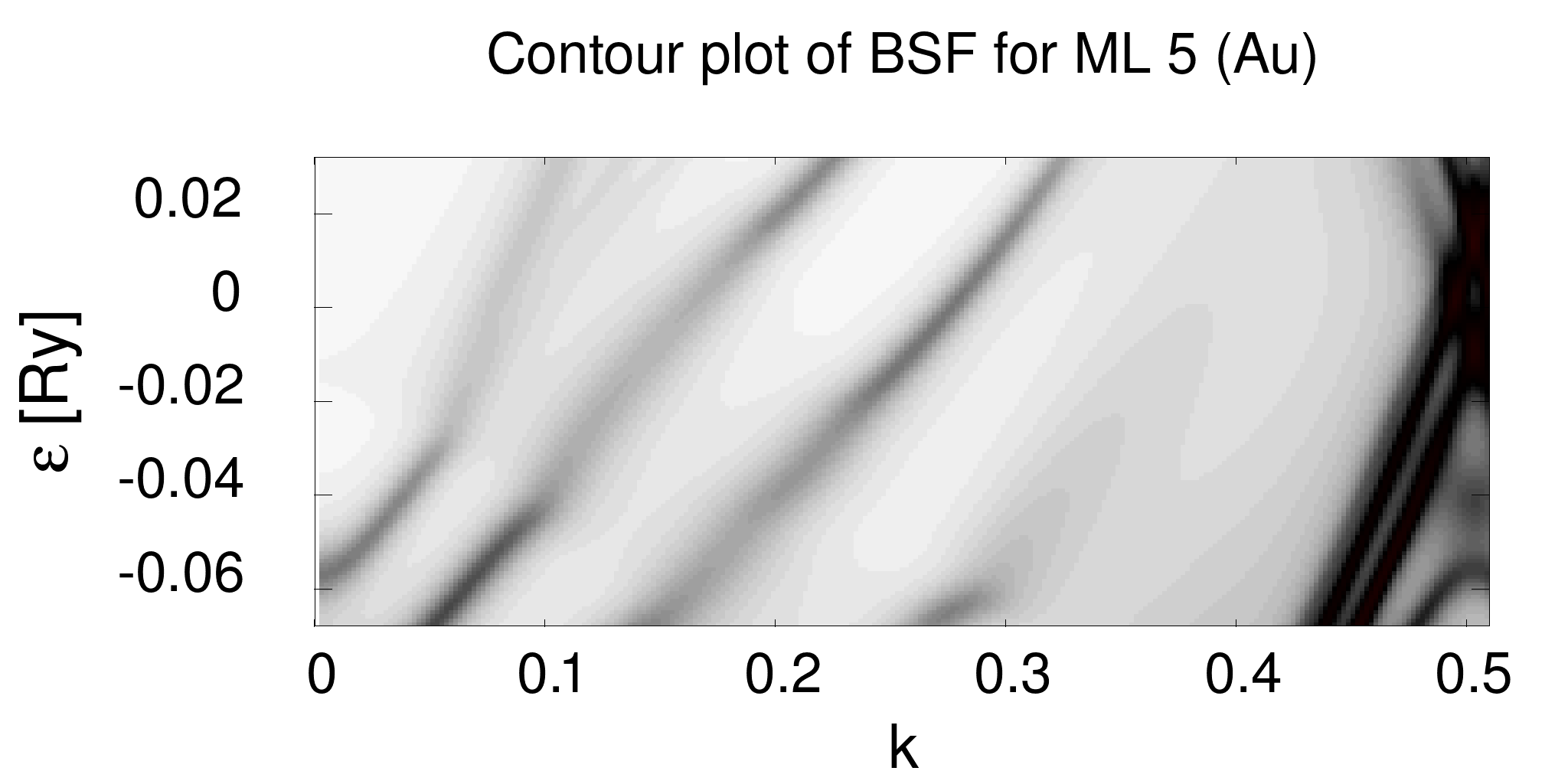}
    \includegraphics[width=0.375\linewidth]{./SupMat_Figures/bsf_l9-5.png}\\
    \includegraphics[width=0.375\linewidth]{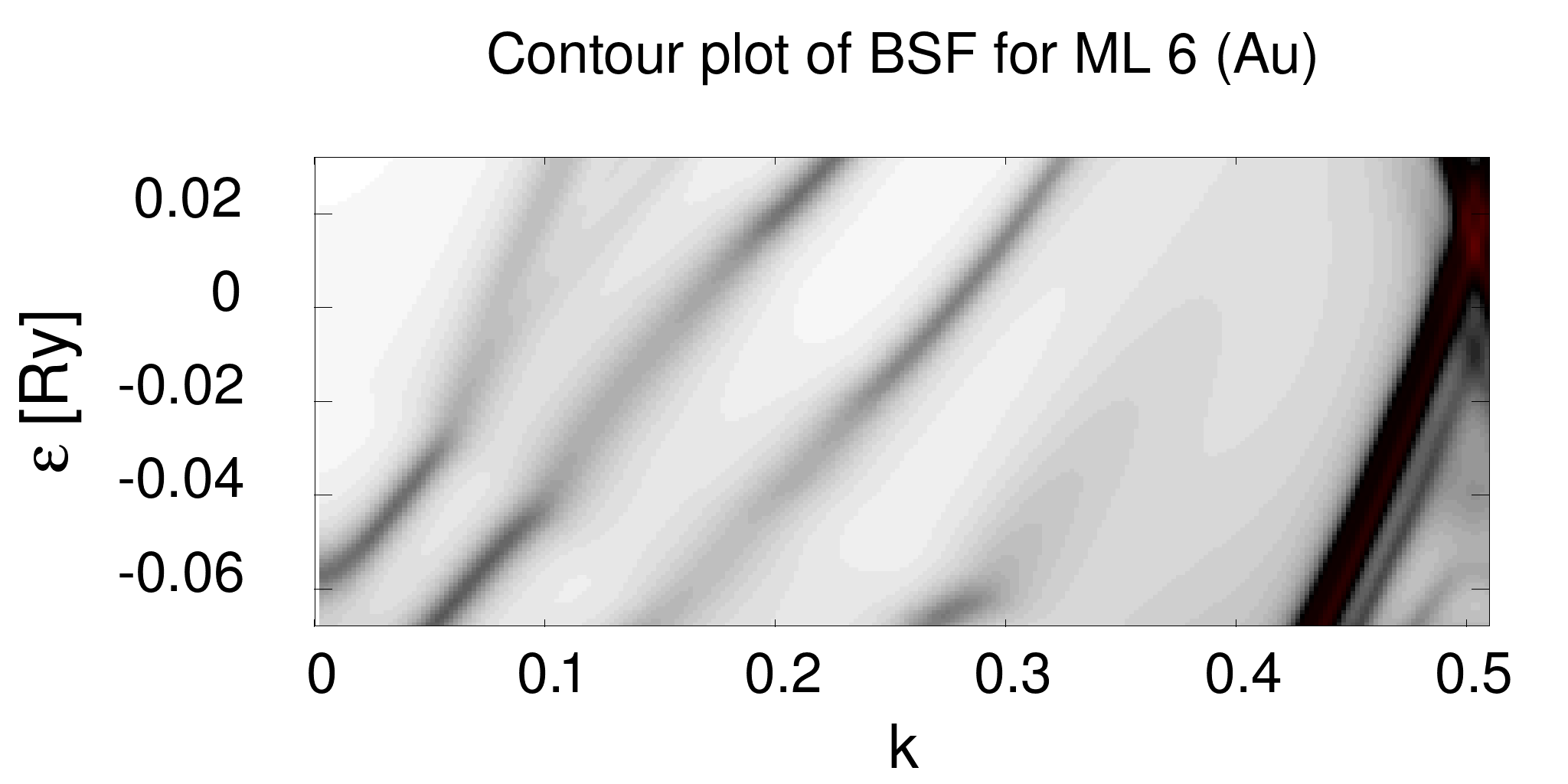}
    \includegraphics[width=0.375\linewidth]{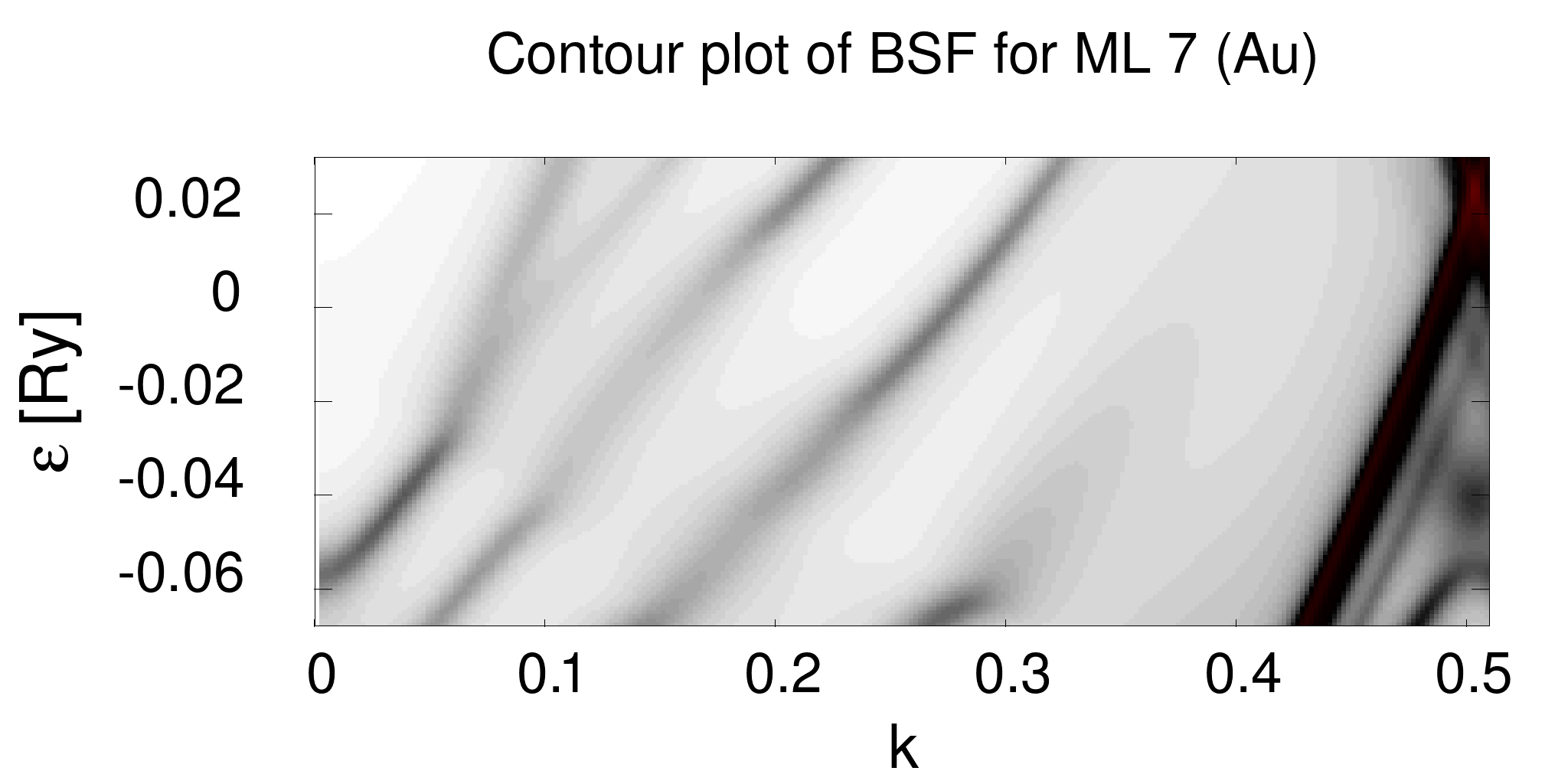}\\
    \includegraphics[width=0.375\linewidth]{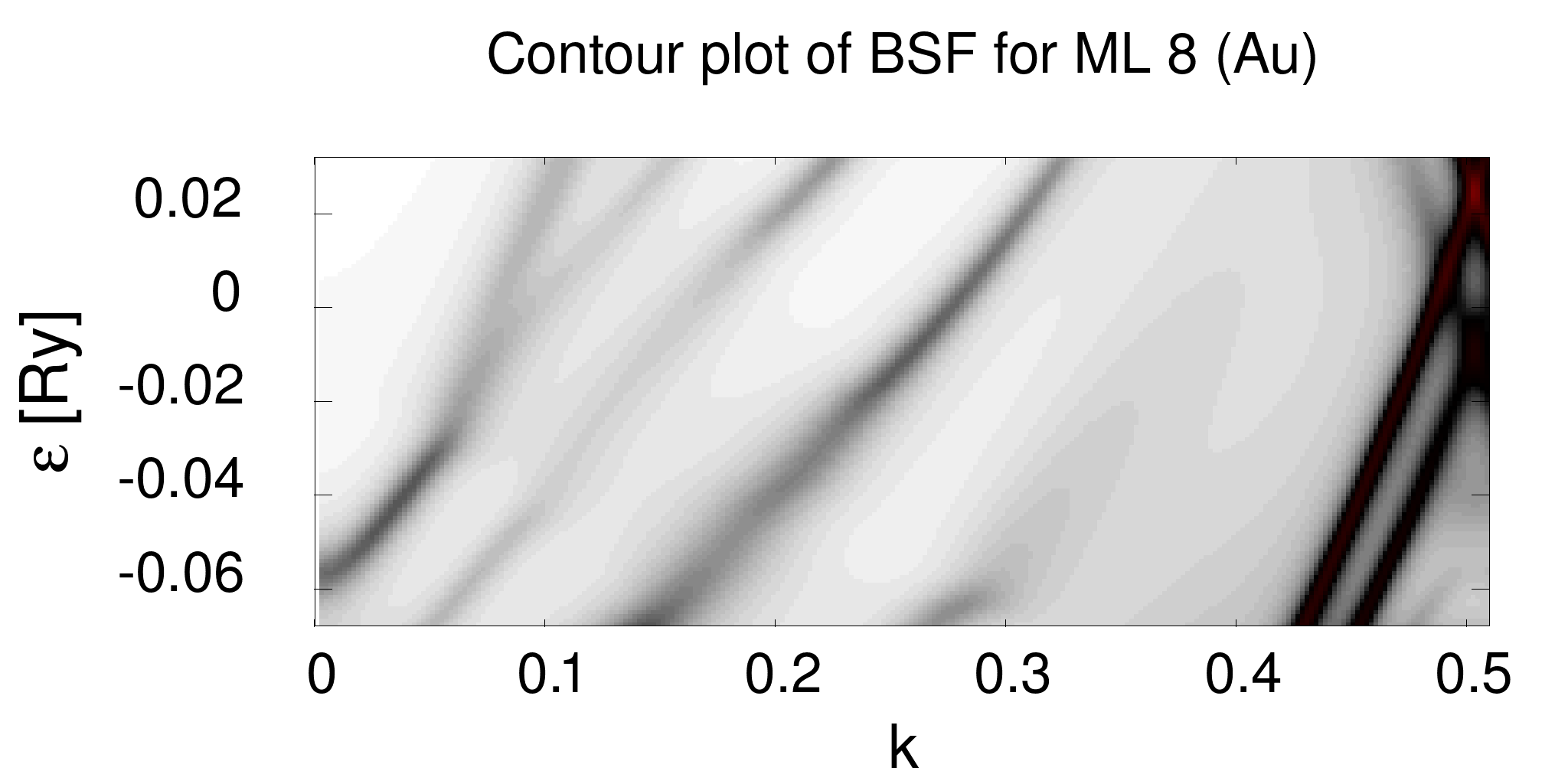}
    \caption{Layer resolved contour plot of Bloch Spectral Function~(BSF) along $k_x=\sqrt{2} k_y$ for Au thickness 9~ML}
    \label{fig:bsf_l9}
\end{figure*}

Here we summarize the main findings of the electronic structure calculations in the normal state.
One can see the contour plot of the layer resolved Bloch Spectral Function (BSF) at the Fermi level in the Nb/Au/Nb system
with $t_\textrm{Au}=$3~ML in Fig.~\ref{fig:fs_l3} and with 9~ML in Fig.~\ref{fig:fs_l9},
In Fig.~\ref{fig:bsf_l3} and Fig.~\ref{fig:bsf_l9} the contour plot of the layer resolved Bloch Spectral Function (BSF) is shown
along $k_x=\sqrt{2} k_y$ where $\varepsilon=0$~Ry corresponds to the Fermi level for systems with 3 and 9~ML Au thickness, respectively.

\begin{figure*}[htb]
    \centering
    \includegraphics[width=0.375\linewidth]{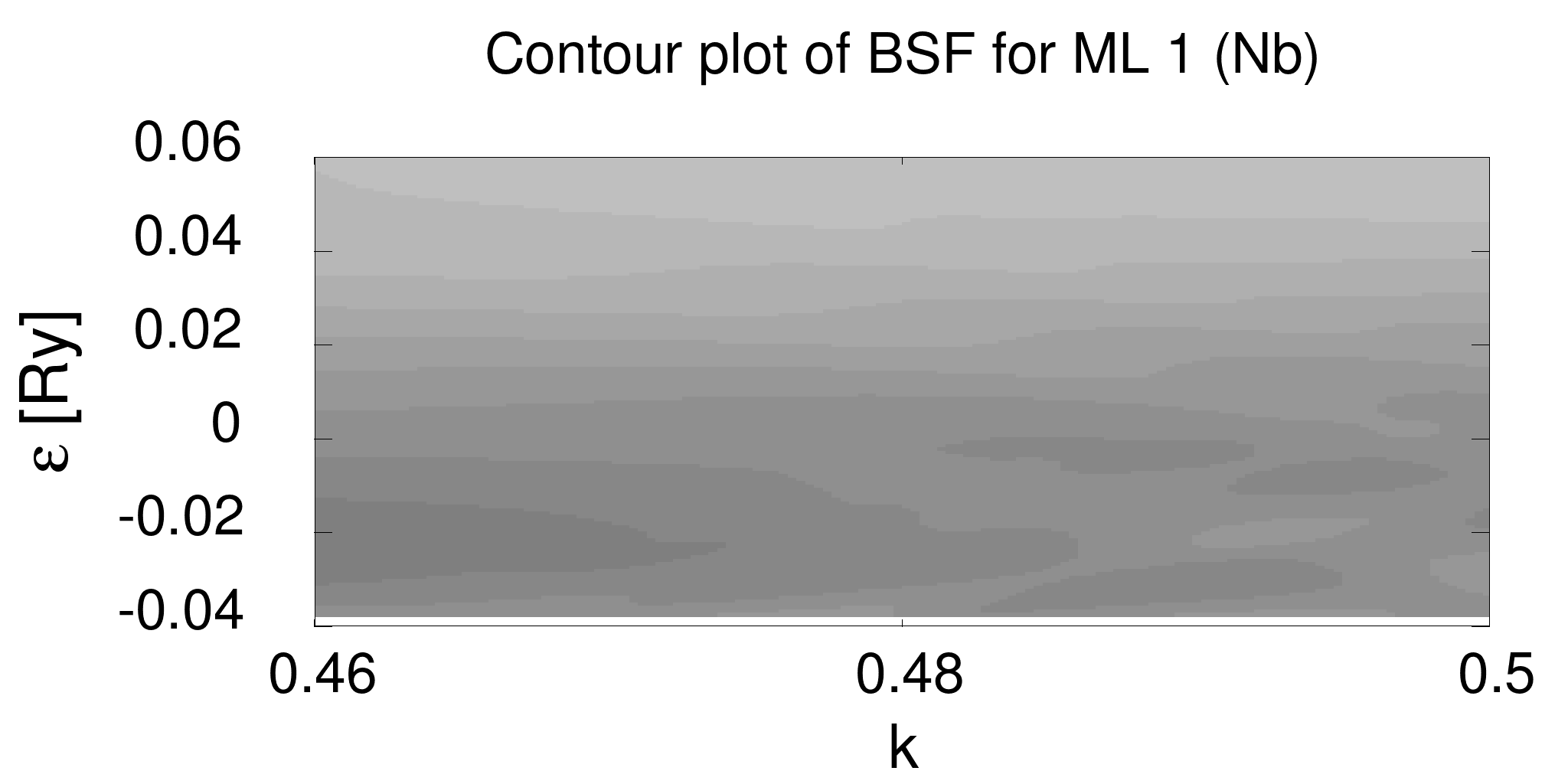}
    \includegraphics[width=0.375\linewidth]{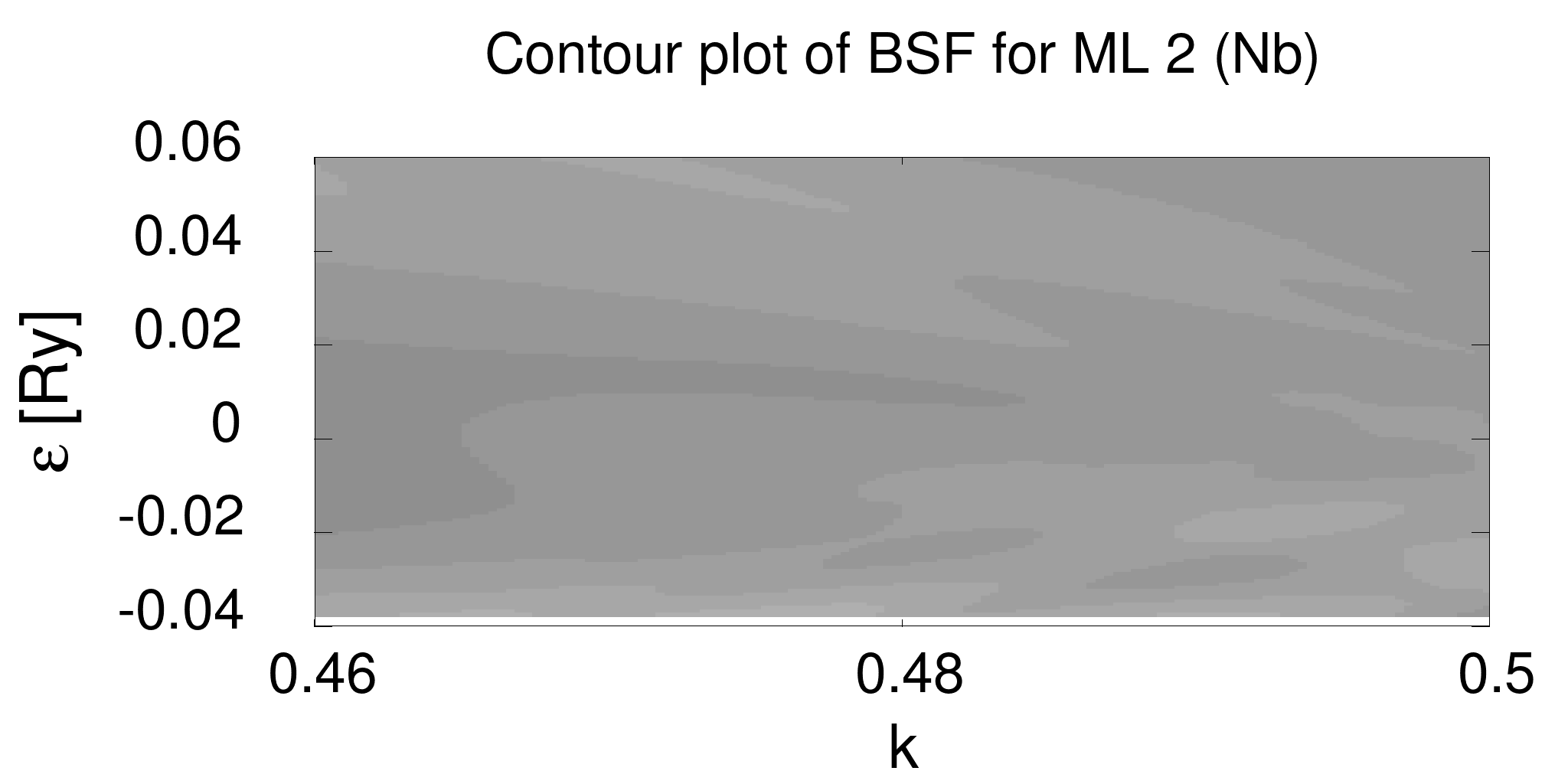}\\
    \includegraphics[width=0.375\linewidth]{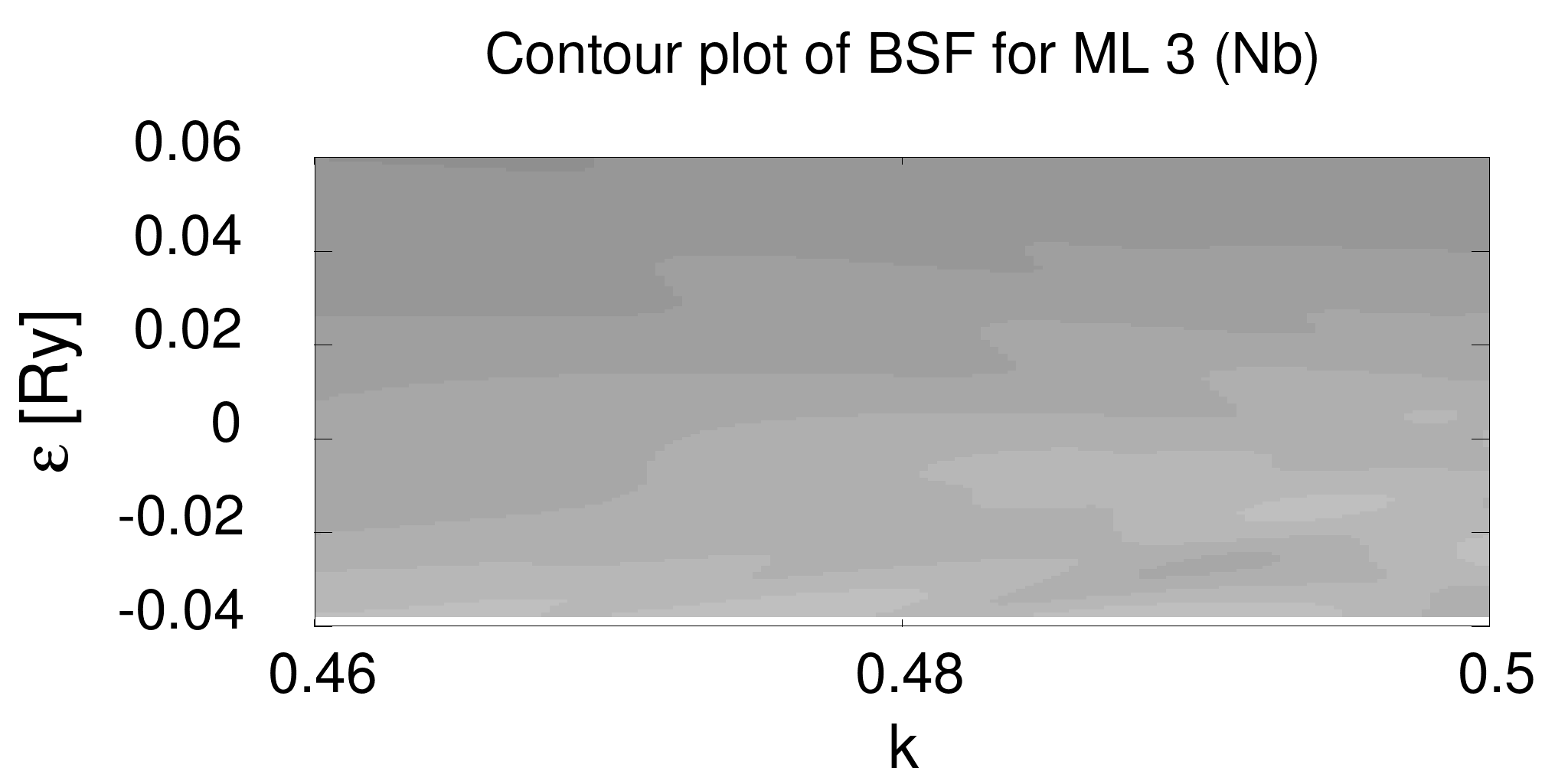}
    \includegraphics[width=0.375\linewidth]{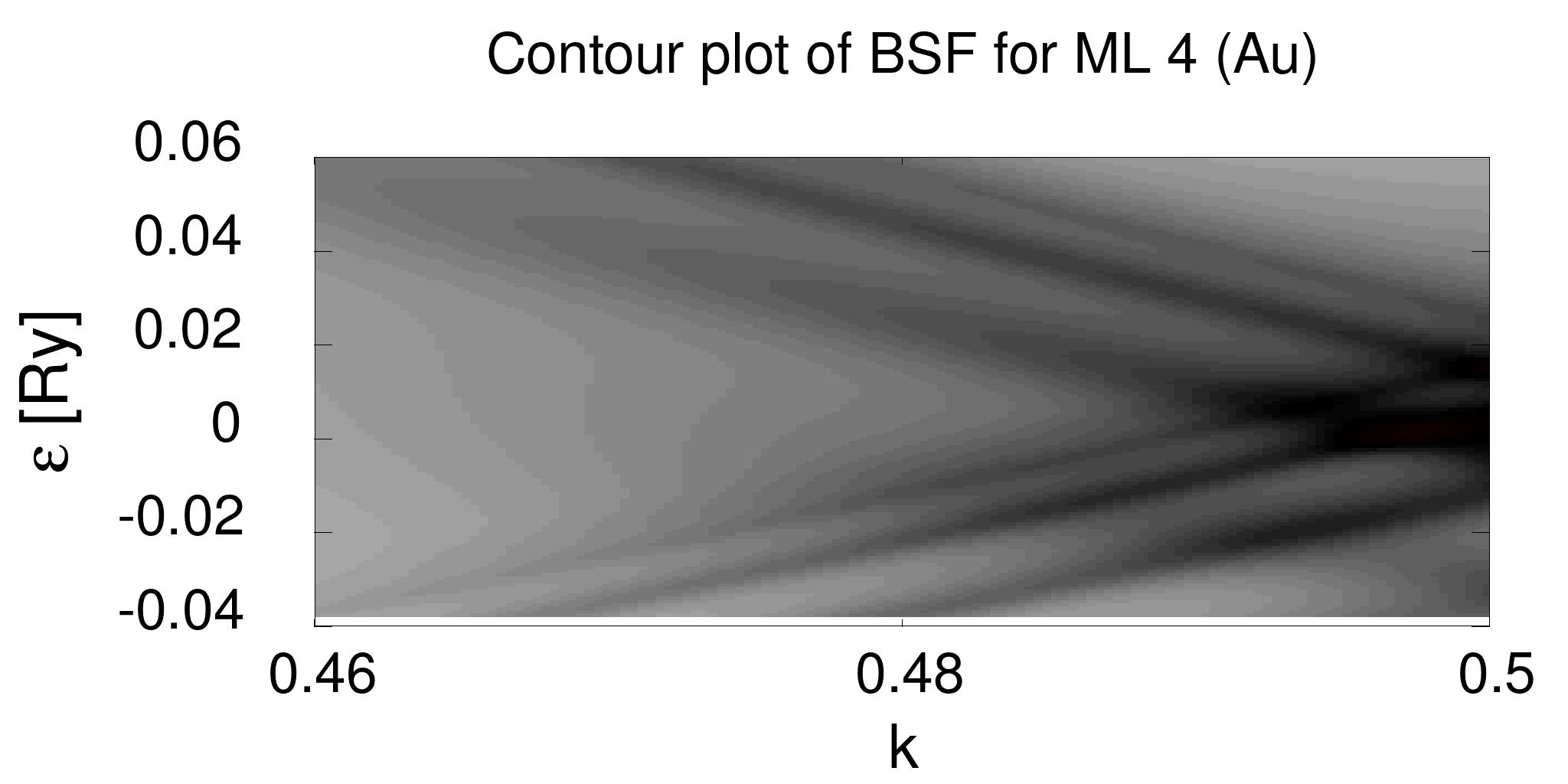}\\
    \includegraphics[width=0.375\linewidth]{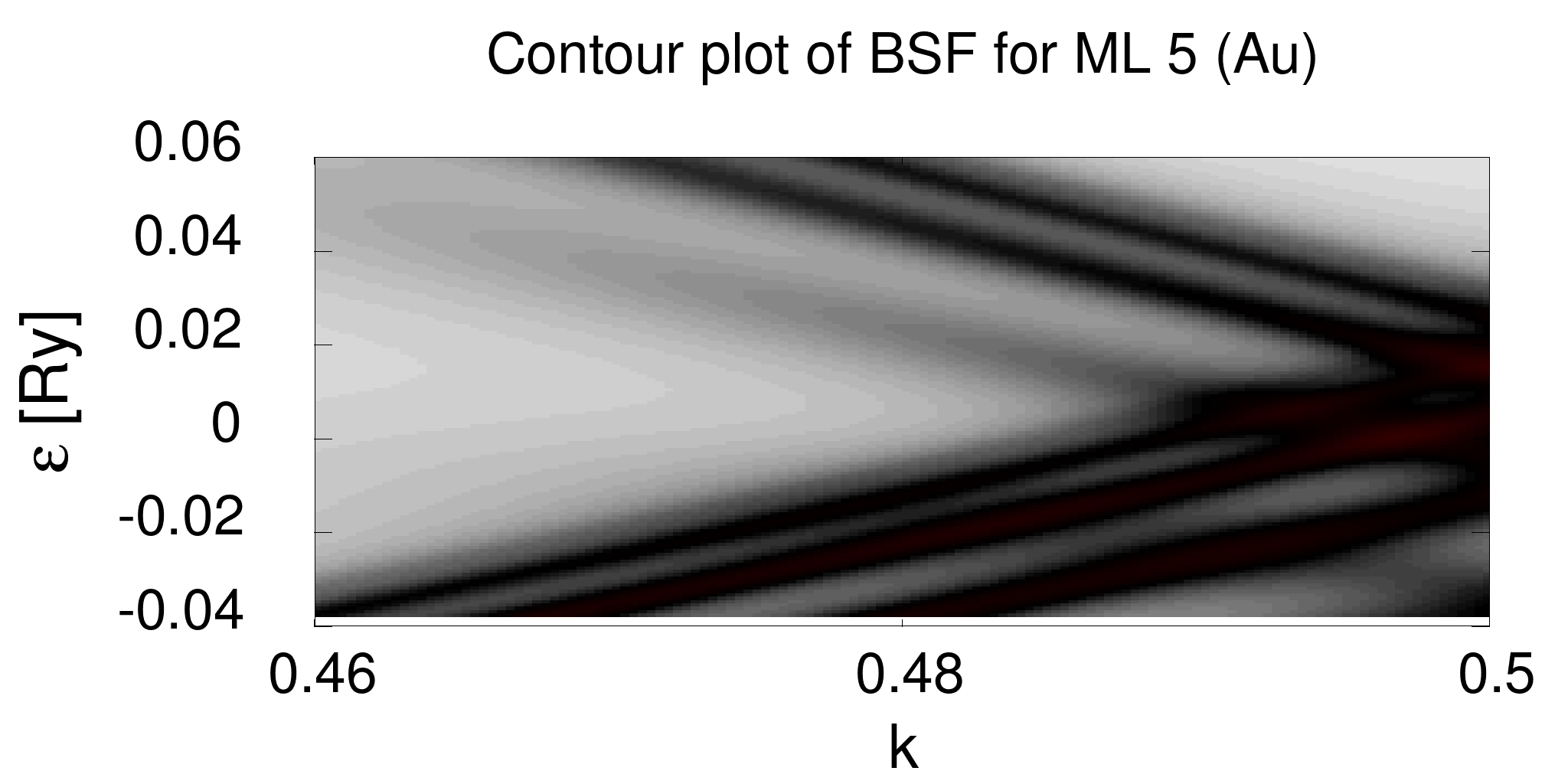}
    \includegraphics[width=0.375\linewidth]{./SupMat_Figures/bsf_l12-5.png}\\
    \includegraphics[width=0.375\linewidth]{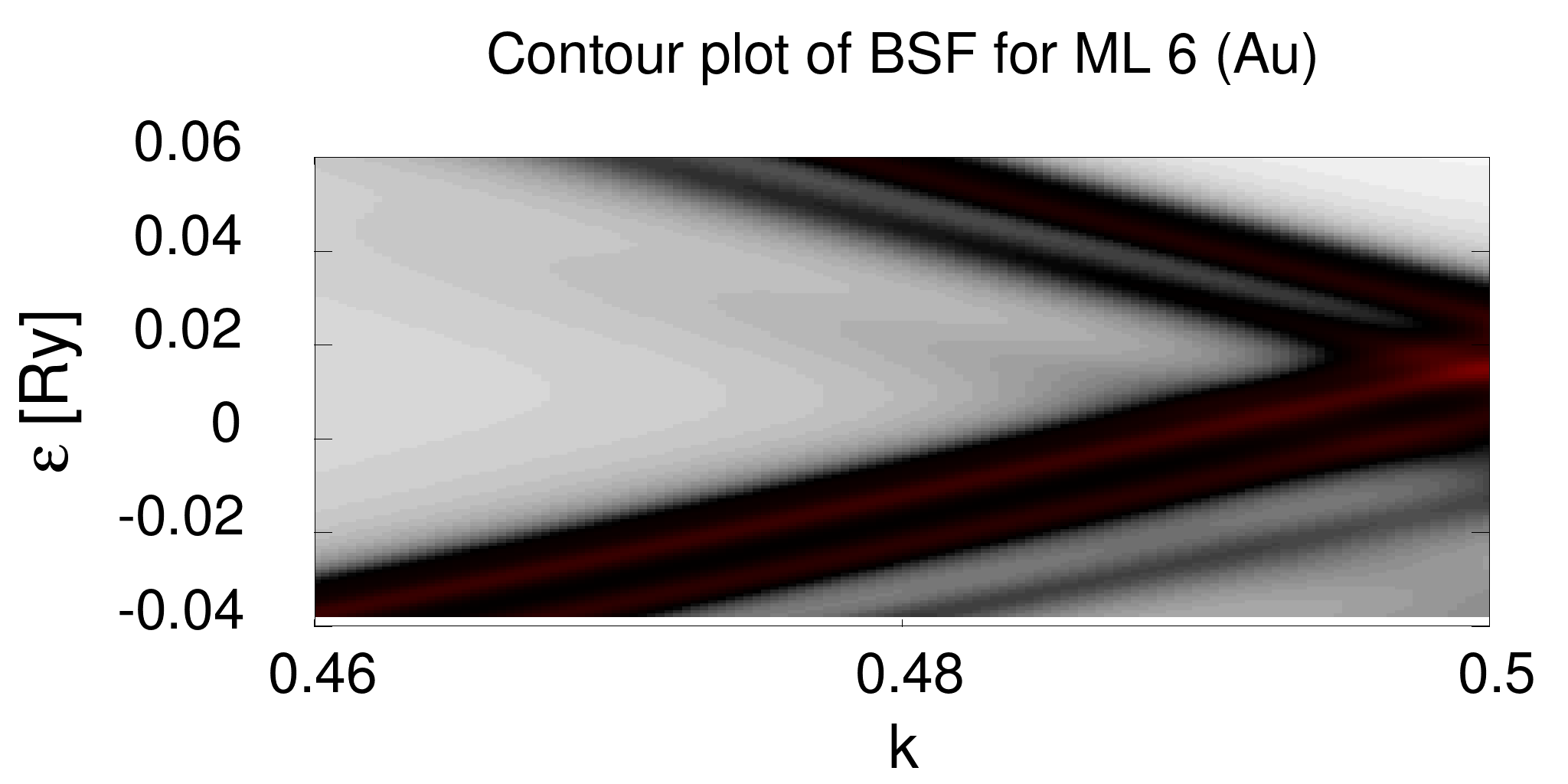}
    \includegraphics[width=0.375\linewidth]{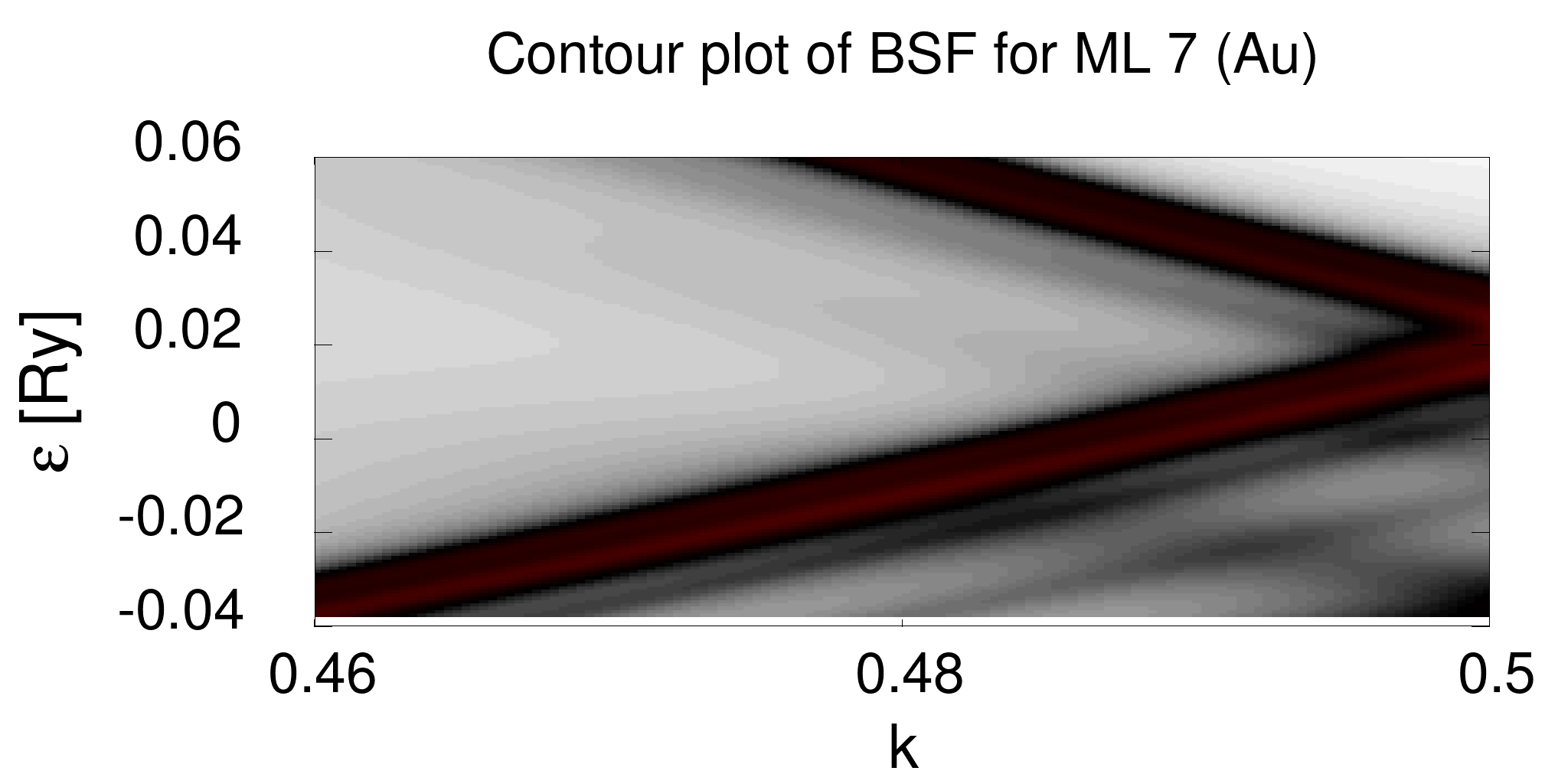}\\
    \includegraphics[width=0.375\linewidth]{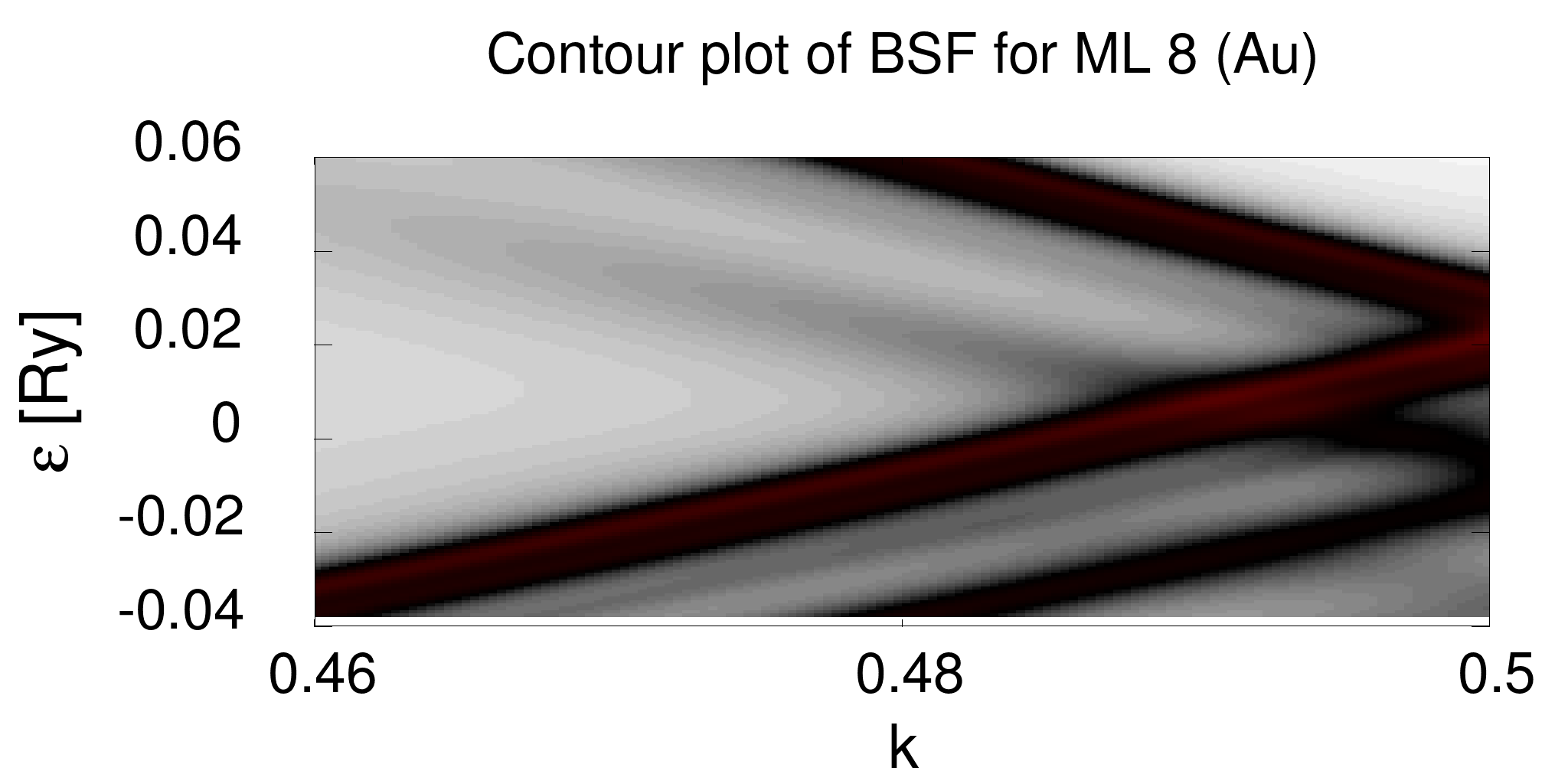}
    \includegraphics[width=0.375\linewidth]{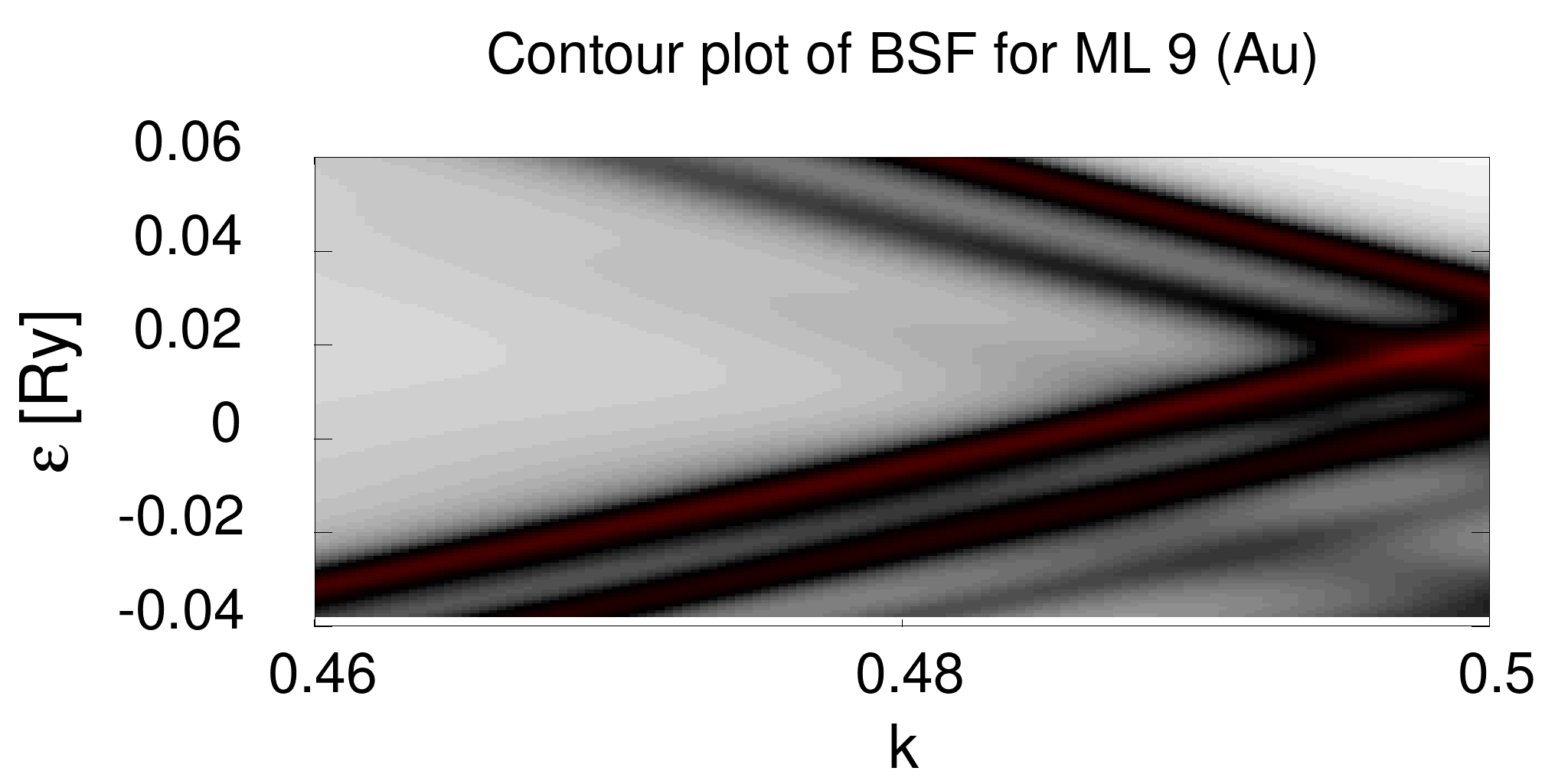}\\
    \includegraphics[width=0.375\linewidth]{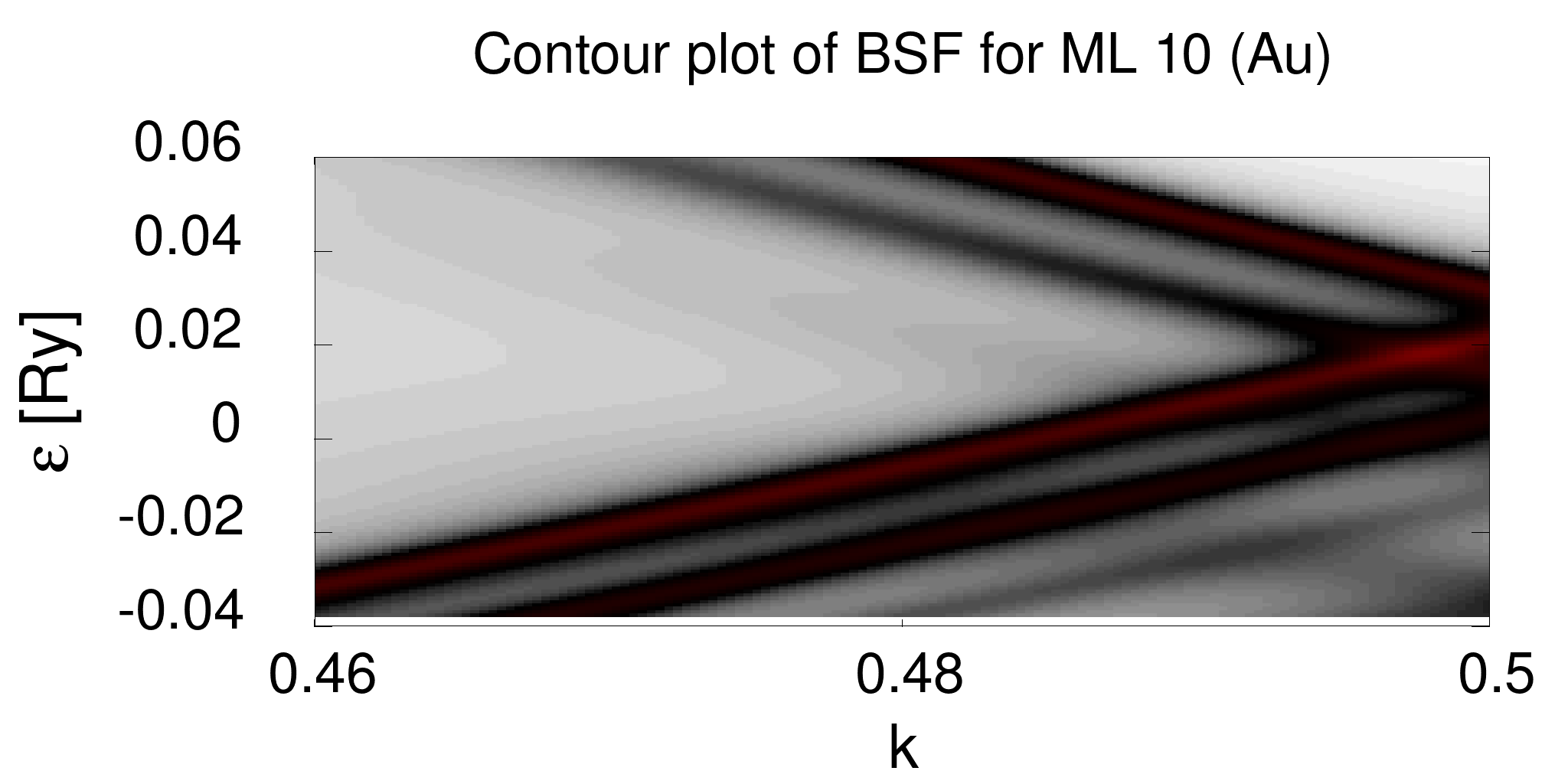}
    \caption{Layer resolved contour plot of Bloch Spectral Function~(BSF) along $k_x=\sqrt{2} k_y$ for Au thickness 12~ML
             at the edge of Brillouin zone showing the complexity and Dirac-like states}
    \label{fig:bsf_l12}
\end{figure*}

Comparing the system with 3 and 9~ML Au thicknesses one can obtain several striking features.
When these figures are viewed as a sequence, one can
recognize the signatures of confinement. Where the DOS in
the bulk niobium is low, the states in the Au layer are confined, as they
cannot scatter into the Nb. In those regions where the DOS is high in
the Nb, the states in the Au layer are smeared out, as here the
appropriate electrons can scatter more easily into the other side
of the interface. The confined states in the Au layer, therefore, can
be regarded as quantum-well (QW) states. 
The quantum confinement manifests itself in a roughly $2\pi/t_{Au}$ sampling 
of the bulk Au band structure (the QW bands become denser as $t_\textrm{Au}$ is increasing).
However, another interesting feature can be observed at the edge of the BZ (see Fig.~\ref{fig:bsf_l12} for the zoomed BSF):
several bands cross each other with opposite Fermi velocities around the Fermi level forming Dirac-like states.
Important to note, that these crossings appear at the Fermi level starting from the Au thickness
$t_\textrm{Au}$=9~ML and persist for larger Au thicknesses.

\section{Quasi-particle spectrum}

The physical mechanism casuing bound-states in superconductor -- normal metal -- superconductor (SNS) junctions
is usually associated with the Andreev reflection of quasi-particles at the superconductor -- normal metal interface
which leads to standing waves of Andreev-reflected particles and holes in the normal metal (due to a pairing potential well).
On the other hand, in Refs.~\onlinecite{Gyorffy1996, Csire2015, Csire2016, Csire2016b} the following understanding of the quasi-particle spectrum has emerged:
the bound states are trapped inside the middle normal region by the same mechanism which determines electronic structure of semiconducting
heterostructures, i.e. the electrons concerned cannot penetrate deep into the superconducting
regions because there are no allowed states available for them. The superconductivity of
the outer layers (Nb) leads to minigaps wherever particle-like and hole-like bands cross making the BCS pairing possible.

\begin{figure*}[htb]
    \centering
    \includegraphics[width=0.25\linewidth]{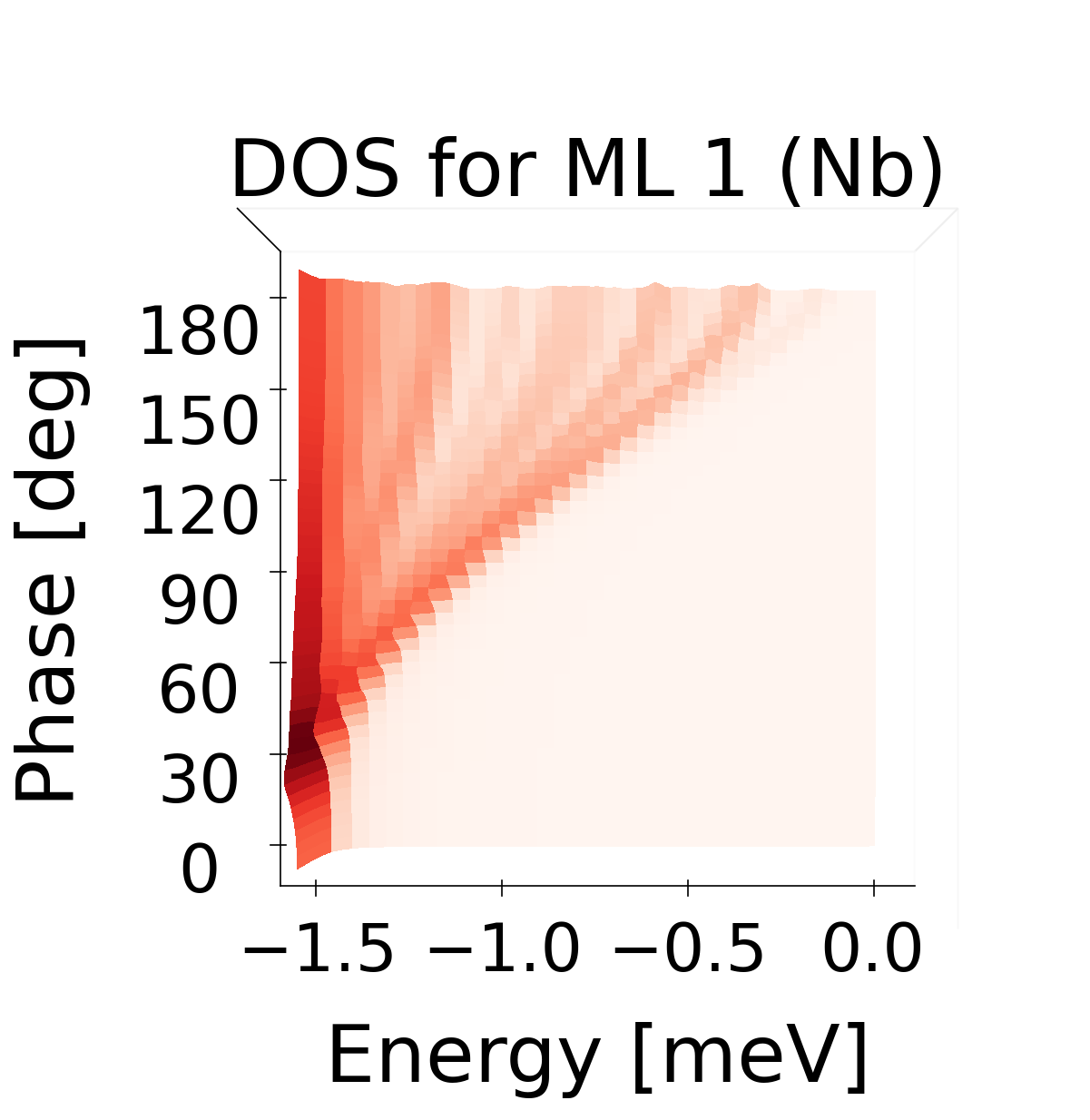}
    \includegraphics[width=0.25\linewidth]{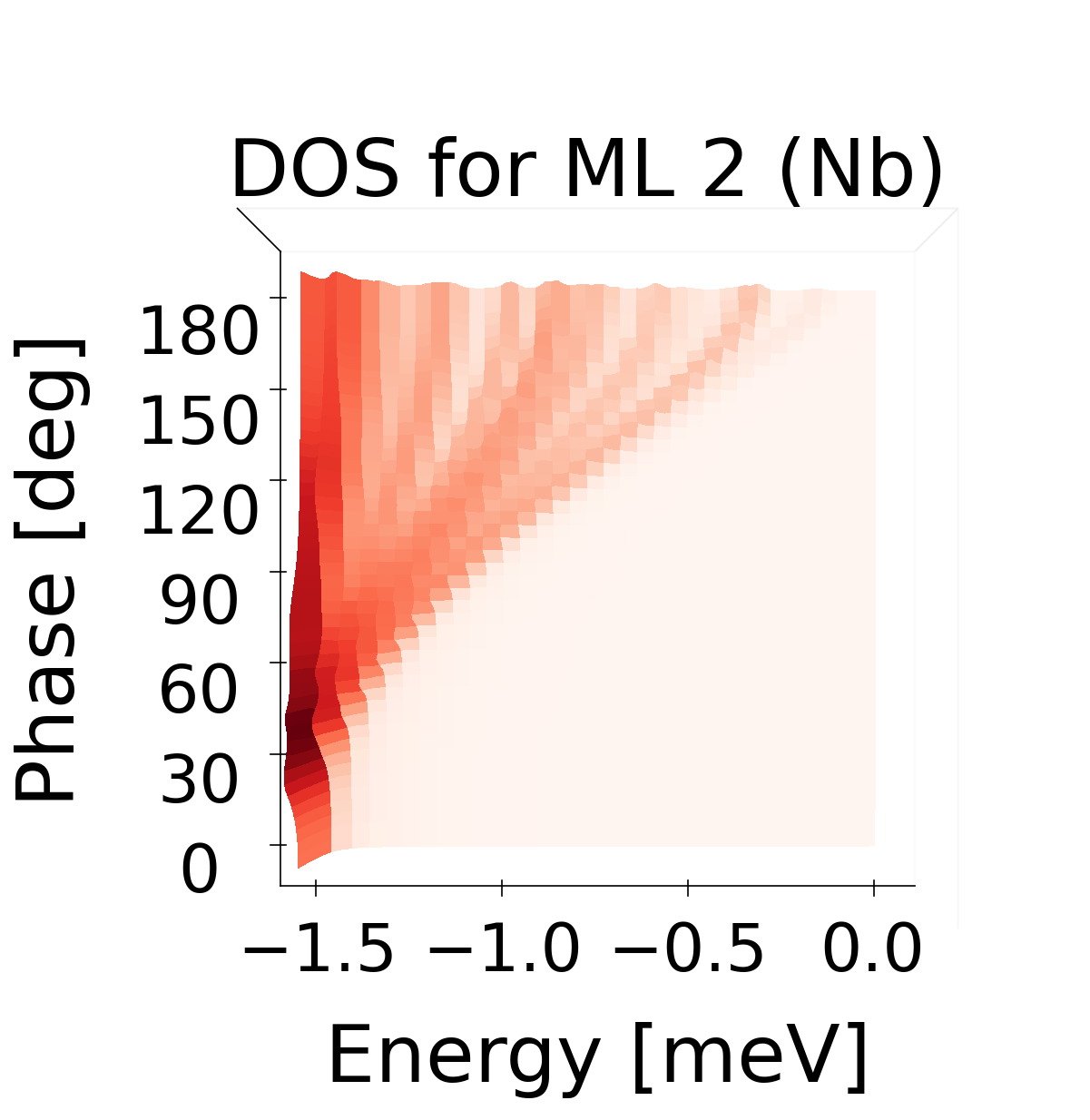}
    \includegraphics[width=0.25\linewidth]{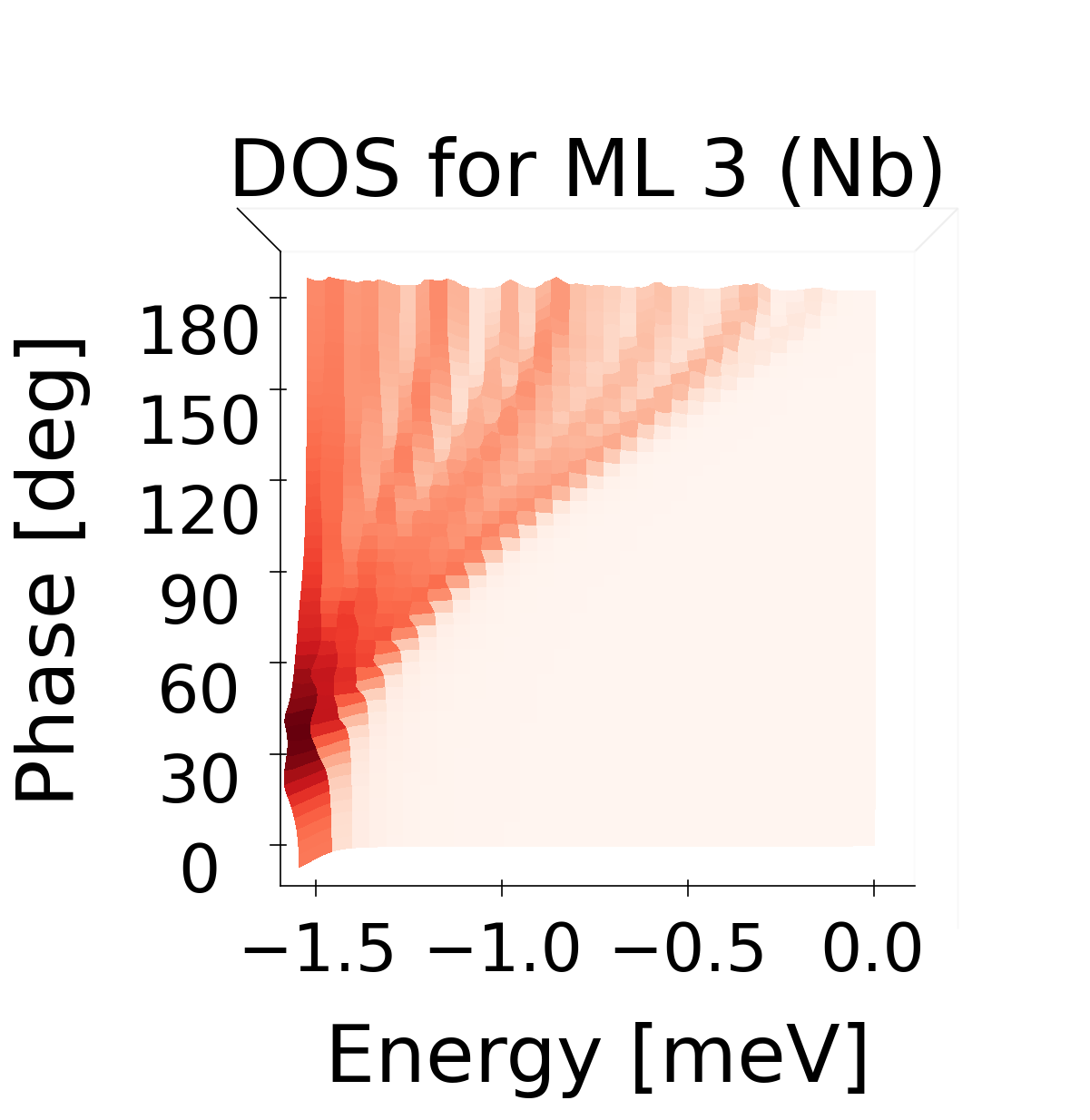}\\
    \includegraphics[width=0.25\linewidth]{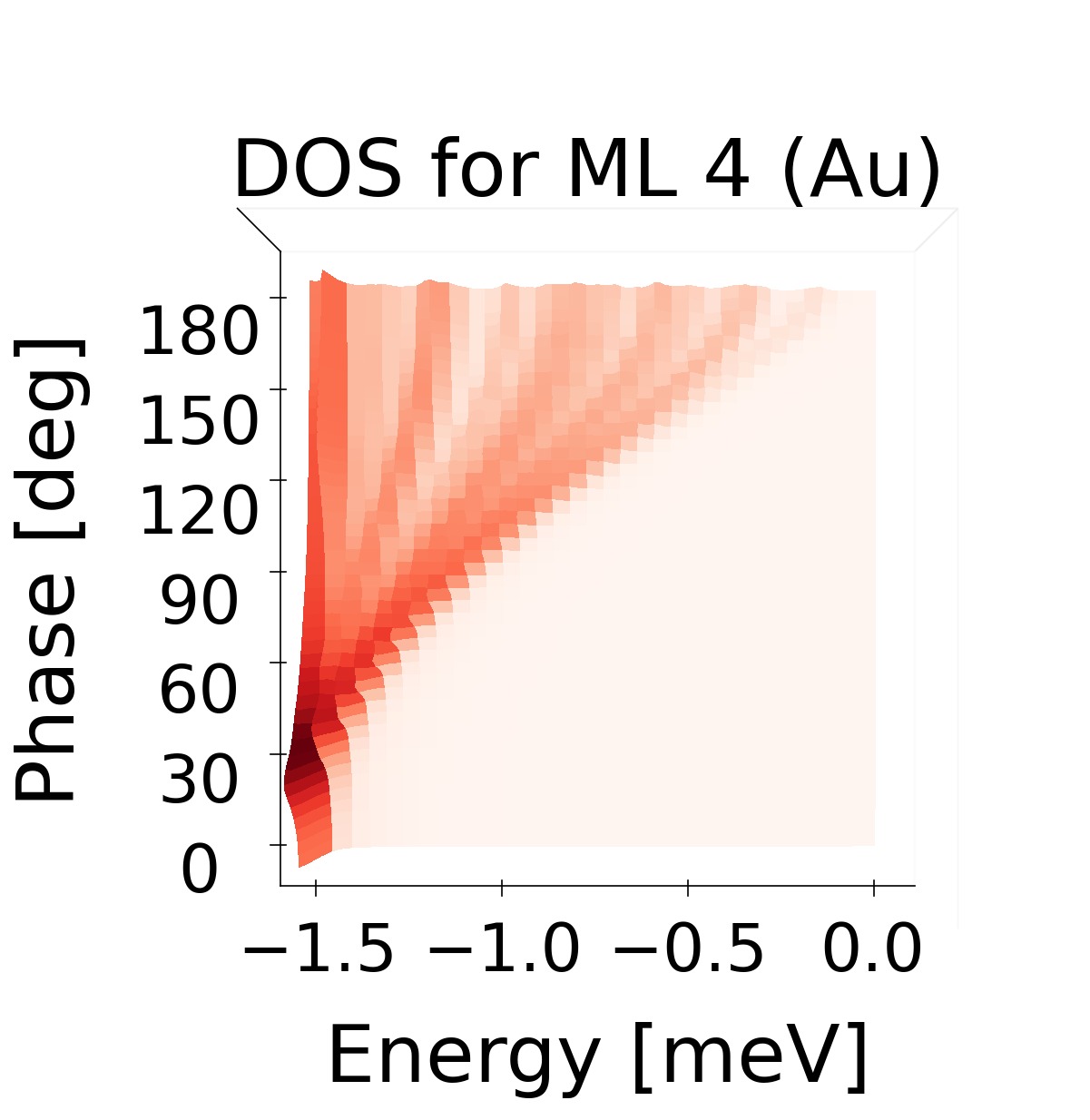}
    \includegraphics[width=0.25\linewidth]{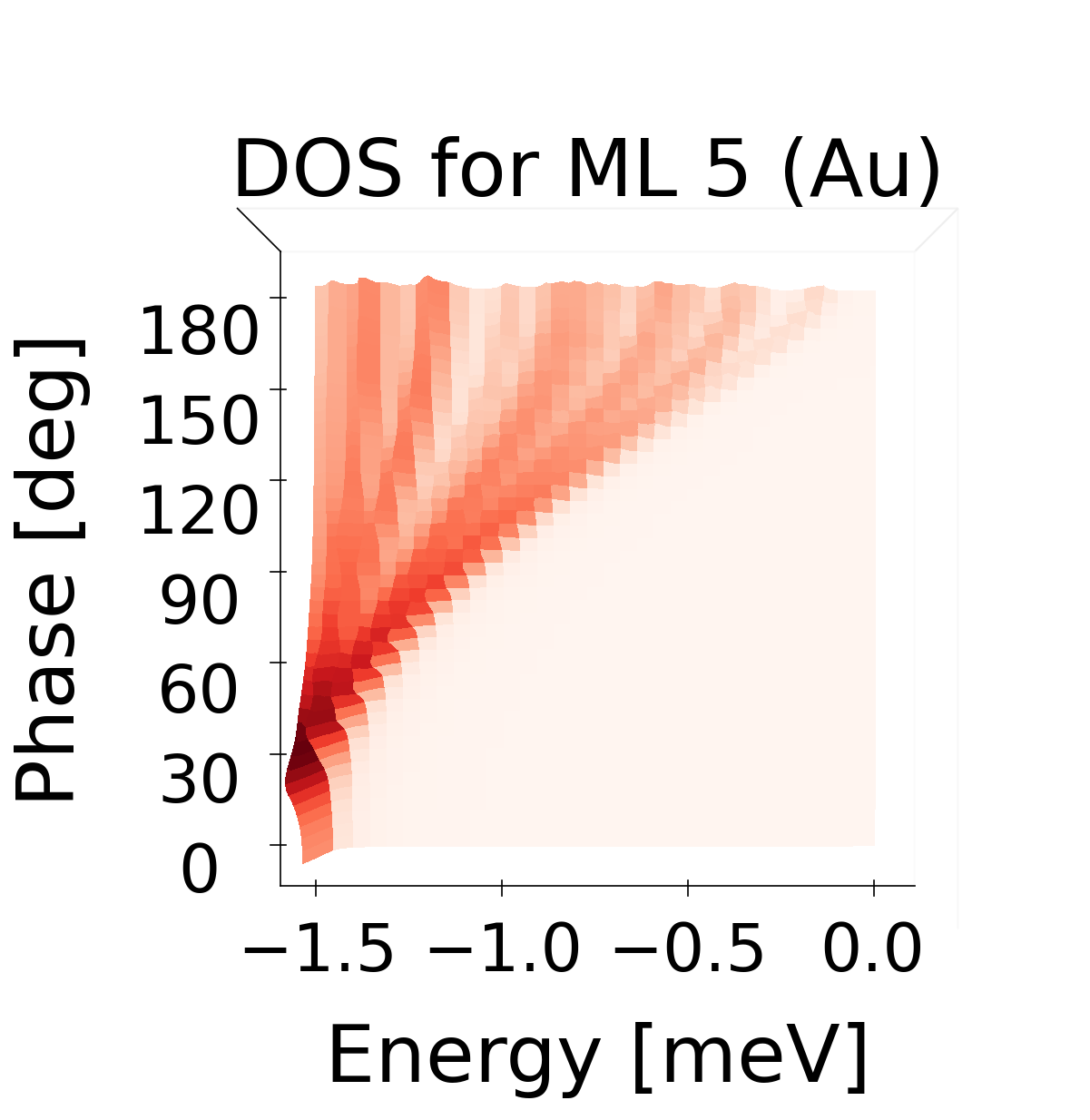}
    \includegraphics[width=0.25\linewidth]{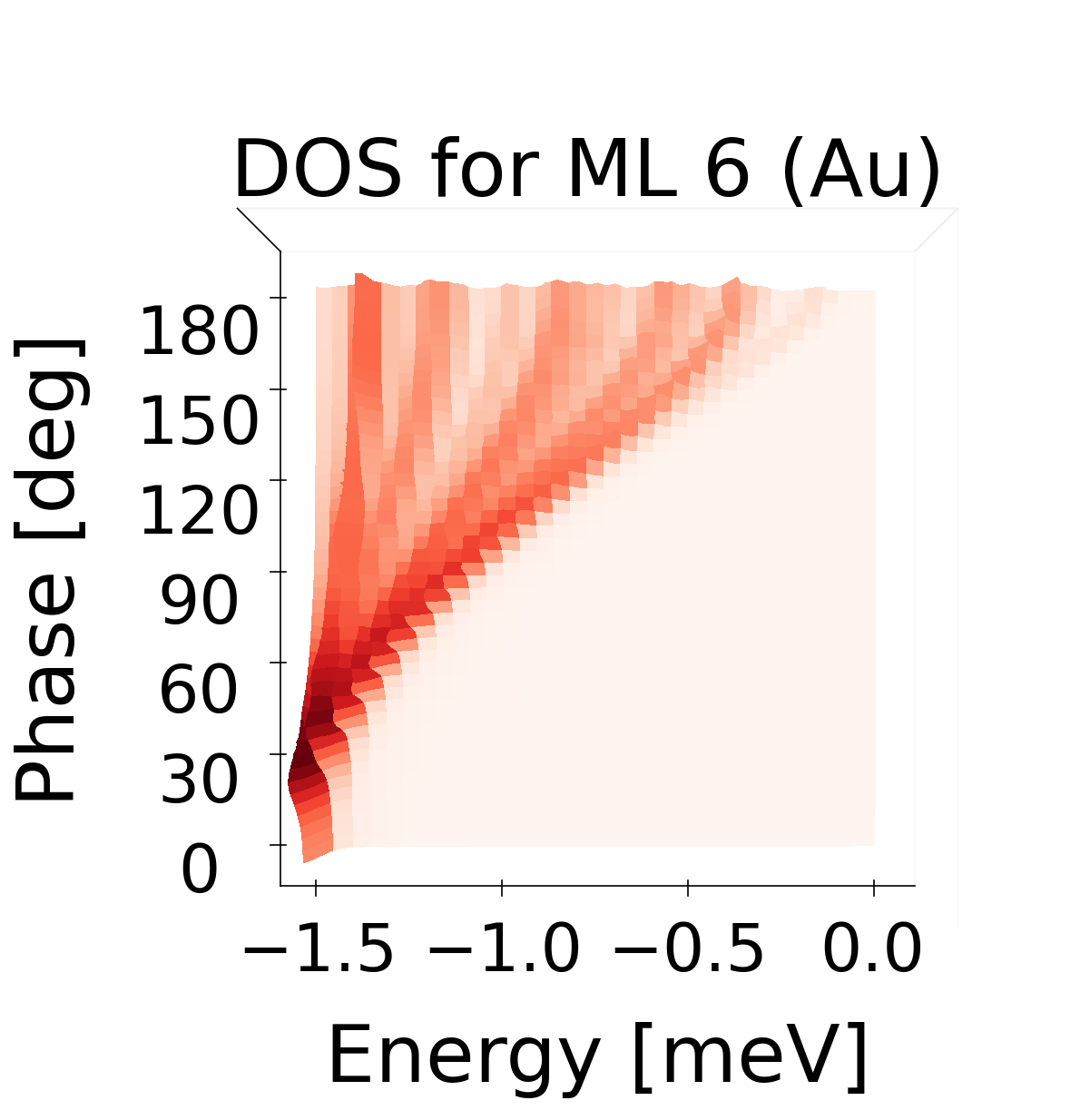}
    \caption{Layer resolved quasi-particle DOS for gold thickness 6~ML (up to the middle of the sample)}
    \label{fig:qp_l6}
\end{figure*}

\begin{figure*}[htb]
    \centering
    \includegraphics[width=0.25\linewidth]{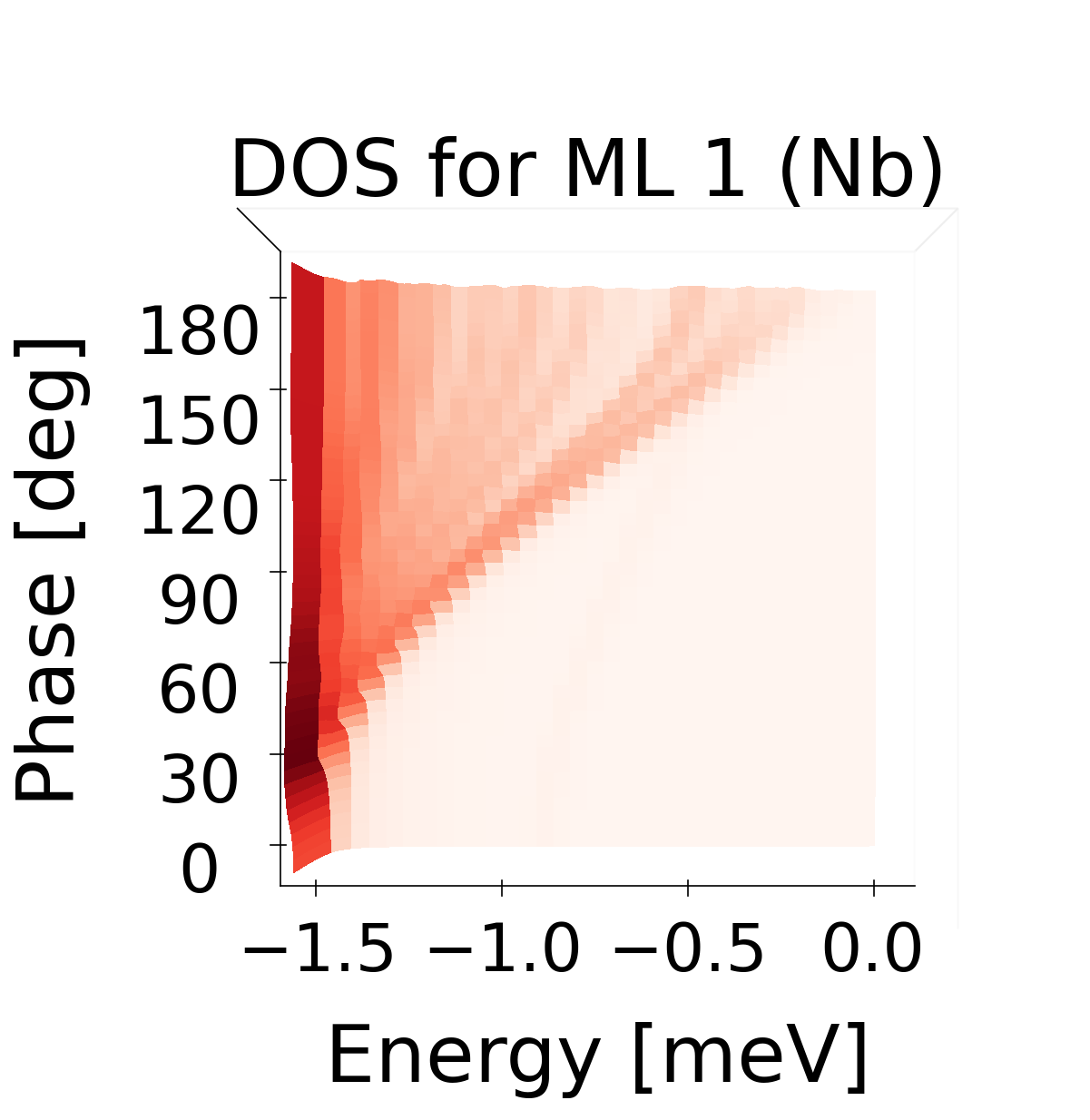}
    \includegraphics[width=0.25\linewidth]{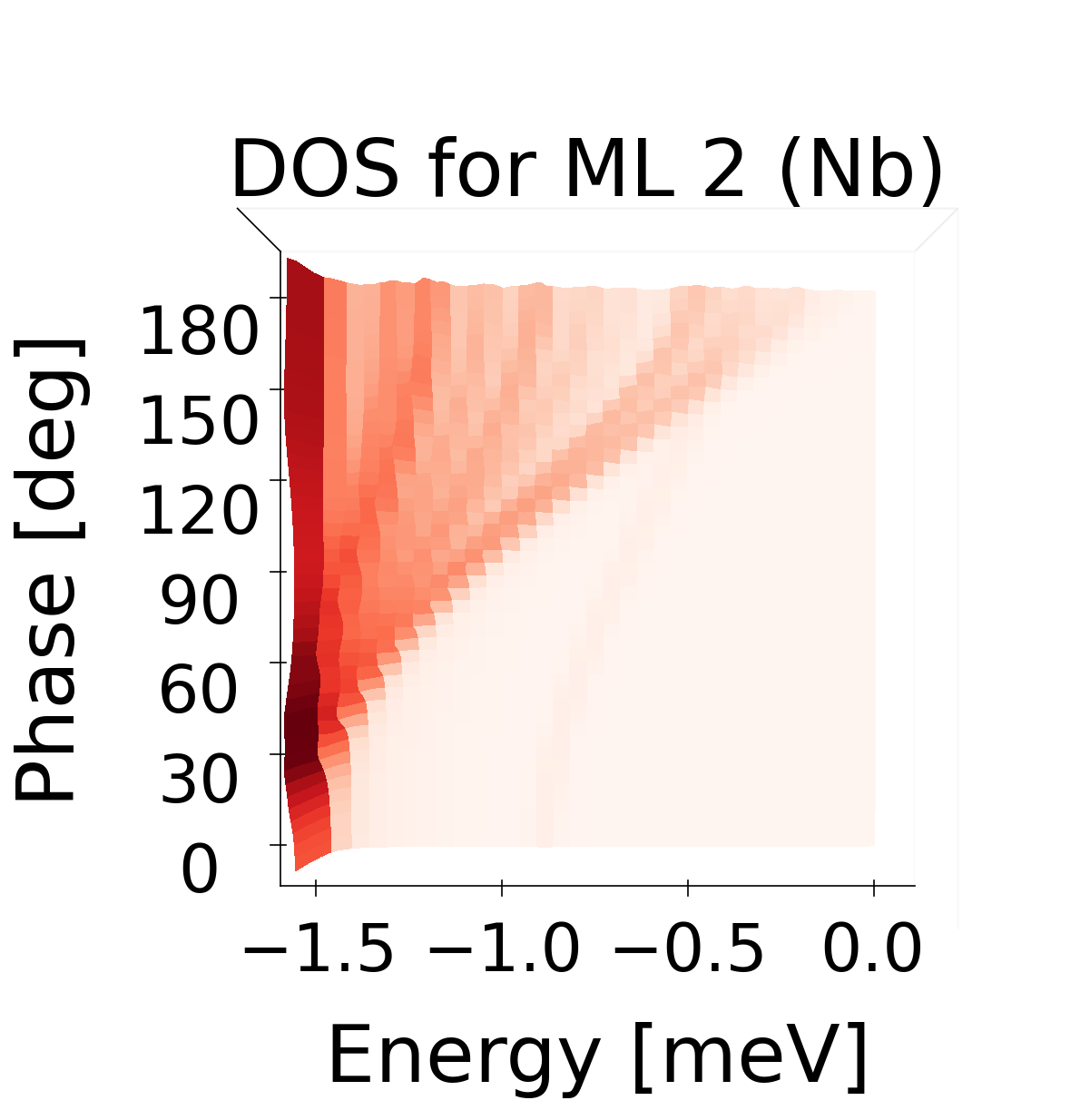}
    \includegraphics[width=0.25\linewidth]{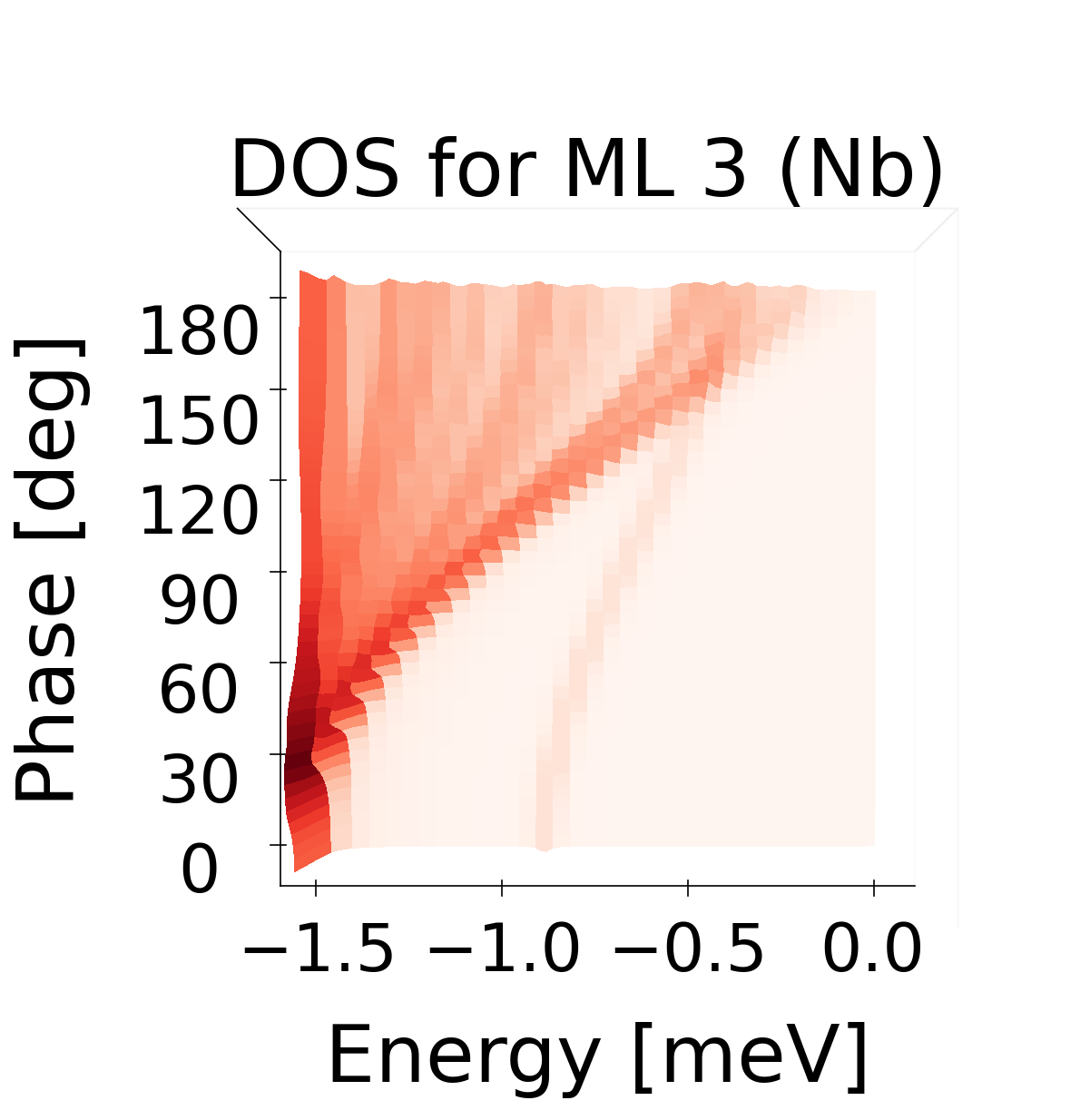}\\
    \includegraphics[width=0.25\linewidth]{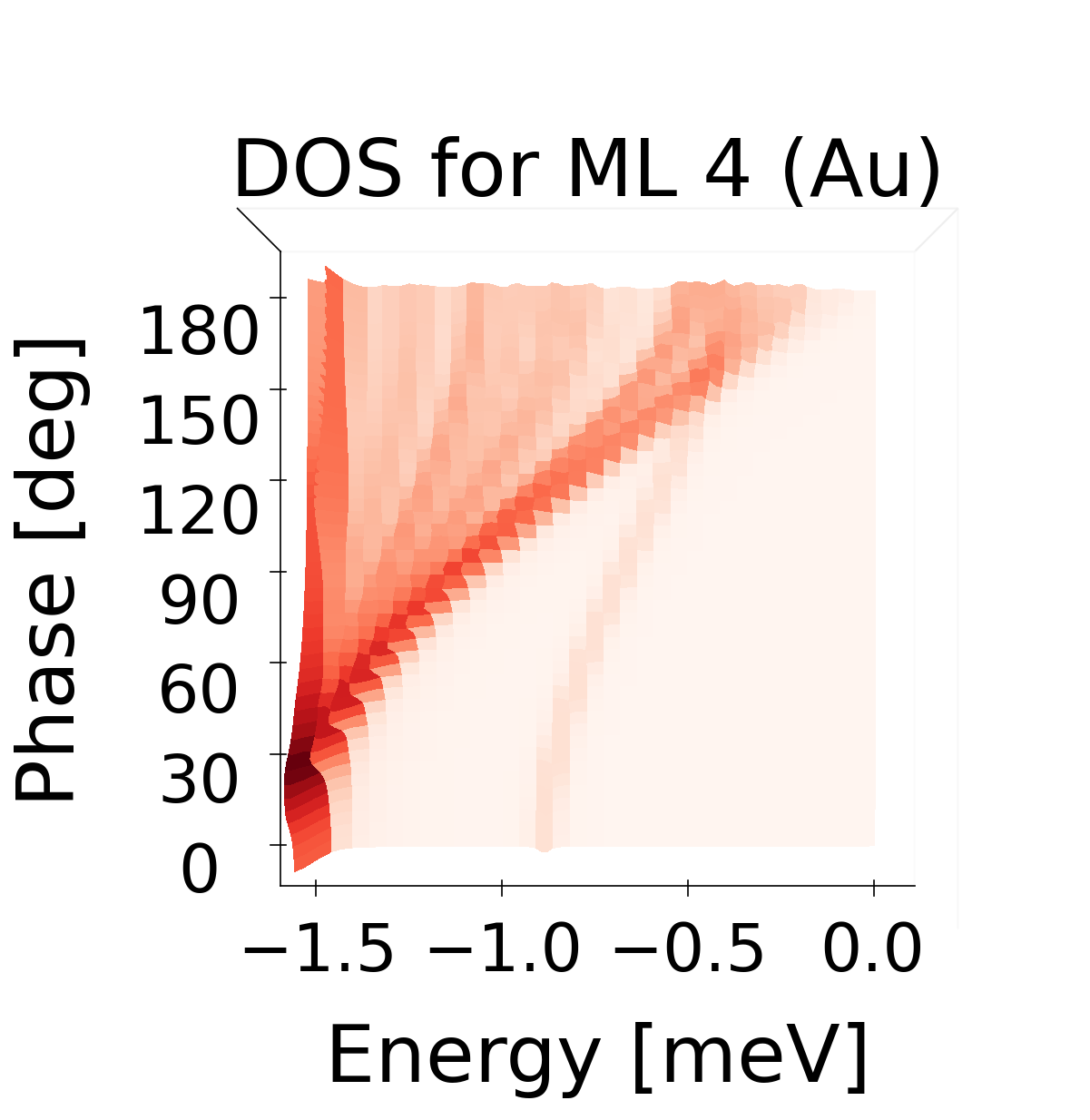}
    \includegraphics[width=0.25\linewidth]{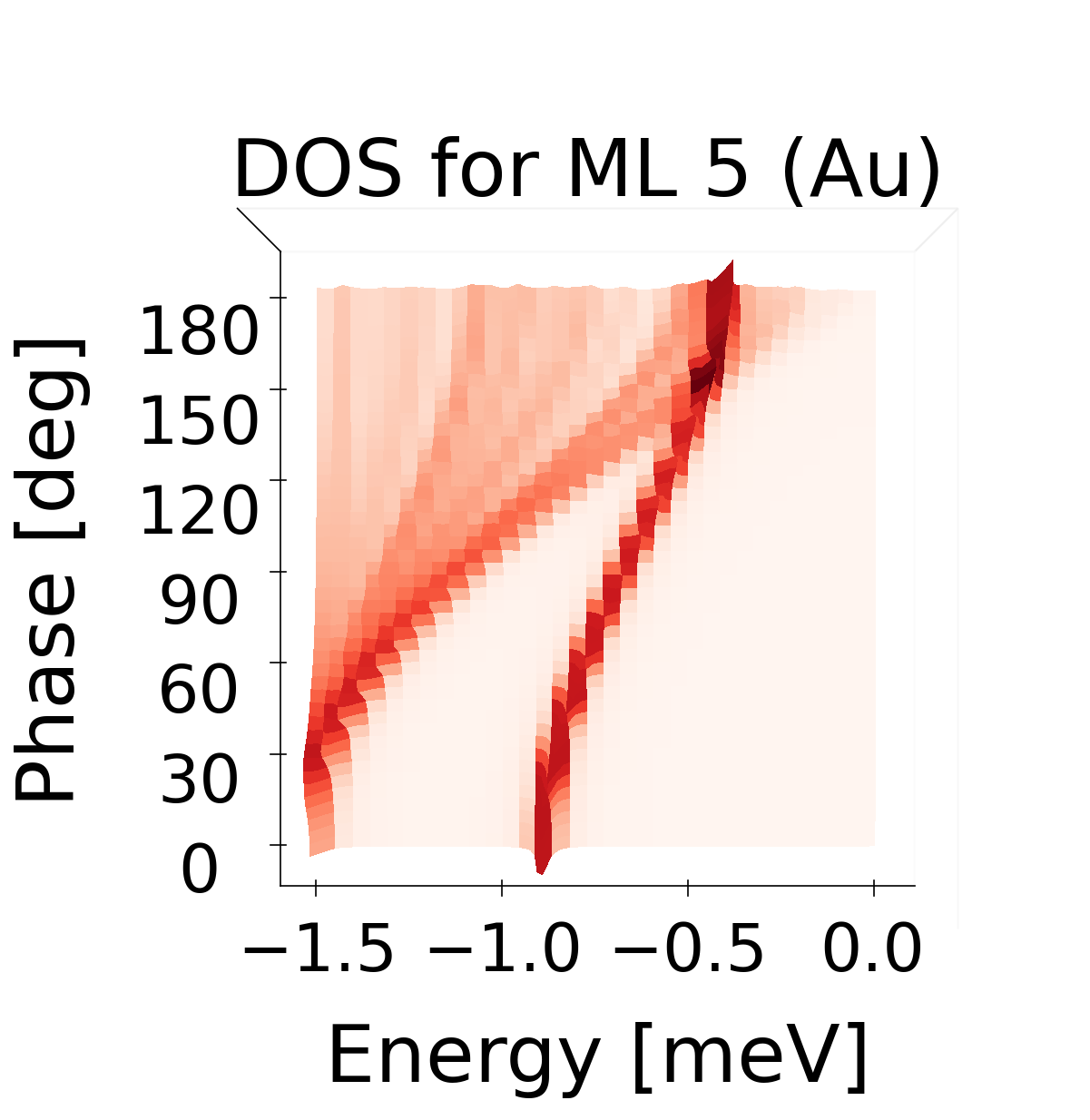}
    \includegraphics[width=0.25\linewidth]{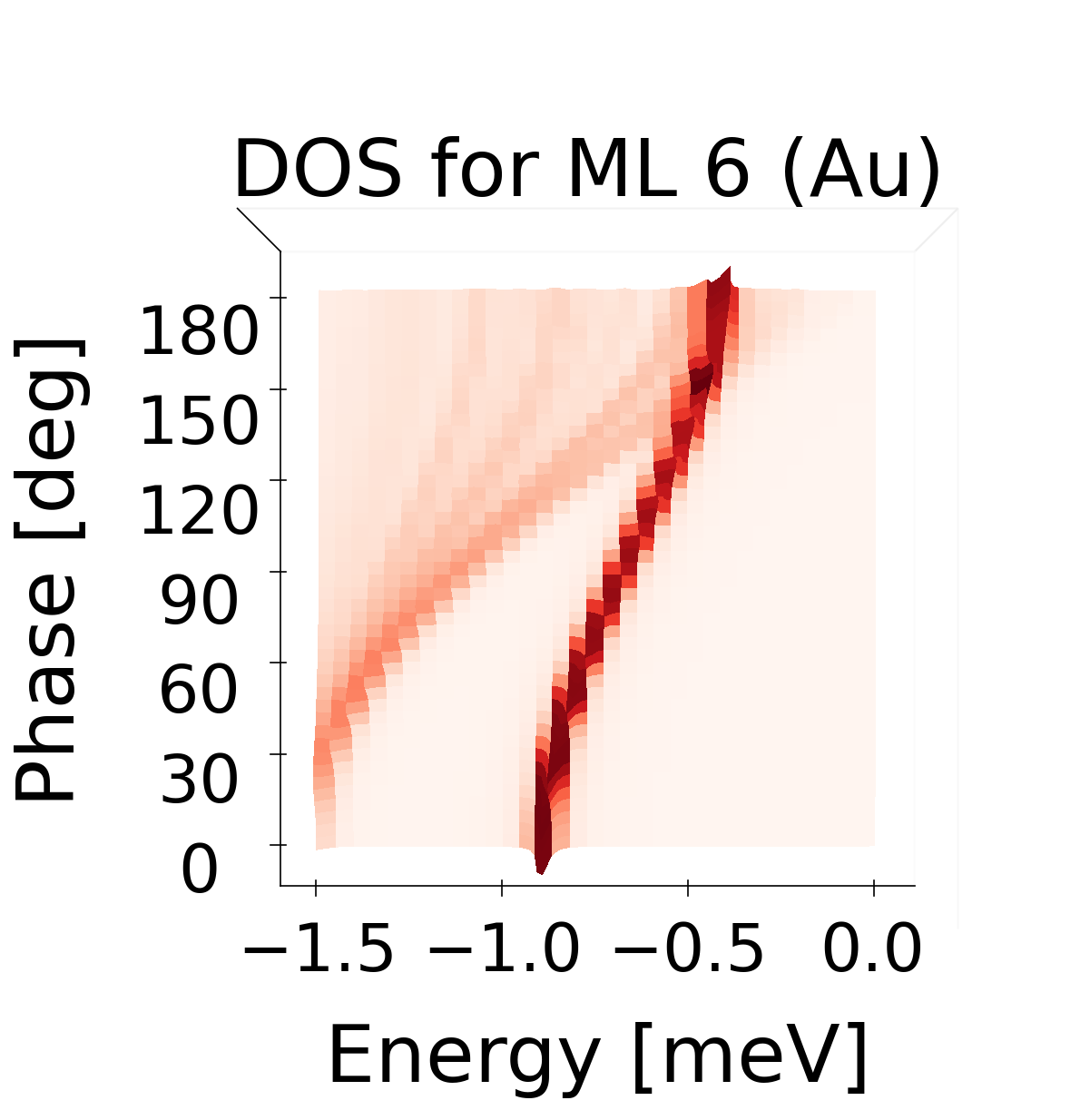}\\
    \includegraphics[width=0.25\linewidth]{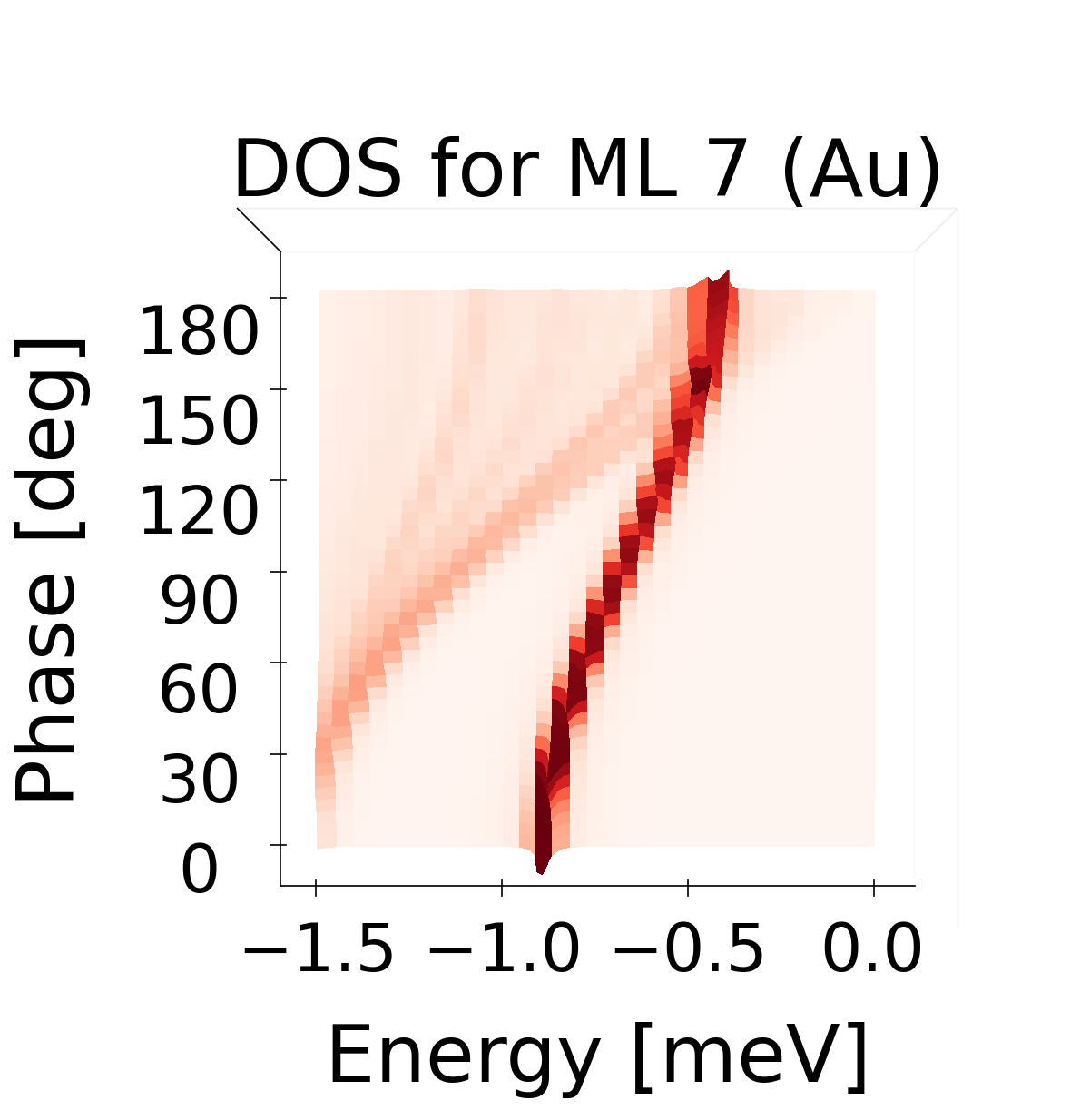}
    \includegraphics[width=0.25\linewidth]{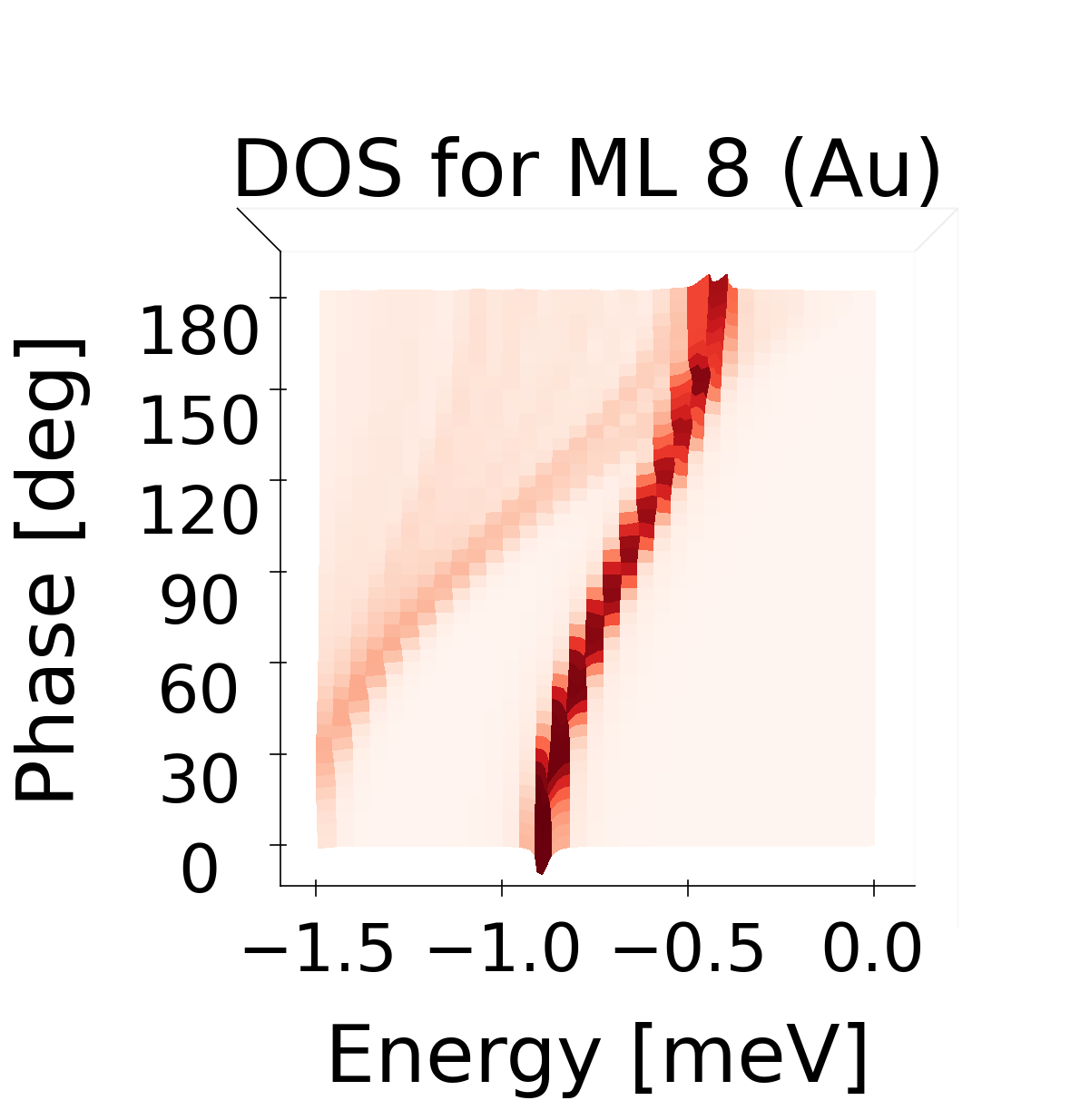}
    \caption{Layer resolved quasi-particle DOS for gold thickness 9~ML (up to the middle of the sample)}
    \label{fig:qp_l9}
\end{figure*}

\begin{figure*}[htb]
    \centering
    \includegraphics[width=0.25\linewidth]{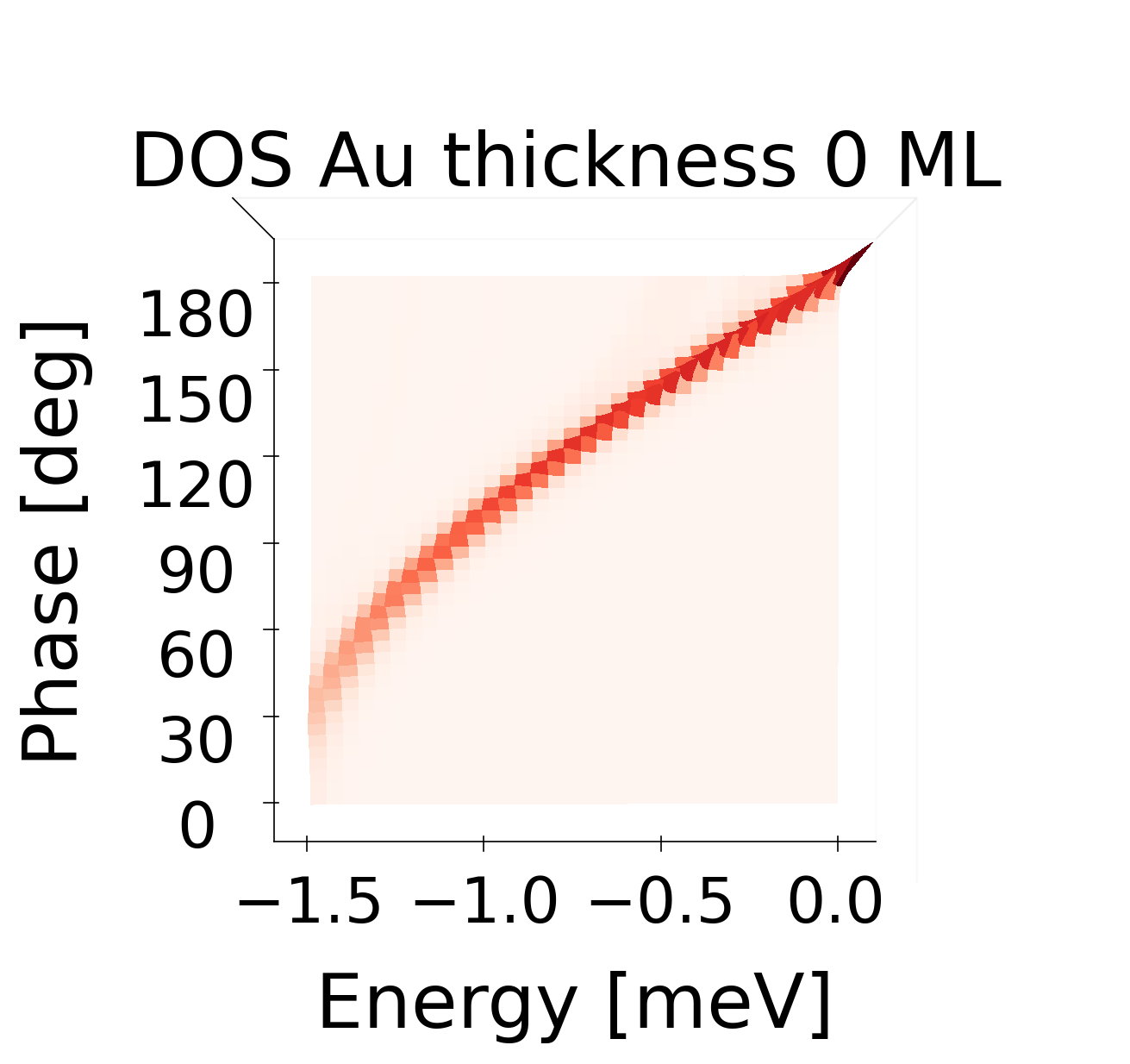}
    \includegraphics[width=0.25\linewidth]{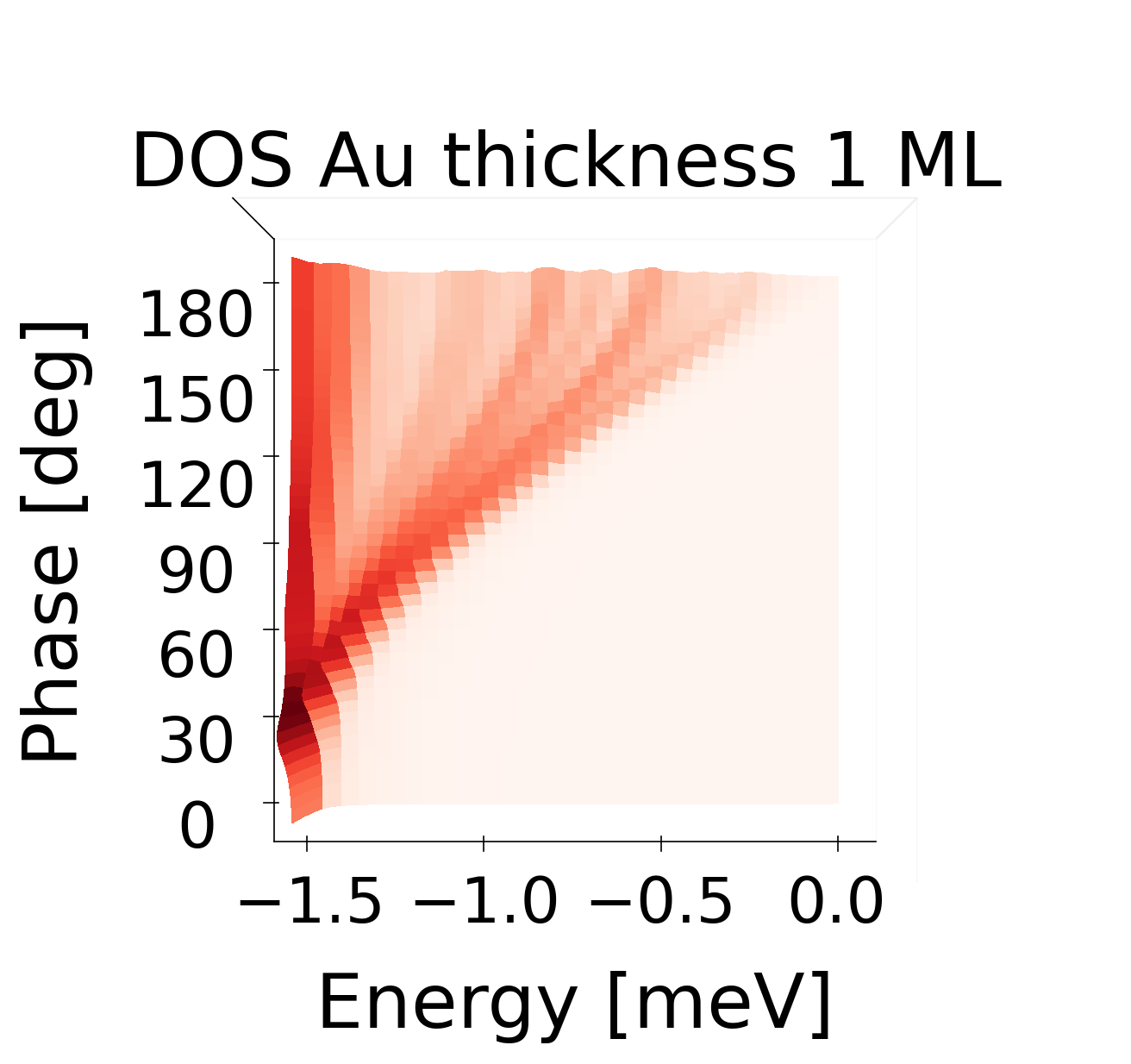}
    \includegraphics[width=0.25\linewidth]{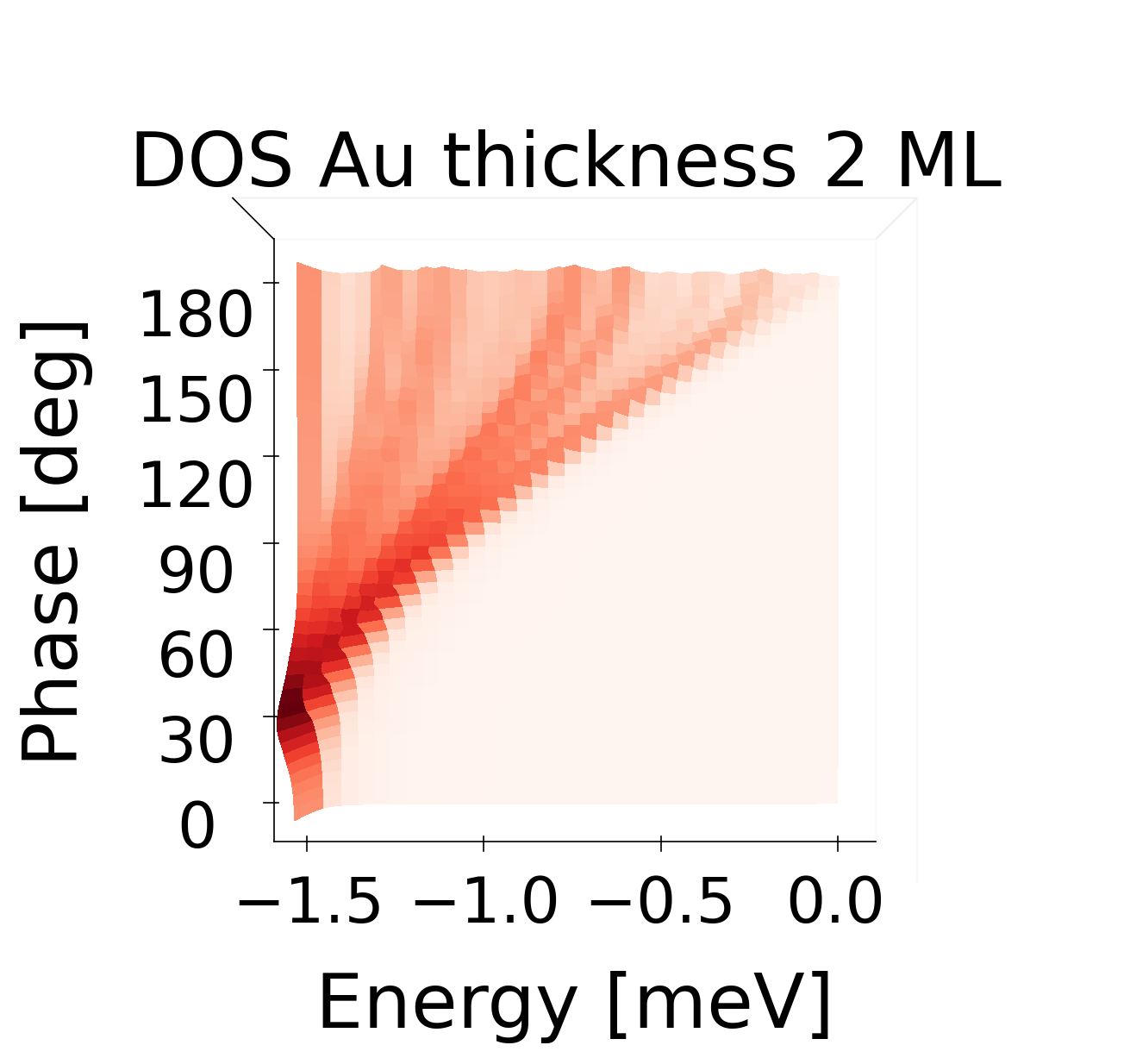}\\
    \includegraphics[width=0.25\linewidth]{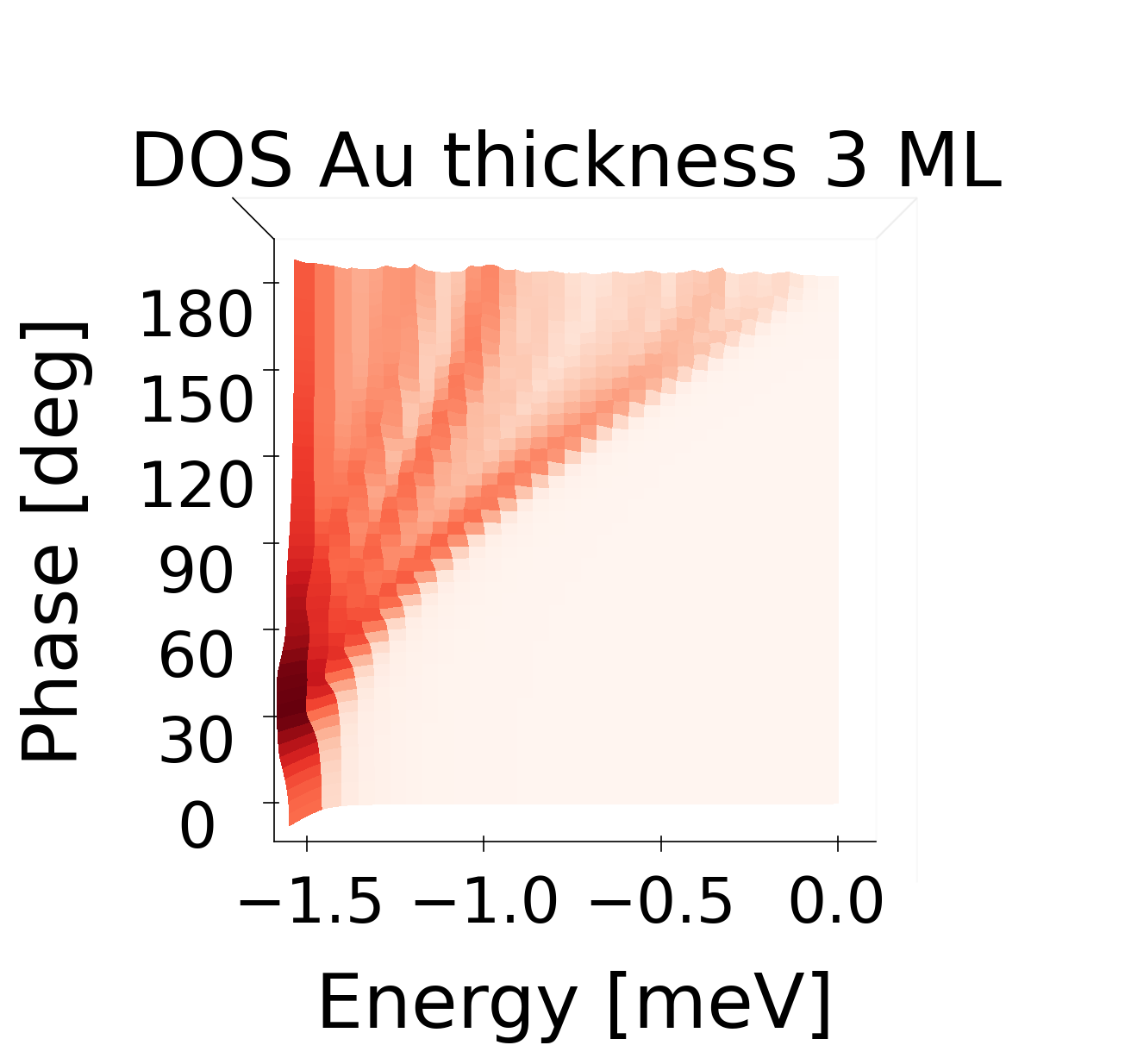}
    \includegraphics[width=0.25\linewidth]{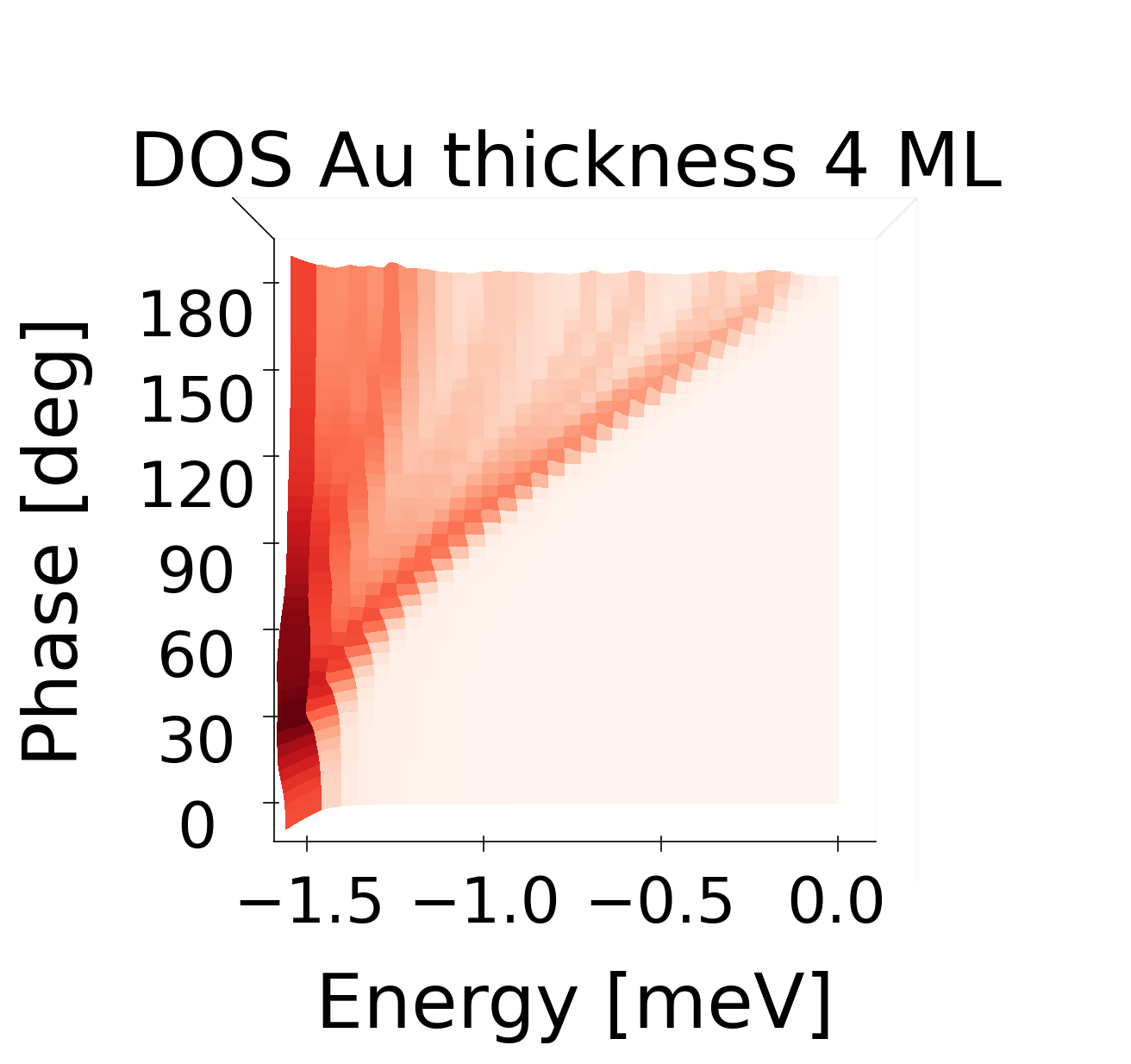}
    \includegraphics[width=0.25\linewidth]{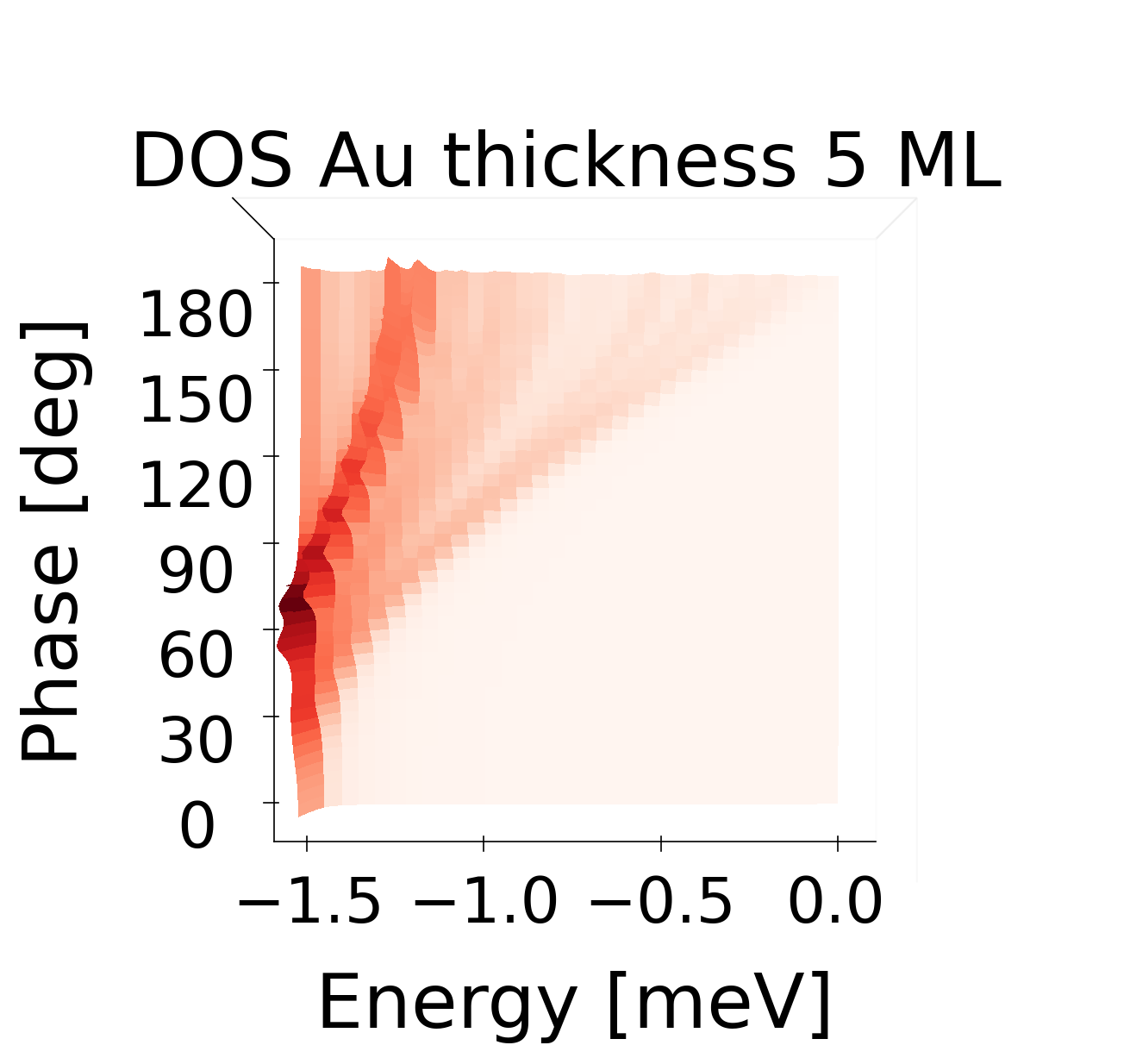}\\
    \includegraphics[width=0.25\linewidth]{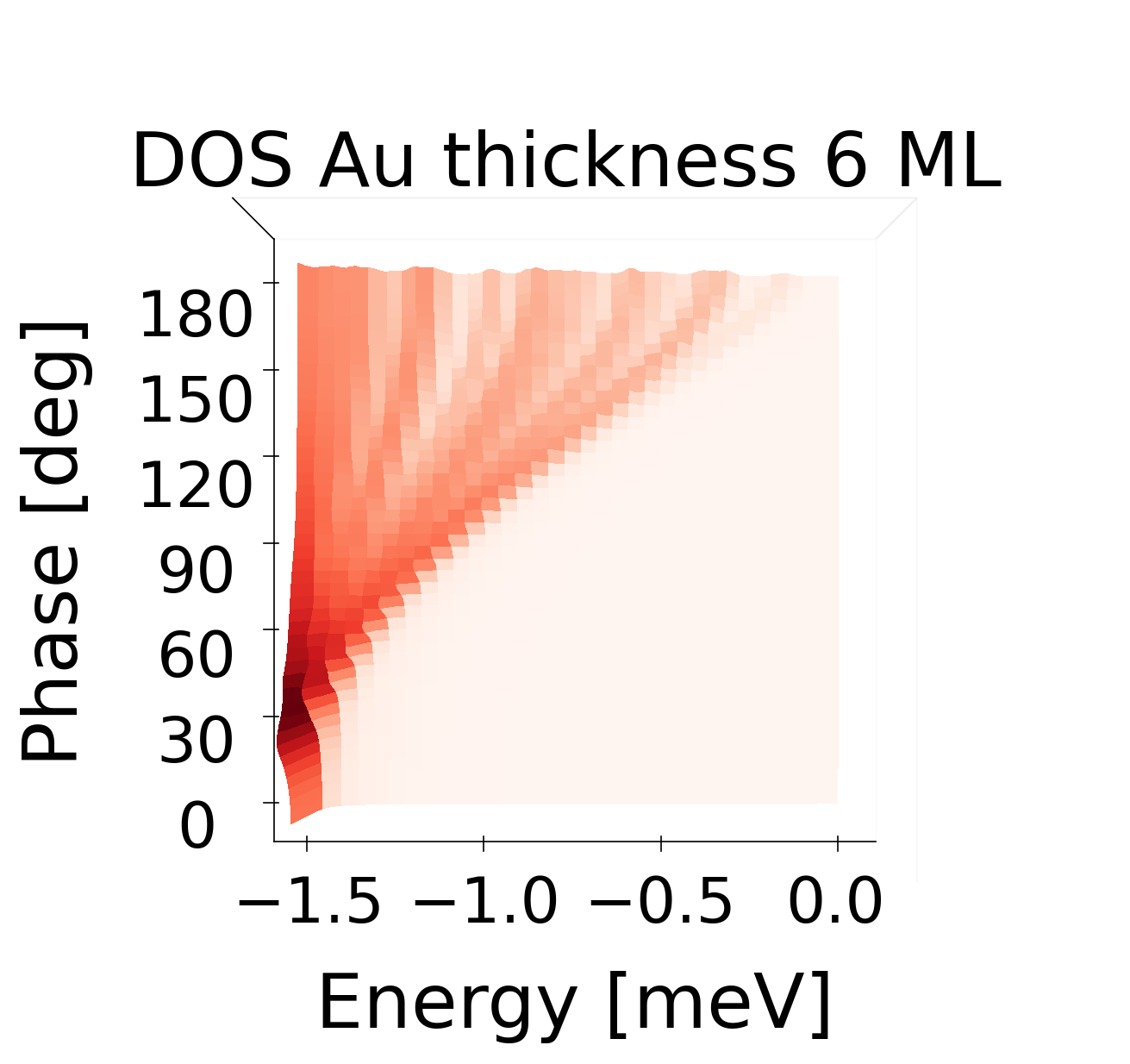}
    \includegraphics[width=0.25\linewidth]{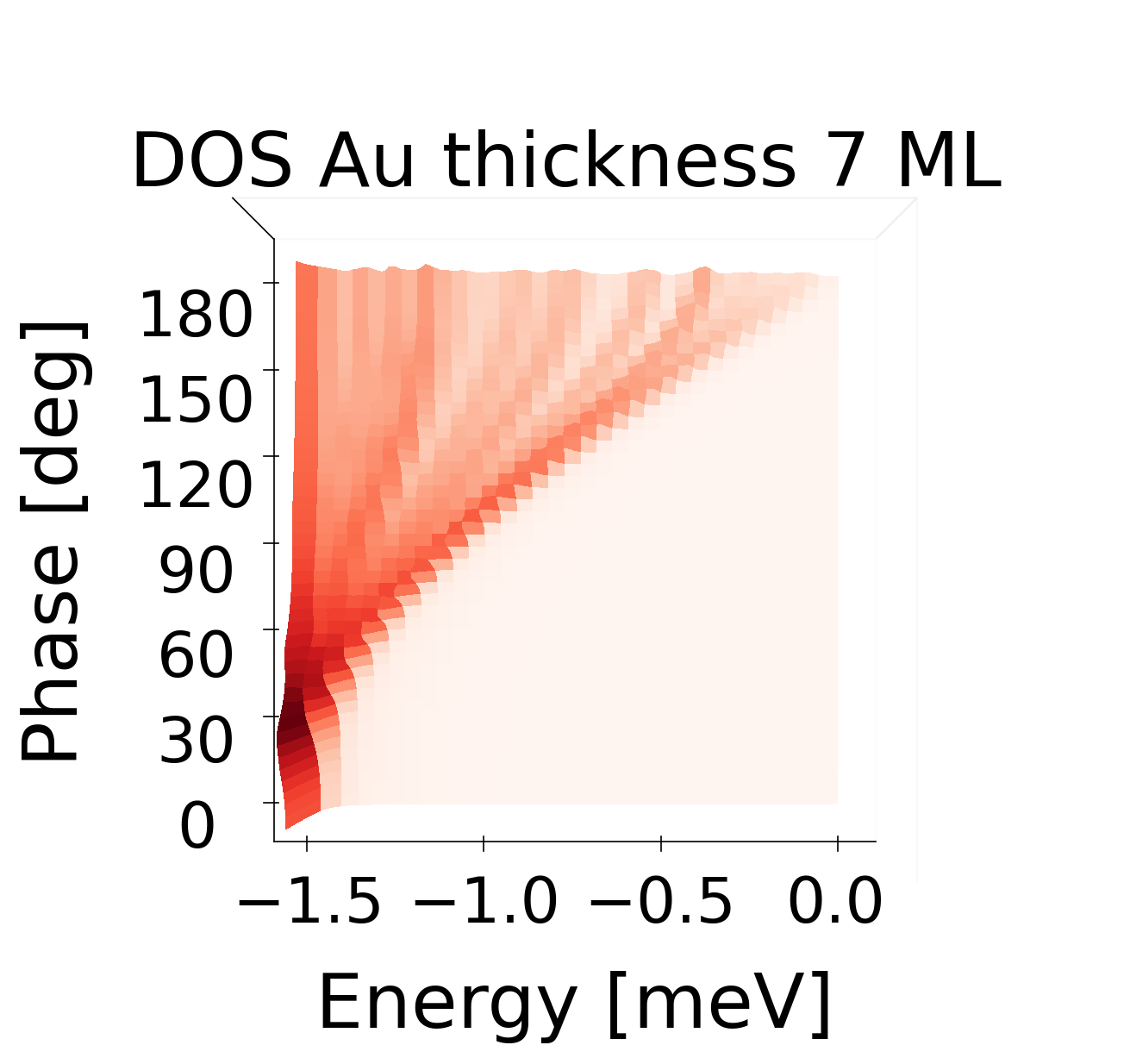}
    \includegraphics[width=0.25\linewidth]{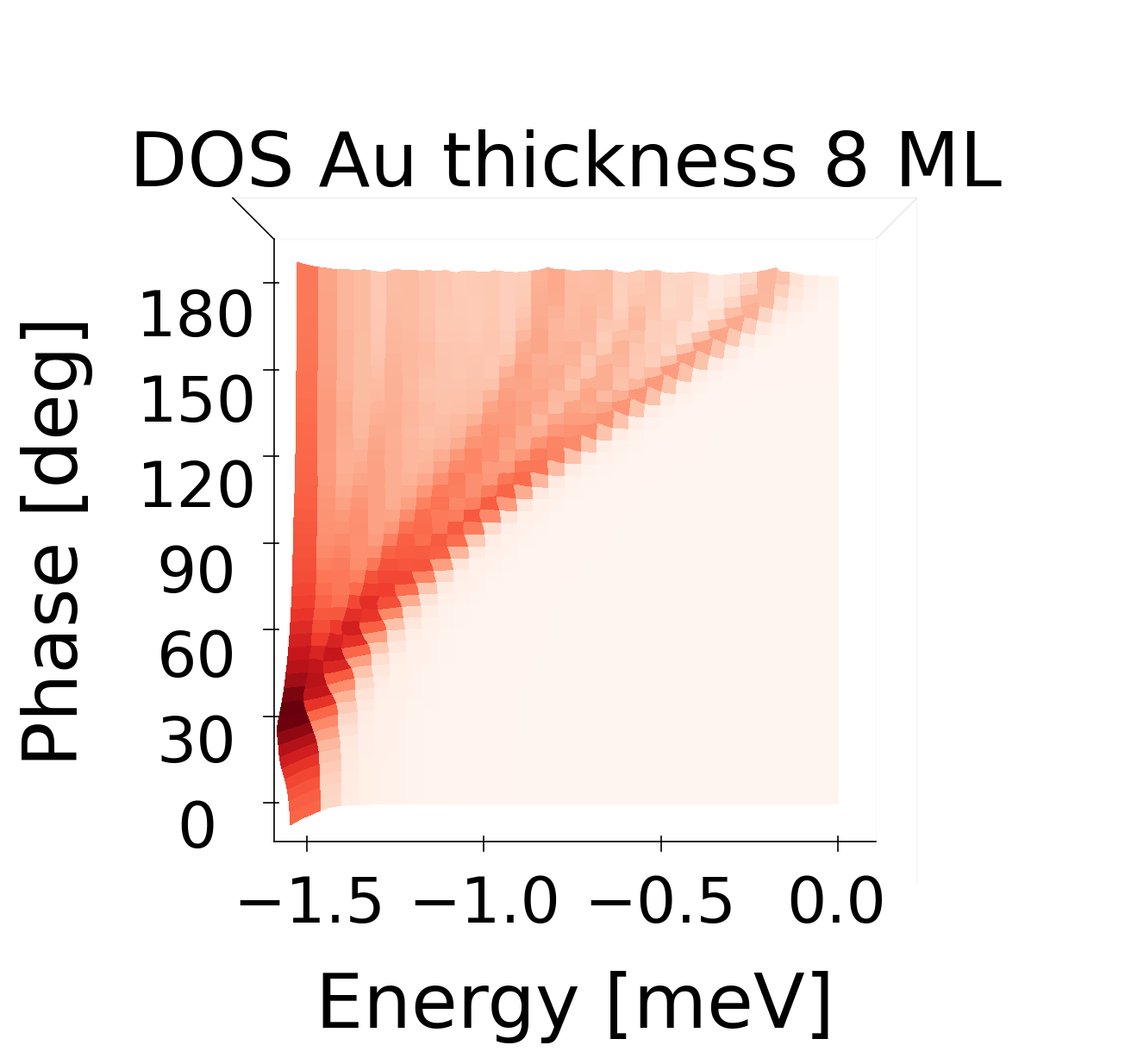}\\
    \includegraphics[width=0.25\linewidth]{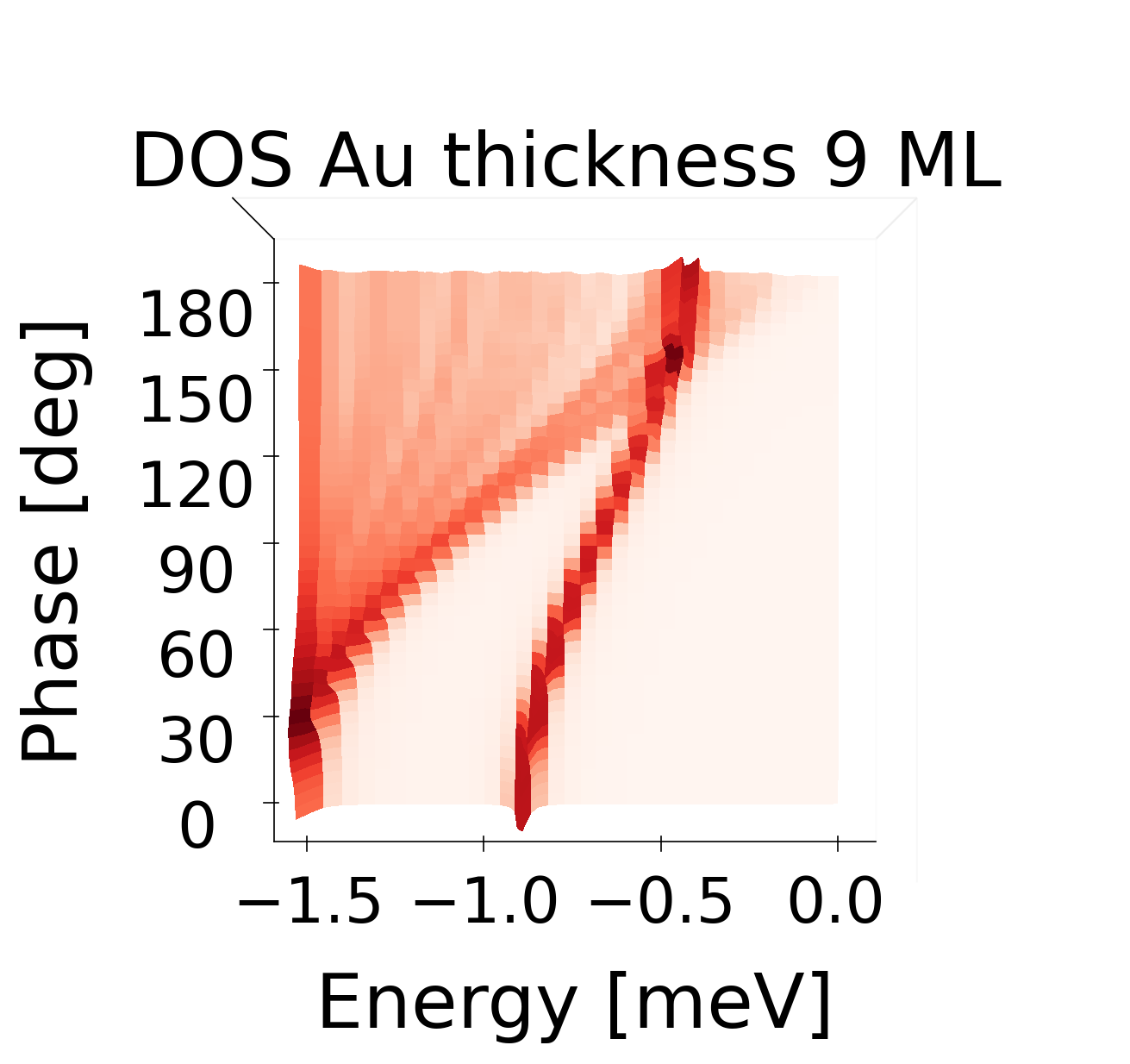}
    \includegraphics[width=0.25\linewidth]{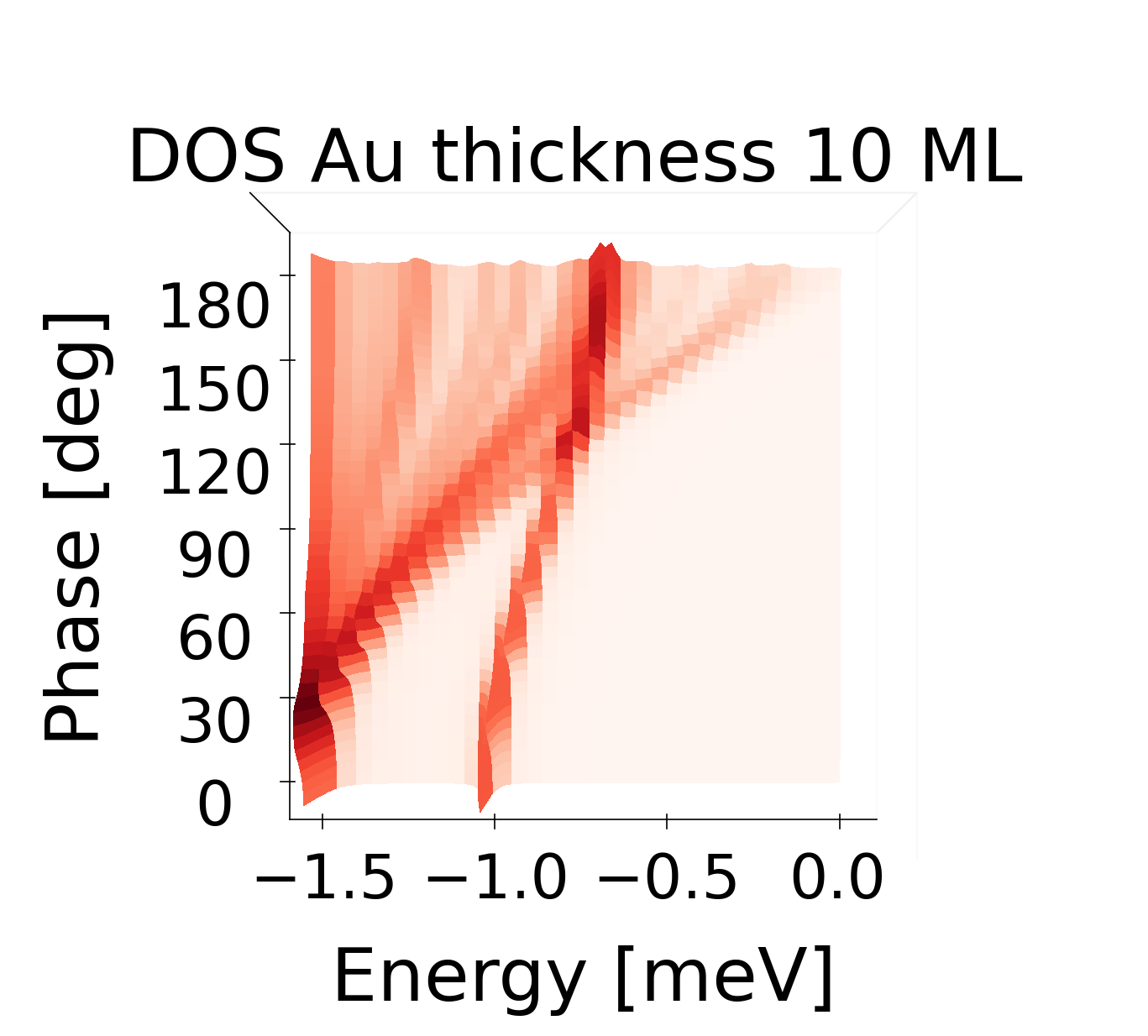}
    \includegraphics[width=0.25\linewidth]{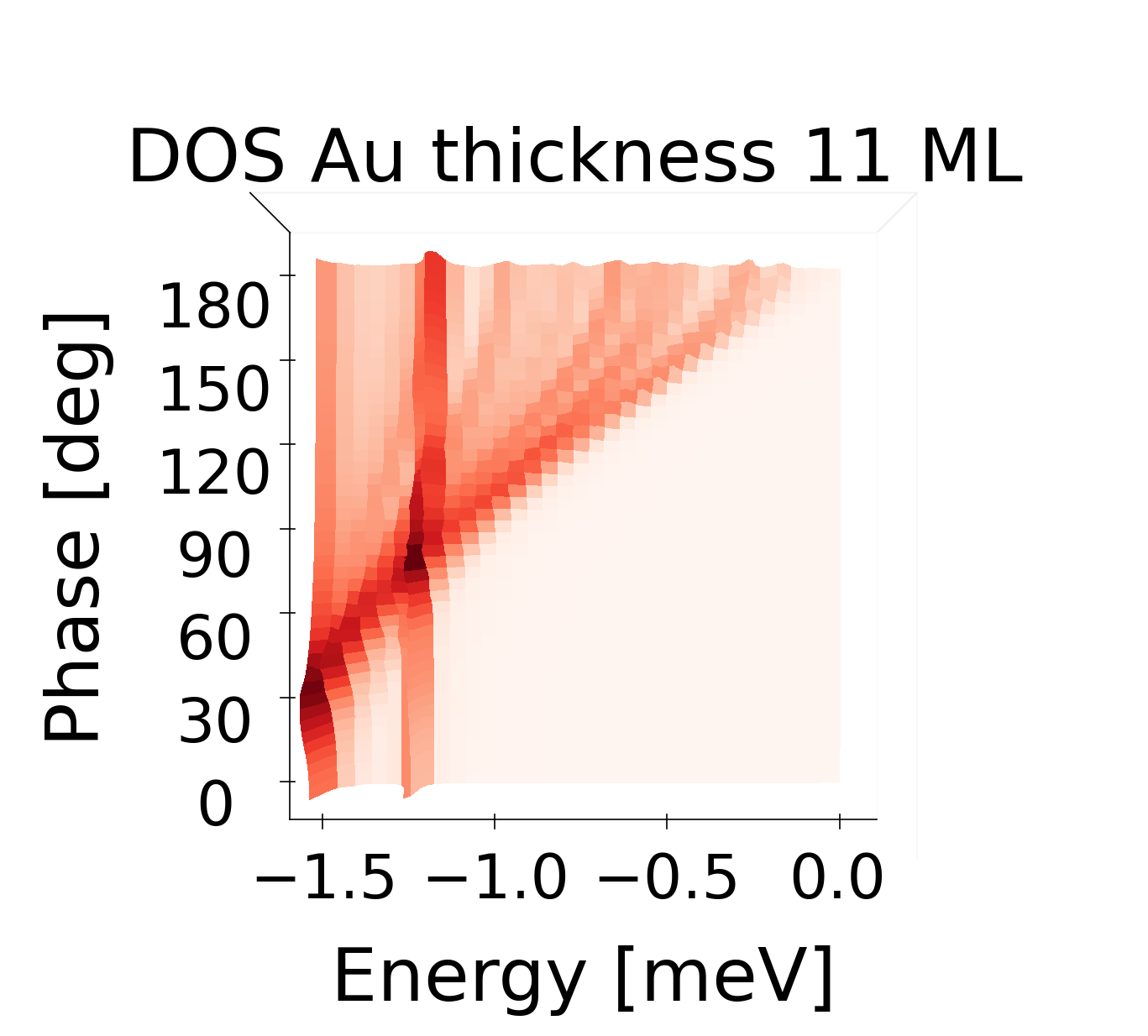}\\
    \includegraphics[width=0.25\linewidth]{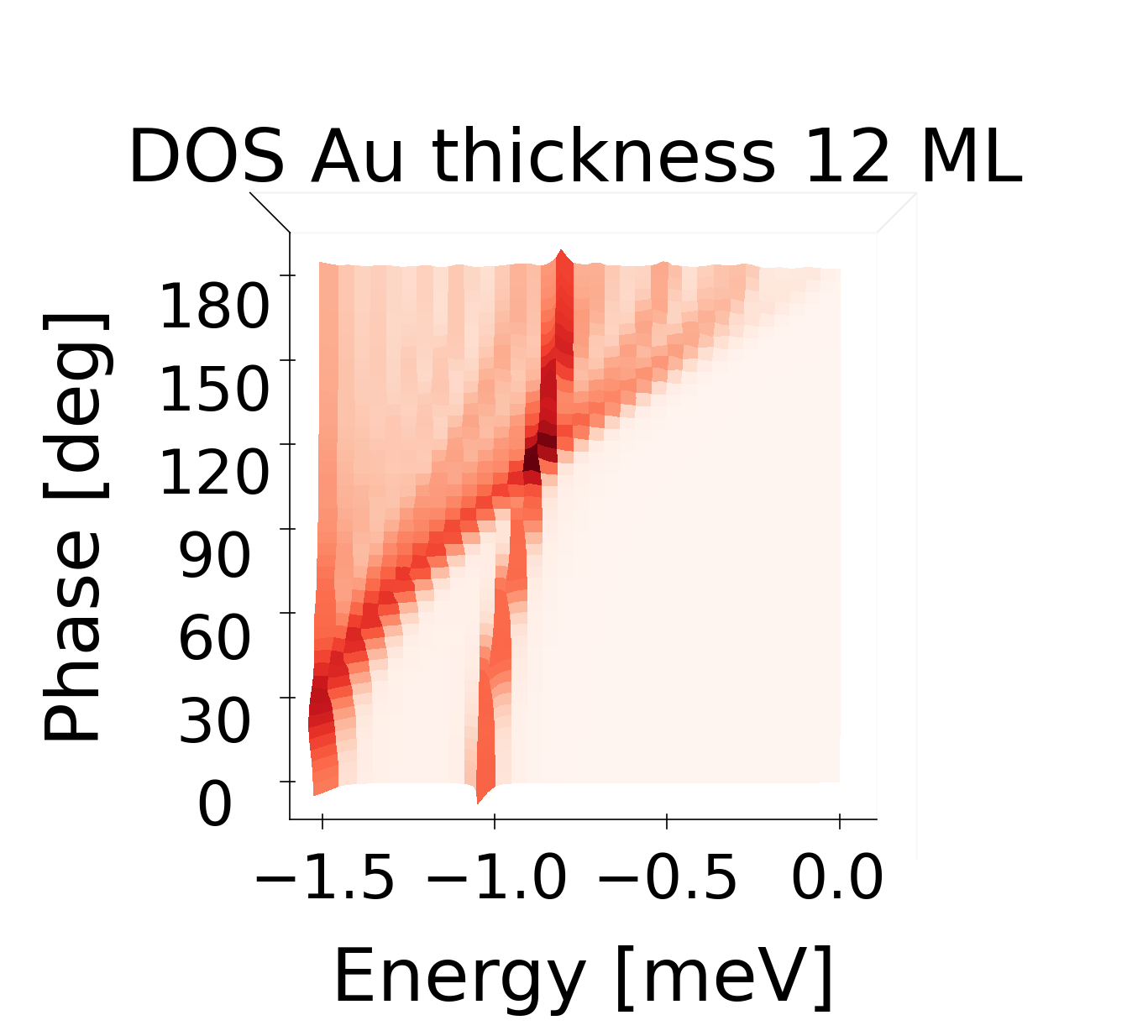}
    \caption{Full quasi-particle DOS for various gold thicknesses}
    \label{fig:qp_vl}
\end{figure*}

Here, instead of the spectral properties, we focus on the integrated quasi-particle DOS.
In Fig.~\ref{fig:qp_l6} and Fig.~\ref{fig:qp_l9} we show the layer resolved DOS
for gold thickness $t_\textrm{Au}$=6~ML and $t_\textrm{Au}$=9~ML, respectively.
For the system with $t_\textrm{Au}$=6~ML no significant change can be observed in the layer resolution if it is viewed as a sequence.
However, the system with $t_\textrm{Au}$=9~ML shows a significant different feature, the emergence of an additional bound state.
As shown in Fig.~\ref{fig:qp_vl}, where the full quasi-particle DOS is plotted for all the investigated Au thicknesses,
this bound state appears exactly at $t_\textrm{Au}$=9~ML and persist for larger Au thicknesses. 
Here one can make a connection with the feature seen for the BSF with $t_\textrm{Au} \geq $9~ML at the edge of the BZ:
the Fermi wave vector mismatch between the many subbands are compensated by the nonzero center-of-mass momentum of the Cooper pairs
(leading to the additional bound state) and causing a non-uniform pairing in the system (this is supported by Fig. 3(f) in the main text).

\clearpage
\section{Isotropy of the critical current density}
As we argued in the main text the critical current density does not show significant anisotropic behavior.
To further test the argument, we have calculated the critical current density from the density-functional DBdG theory by introducing
the vector potential $\vec{A}=(A_x,0,0)$ in the in-plane direction as described in Ref.~\cite{Csire2018}.
A similar calculation of the current density has been carried out within the tight-binding approach\cite{Krawiec2002,Krawiec2004},
while here we extend the approach to the ab initio framework.

\begin{figure}[htb!]
    \centering
    \includegraphics[width=0.95\linewidth]{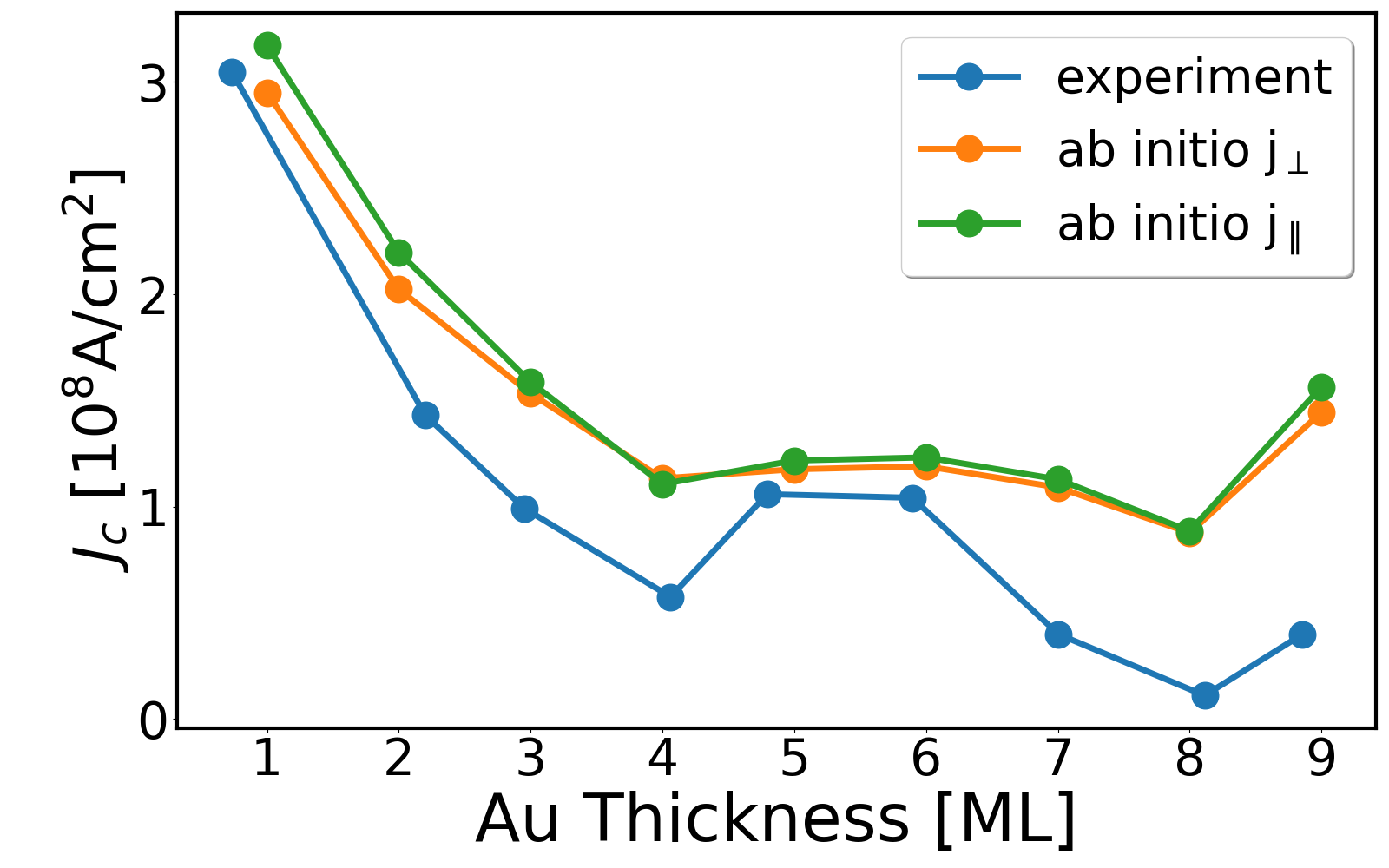}
    \caption{The calculated in-plane and out-of-plane critical current densities resembling isotropic behavior compared to the critical current estimated based on Bean's model from the experiments ($T=6$~K).}
    \label{fig:cc}
\end{figure}

The critical current density can be calculated from the following expression of the quasi-particle current
\begin{align}
\nonumber
   j^{x} = 
 -&\frac{e c}{\pi} \operatorname{Im} \operatorname{Tr} \int_{\text{WS}} \!\!\! d^3r  \int_{-\infty}^{\infty} d\varepsilon \, (\overrightarrow{\boldsymbol \alpha}^x - \overleftarrow{\boldsymbol \alpha}^x)\\ &\left[ f(\varepsilon) G^{ee}(\mathbf{r}, \mathbf{r}',\varepsilon) \Big|_{\mathbf{r}'= \mathbf{r}}  -(1-f(\varepsilon))G^{hh}(\mathbf{r}, \mathbf{r}',\varepsilon) \Big|_{\mathbf{r}'= \mathbf{r}} \right] .
\end{align}
when superconductivity vanishes due to the effect of the in-plane vector potential.
For the calculation of matrix-elements considering the $\boldsymbol \alpha$ matrices, see Ref.~\cite{Banhart1998}.
The result is shown in Fig.~\ref{fig:cc}, where we compare the in-plane and perpendicular critical current densities and find them to be in excellent agreement, confirming that the anisotropy is negligible for practical considerations.

\end{document}